\documentclass[preprintnumbers,prd,showpacs,floatfix,superscriptaddress,nofootinbib,twocolumn, letterpaper]{revtex4-1}
\pdfoutput=1

\usepackage{longtable}
\usepackage{bm}
\usepackage{relsize}
\usepackage{amsfonts}
\usepackage{amsmath}
\usepackage{amssymb,epsf}
\usepackage{latexsym}
\usepackage{graphicx,epsfig}
\usepackage{amssymb}
\usepackage{float}
\usepackage{subfigure}
\usepackage{epstopdf}
\usepackage[colorlinks=true,citecolor=blue,linkcolor=blue,urlcolor=black]{hyperref}
\usepackage{dcolumn}
\usepackage{psfrag}
\usepackage{wrapfig}
\usepackage{makeidx}
\usepackage{epsf}
\usepackage{color}
\usepackage{multirow}
\usepackage{mathtools}

\begin{document}

\title{Observational properties of relativistic fluid spheres with thin accretion disks}

\author{João Luís Rosa}
\email{joaoluis92@gmail.com}
\affiliation{Institute of Physics, University of Tartu, W. Ostwaldi 1, 50411 Tartu, Estonia}
\affiliation{University of Gda\'{n}sk, Jana Ba\.{z}y\'{n}skiego 8, 80-309 Gda\'{n}sk, Poland}

\date{\today}

\begin{abstract} 
In this work we analyze the observational properties of incompressible relativistic fluid spheres with and without thin-shells, when surrounded by thin accretion disks. We consider a set of six configurations with different combinations of the star radius $R$ and the thin-shell radius $r_\Sigma$ to produce solutions with neither thin-shells nor light-rings, with either of those features, and with both. Furthermore, we consider three different models for the intensity profile of the accretion disk, based on the Gralla-Lupsasca-Marrone (GLM) disk model, for which the peaks of intensity occur at the Innermost Stable Circular Orbit (ISCO), the Light-Ring (LR), and the center of the star. The observed images and intensity profiles for an asymptotic observer are produced using a Mathematica-based ray-tracing code. Our results indicate that, in the absence of a light-ring, the presence of a thin-shell produces a negligible effect in the observational properties of the stars. However, when the spacetime features a light-ring, the portion of the mass of the star that is stored in the thin-shell has a strong effect on its observational properties, particularly in the magnitude of the central gravitational redshift effect responsible for causing a central shadow-like dimming in the observed images. A comparison with the Schwarzschild spacetime is also provided and the most compact configurations are shown to produce observational imprints similar to those of black-hole solutions, with subtle qualitative differences, most notably extra secondary image components that decrease the radius of the shadow and are potentially observable.
\end{abstract}


\maketitle

\section{Introduction}\label{sec:intro}

In the last few years, several high-precision experiments in gravitational physics, namely the LIGO/Virgo gravitational wave detectors \cite{LIGOScientific:2020ibl,LIGOScientific:2016aoc} and long baseline interferometers e.g. the Event Horizon Telescope (EHT) \cite{EventHorizonTelescope:2019dse,EventHorizonTelescope:2022wkp,EventHorizonTelescope:2020qrl} and the GRAVITY instrument of the European Southern Observatory (ESO) \cite{GRAVITY:2020gka,GRAVITY:2020lpa}, have contributed massively to a deeper understanding of the strong-field regime of gravity, providing a formidable framework on which to analyze several unsolved issues in modern physics, see \cite{Barack:2018yly} for a review. In particular, the hypothesis that a full gravitational collapse leads to the formation of a black-hole (BH) \cite{Oppenheimer:1939ue} can be analyzed with the help of these observations.

The observations mentioned above seem to be consistent with the hypothesis that the outcome of a full gravitational collapse settles down under a Kerr BH described my a given mass and angular momentum. Indeed, photons that approach a BH with a small enough impact parameter will be captured by the event horizon and never reach an asymptotic observer, thus resulting in the appearance of a shadow \cite{Luminet:1979nyg,Falcke:1999pj,Gralla:2020srx,Cunha:2018acu,Cardoso:2021sip,Vincent:2022fwj}, a feature that the EHT collaboration has successfully observed experimentally \cite{EventHorizonTelescope:2019dse,EventHorizonTelescope:2022wkp}. However, the collapse of matter into a BH unavoidably leads to the formation of singularities \cite{Penrose:1964wq,Penrose:1969pc}, i.e., geodesic incomplete regions of the spacetime. Singularities are poorly understood at a fundamental level and represent an important drawback of the BH hypothesis. It is thus natural to ask the question: are there any alternatives to the BH hypothesis that reproduce the same observation but do not feature this inconvenient property?

Several alternatives to the black-hole hypothesis have been proposed, including self gravitating fundamental fields \cite{Liebling:2012fv,Macedo:2013jja,Brito:2015pxa,Berti:2019wnn,Carloni:2019cyo}, relativistic perfect fluids \cite{Buchdahl:1959zz,Rosa:2020hex,Raposo:2018rjn,Cardoso:2015zqa}, gravastars \cite{Mazur:2004fk,Chirenti:2007mk}, black bounces \cite{Guerrero:2021ues,Olmo:2021piq}, wormholes \cite{Guerrero:2022qkh}, among others. We refer to \cite{Cardoso:2019rvt} for an extensive review. Interestingly, several of these so-called BH mimickers were shown to feature observational properties similar to those of BHs by e.g. casting a shadow \cite{Vincent:2015xta,Rosa:2022tfv,Guerrero:2022msp} or via astrometric observables \cite{Rosa:2022toh}. The properties of the observed shadow depend strongly on both the geometry of the background spacetime and the astrophysical properties of the accretion disk surrounding the central object \cite{Lara:2021zth,Wielgus:2021peu,Vincent:2020dij}, and thus provide an ideal framework to test the viability of BH mimickers against the BH hypothesis.

In this work we study a particular kind of BH mimicker belonging to the class of relativistic perfect fluids mentioned above, consisting of a family of incompressible fluid spheres supported by thin-shells \cite{Rosa:2020hex}. Configurations belonging to this family of solutions with a compactness arbitrarily close to that of a black hole, and thus featuring similar properties e.g. an unstable light-ring and Innermost Stable Circular Orbit (ISCO), have been shown to be linearly stable against radial perturbations and to be composed of physically relevant (non-exotic) matter. The study of the observational properties of these solutions allows us not only to assert their relevance as suitable BH mimickers but also to analyze how the presence of a thin-shell affects the observables of a compact object.

This paper is organized as follows. In Sec.\ref{sec:theory} we briefly review the family of solutions considered and analyze their geodesic properties; in Sec. \ref{sec:shadows} we introduce the models for the intensity profiles of the accretion disks, produce the observed shadow images and observed intensity profiles for several combinations of configurations with and without thin-shells and light-rings, and compare the results with the same predictions for the Schwarzschild spacetime; and in Sec.\ref{sec:concl} we trace our conclusions. A more detailed analysis of the effects of the thin-shell and the light-ring on the observables is provided in Appendix \ref{sec:appendixA}. We adopt a system of geometrized units for which $G=c=1$, where $G$ is the gravitational constant and $c$ is the speed of light

\section{Theory and framework}\label{sec:theory}

\subsection{Geometry and matter contents}

Relativistic fluid spheres have been studied under several frameworks and assumptions. In this section, we summarize our own assumptions and introduce the models for the configurations analyzed in the following sections. In what follows, we consider the usual spherical coordinates $x^\mu=\left(t,r,\theta,\phi\right)$.

In this work we restrict our analysis to static and spherically symmetric spheres of incompressible fluid. These configurations consist of two regions: an interior region populated by the relativistic perfect fluid in the range $r<r_\Sigma$, and an exterior vacuum region in the range $r>r_\Sigma$, where $r_\Sigma$ denotes the radius of the spherical hypersurface that separates the two regions. A well-known example of such a model is the Schwarzschild fluid star \cite{Buchdahl:1959zz}, although in this work we extend the analysis to more complicated configurations, as clarified in what follows. The interior region is described by the line element $ds^2_-$ as
\begin{eqnarray}\label{eq:metric_int}
    ds^2_ -&=&-\frac{1}{4}\left(3\sqrt{1-\frac{2M}{R}}-\sqrt{1-\frac{2r^2M}{R^3}}\right)^2dt^2+\\
    &+&\left(1-\frac{2r^2M}{R^3}\right)^{-1}dr^2+r^2\left(d\theta^2+\sin^2\theta d\phi^2\right),\nonumber
\end{eqnarray}
where $M$ is the total mass of the fluid star and $R$ is the radius of the star. For the Schwarzschild fluid star, the radius of the star $R$ coincides with the radius of the separation $r_\Sigma$. In the interior region, the matter contents are described by an isotropic relativistic perfect fluid with a stress-energy tensor $T_{ab}$ of the form
\begin{equation}\label{eq:def_tab}
    T_a^b=\text{diag}\left(-\rho,p,p,p\right),
\end{equation}
where $\rho=3M/(4\pi R^3)$ is the constant energy density, consistently with the assumption of incompressibility of the fluid, and $p=p\left(r\right)$ is the isotropic pressure given in terms of the radial coordinate as
\begin{equation}\label{eq:pressure}
    p\left(r\right)=\rho\frac{\sqrt{1-\frac{2r^2M}{R^3}}-\sqrt{1-\frac{2M}{R}}}{3\sqrt{1-\frac{2M}{R}}-\sqrt{1-\frac{2r^2M}{R^3}}}.
\end{equation}
Note that at the surface $r=R$ the pressure $p\left(R\right)=0$ vanishes, as it happens for the Schwarzschild fluid star. An interesting property of Eq.\eqref{eq:pressure} is the fact that the central pressure $p_c\equiv p\left(0\right)$ diverges for a star radius $R=9M/4\equiv R_b$, which corresponds to a curvature singularity of the Ricci scalar. This property is known as the Buchdahl limit, and $R_b$ is known as the Buchdahl radius \cite{Buchdahl:1959zz}.

Regarding the exterior vacuum region, it is well-described by the Schwarzschild spacetime, i.e., the line-element $ds^2_+$ is given by
\begin{eqnarray}\label{eq:metric_ext}
ds^2_+&=&-\left(1-\frac{2M}{r}\right)dt^2+\left(1-\frac{2M}{r}\right)^{-1}dr^2+\\
&+&r^2\left(d\theta^2+\sin^2\theta d\phi^2\right).\nonumber
\end{eqnarray}
To perform the matching between the interior and the exterior spacetimes, one recurs to the junction conditions \cite{darmois,Israel:1966rt}, a set of conditions that the geometrical properties and matter fields of both spacetimes must satisfy to guarantee that their union is itself a solution of the field equations. In General Relativity (GR), there are two junction conditions: the induced metric at the separation hypersurface, $h_{ab}=g_{\mu\nu}e^\mu_a e^{\nu}_b$, where $e^\mu_a= dx^\mu/dy^a$ are the projection vectors from the four-dimensional manifold described by a set of coordinates $x^\mu$ into a three-dimensional hypersurface described by a set of coordinates $y^a$, and the extrinsic curvature of the separation hypersurface, $K_{ab}=e^\mu_a e^{\nu}_b\nabla_\mu n_\nu$, where $\nabla_\mu$ denotes covariant derivatives and $n_\nu$ is the normal vector to the separation hypersurface, must be continuous. If the latter condition is not satisfied, a thin-shell of matter is necessary at the separation hypersurface to guarantee the validity of the matching.

\begin{figure}
    \centering
    \includegraphics[scale=0.35]{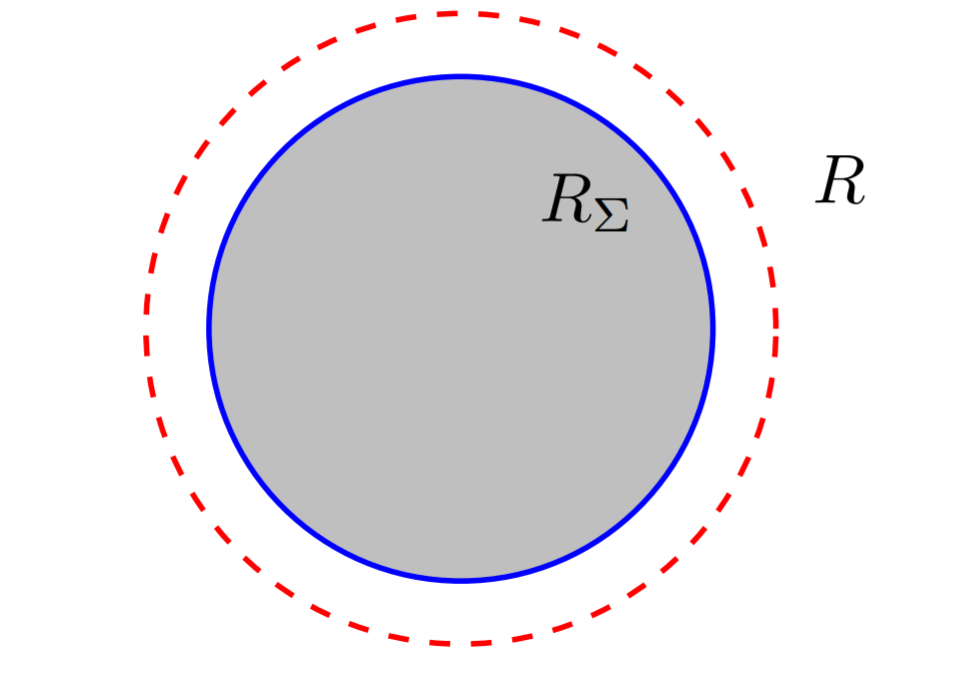}
    \caption{Schematic representation of the relativistic fluid stars supported by thin-shells. The interior fluid region (gray) is separated from the exterior vacuum region (white) by a thin shell of matter (solid blue line) at $r=r_\Sigma<R$, where $R$ is the initial radius of the star (dashed red line).}
    \label{fig:star}
\end{figure}

The two metrics in Eqs.\eqref{eq:metric_int} and \eqref{eq:metric_ext}, as well as their respective Lie derivatives, are continuous at $r=R$. Hence, the matching between the interior and the exterior regions for the Schwarzschild star is smooth, i.e., no thin-shell of matter at the separation hypersurface $R=r_\Sigma$ is necessary to preserve the regularity of the full spacetime solution. However, the situation $R=r_\Sigma$ is a mere particular case of a broader family of relativistic fluid star solutions with $r_\Sigma<R$. In this case, the total mass of the star $M$ and the interior fluid density $\rho$ remain the same, but the exterior layers of the star are compressed into a thin-shell standing at the radius $r=r_\Sigma$. A schematic representation of these configurations if provided in Fig. \ref{fig:star}.

In a previous work \cite{Rosa:2020hex} the broader class of solutions with $r_\Sigma<R$ was analyzed and several important conclusions were traced, namely: (i) for a wide region of the parameter space, these solutions were found to satisfy all of the energy conditions; (ii) for a wide region of the parameter space, the configurations supported by thin-shells, i.e., with $R\neq r_\Sigma$, were proven to be linearly stable against radial perturbations; and (iii) the radius of the separation hypersurface $r_\Sigma$ can be arbitrarily close to the Schwarzschild radius $r_s=2M$ without developing an interior curvature singularity, thus avoiding the previously mentioned Buchdahl limit. These three properties emphasize the validity of these family of solutions as suitable and physically relevant alternatives to the black-hole scenario, and thus motivates further study on their observational properties. In the following sections, we analyze the appearance of optically thin accretion disk models, i.e., models that take as an assumption that the accretion disk is transparent to its own radiation, around relativistic fluid star configurations belonging to this family of solutions.

\subsection{Geodesic motion and ray-tracing}

The motion of test particles, either massless (photons) or massive, in a background spacetime geometry is described by the geodesic equation. This equation can be obtained via the variational method from a Lagrangian density of the form $\mathcal L=g_{\mu\nu}\dot x^\mu\dot x^\nu=-\delta$, where a dot denotes a derivative with respect to the affine parameter $\lambda$ along the geodesics and the constant $\delta$ assumes the values $\delta=1$ for timelike (massive) test particles or $\delta=0$ for null (massless) test particles. In the particular case of spherically symmetric background spacetimes like the ones considered in this work, one can consider the geodesic movement solely along the equatorial plane, i.e., the plane defined by $\theta=\pi/2$, without loss of generality, which identically solves the $\theta$ component of the geodesic equation. Under this assumption, the $t$ and $\phi$ components of the geodesic equation define two conserved quantities, namely the energy per unit mass $E=-g_{tt}\dot t$, and the angular momentum per unit mass $L=r^2\dot \phi$, respectively. Finally, under an appropriate re-scaling of the affine parameter along the geodesics, the radial component of the geodesic equation may be written in the form of the equation of motion of a particle moving along a one-dimensional potential $V\left(r\right)$ as
\begin{equation}\label{eq:eom_geodesic}
\dot r^2=\frac{V\left(r\right)}{\sqrt{-g_{tt}g_{rr}}},
\end{equation}
\begin{equation}\label{eq:def_potential}
V\left(r\right)=E^2+g_{tt}\left(\frac{L^2}{r^2}+\delta\right).
\end{equation}
Equations \eqref{eq:eom_geodesic} and \eqref{eq:def_potential} allow one to deduce several interesting properties of the geodesic structure of the background geometry. In particular, one can analyze the stability regimes of the circular orbital motion and verify if the background spacetime features privileged circular curves, like Innermost Stable Circular Orbits (ISCOs), i.e., a marginally stable circular orbit for massive test particles at a radius $r=r_{ISCO}$, which marks a transition point between stable $r>r_{ISCO}$ and unstable $r\lesssim r_{ISCO}$ orbits, and Light-Rings (LRs), i.e., circular orbits for massless test particles at a radius $r=r_{LR}$.

\subsubsection{Circular orbits for massive test particles}

\begin{figure}
    \centering
    \includegraphics[scale=0.8]{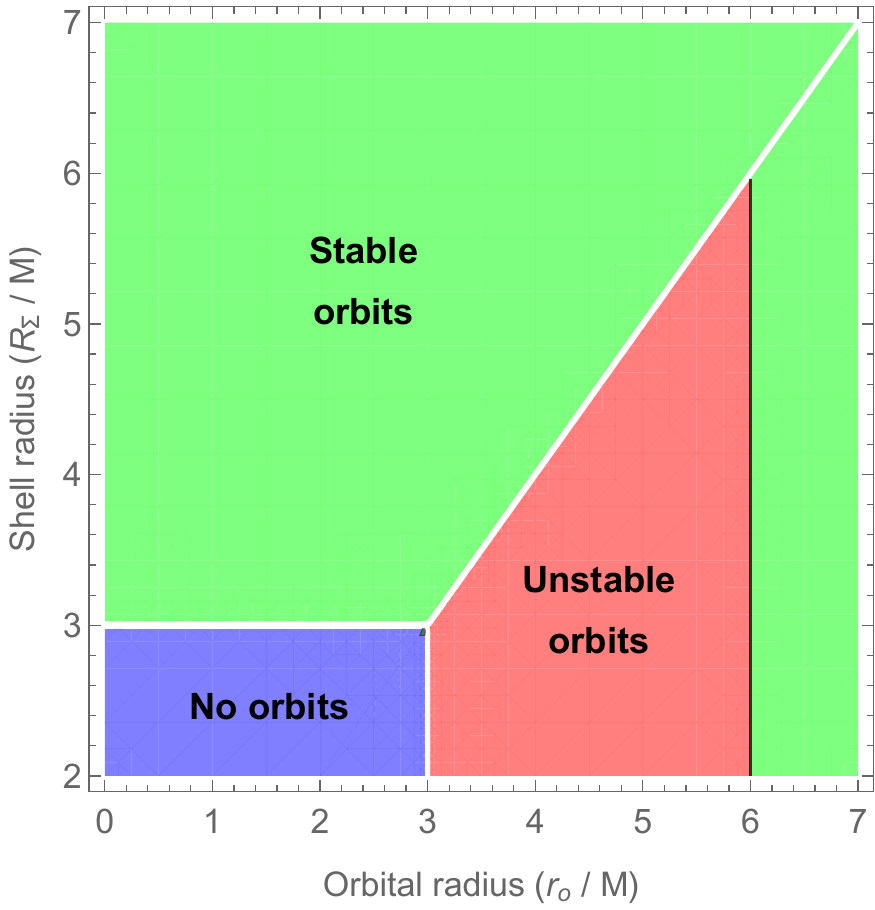}
    \caption{Existence and stability of circular orbits for massive test particles as a function of the shell radius $r_\Sigma$ and the orbital radius $r_o$. Circular orbital motion is possible in the green and red regions, being respectively stable and unstable in these regions. Circular orbital motion is not possible in the blue region.}
    \label{fig:orbits}
\end{figure}

Circular orbits are characterized by $\dot r=\ddot r = 0$ which, upon a replacement into Eq.\eqref{eq:eom_geodesic} and its first derivative with respect to $\lambda$, corresponds to $V\left(r\right)=0$ and $V'\left(r\right)=0$. Using Eq.\eqref{eq:def_potential}, these two constraints on $V\left(r\right)$ allow one to obtain the values of $E$ and $L$ consistent with a circular orbit. Finally, taking $\delta=1$, stable orbits for massive test particles are defined by the condition $V''\left(r\right)>0$. In the Schwarzschild spacetime, the quantity $V''\left(r\right)$ is positive for $r>6M$ and negative for $3M<r<6M$, thus implying that $r_{ISCO}=6M$. However, in the relativistic fluid sphere spacetimes that we are interested in, one can find several alternatives to this scenario, namely:
\begin{enumerate}
    \item If $r_\Sigma>6M$, one verifies that $V''\left(r\right)>0$ for the entire range of the radial coordinate $r$. This implies that these solutions do not feature an ISCO, and all circular orbits for massive test particles are stable independently of the orbital radius;
    \item If $r_\Sigma < 3M$, one verifies that $V''\left(r\right)$ is positive for $r>6M$, negative for $3M<r<6M$, and complex for $r<3M$. This implies that the spacetime features an ISCO at $r_{ISCO}=6M$, orbits for massive test particles are stable for $r>6M$, unstable in the interval $3M<r<6M$, and nonexistent for $r<3M$ ,i.e., the same circular orbit regimes as in the Schwarzschild spacetime;
    \item If $3M<r_\Sigma<6M$, one verifies that $V''\left(r\right)>0$ in two regions, namely for $r>6M$, and for $r<r_\Sigma$. This implies that although there is an ISCO at $r_{ISCO}=6M$, circular orbital motion for massive test particles is allowed in the entire complementary region $r<6M$, and these circular orbits are only unstable in the interval $r_\Sigma<r<6M$.
\end{enumerate}
A summary of the orbital properties for different values of $r_\Sigma$ and the orbital radius $r_o$ is provided in Fig. \ref{fig:orbits}. These orbital properties are independent of the initial radius of the star $R$, and are fully characterized by the radius of the shell $r_\Sigma$. The limit $r_\Sigma=R$ follows naturally and preserves the results. Furthermore, note that in this analysis we have neglected a possible non-conservative interaction between the test particles and the fluid, which could render unstable any orbit with an orbital radius $r_o<r_\Sigma$.

\subsubsection{Circular orbits for massless test particles}

\begin{figure}
    \centering
    \includegraphics[scale=0.8]{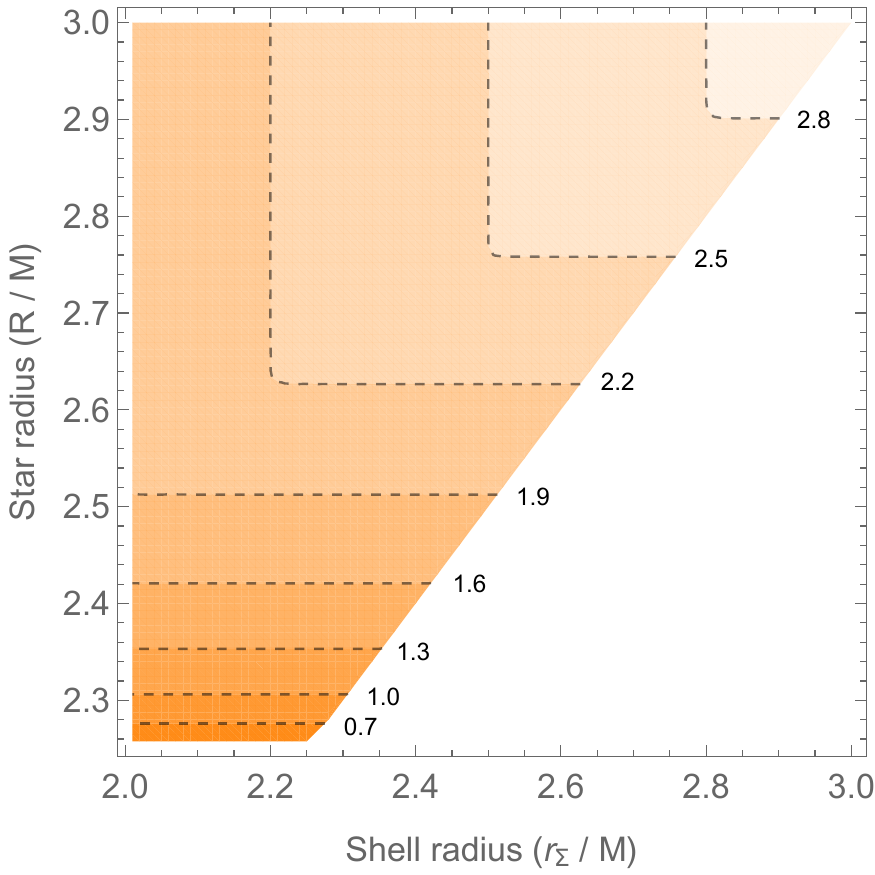}
    \caption{Radius of the stable light-ring $\bar r_{LR} / M$ as a function of the radius of the star $R$ and the radius of the thin-shell $r_\Sigma$.}
    \label{fig:lightrings}
\end{figure}

Considering now massles test particles, i,.e., $\delta=0$, the radii of the LRs, if any exist, can be obtained by taking a derivative of Eq. \eqref{eq:def_potential}, impose the constraint $V'\left(r\right)=0$ and solving for $r$. In the Schwarzschild spacetime, a single LR with a radius $r_{LR}=3M$ is present. A recent theorem published in Ref.\cite{Cunha:2017qtt} proves that in regular ultracompact spacetimes spacetimes, i.e., spacetimes on which event horizons and singularities are absent, LRs manifest in possibly degenerate pairs. Indeed, the relativistic fluid sphere spacetimes considered in this work are an example of such a case, and several alternatives might arise:
\begin{enumerate}
    \item If $r_\Sigma>3M$, one verifies that the condition $V'\left(r\right)=0$ has an empty set of solutions. As such, the spacetime does not feature any LRs and circular orbits for massless test particles are nonexistent;
    \item If $r_\Sigma=3M$, one verifies that the condition $V'\left(r\right)=0$ features a single solution corresponding to a LR at a radius $r_{LR}=3M$, which corresponds to a saddle point of the potential $V\left(r\right)$. This corresponds to the particular case for which the pair of LRs mentioned previously are degenerate, and a single unstable circular orbit exists for massless test particles;
    \item If $2M<r_\Sigma<3M$ and $R>R_b$, one verifies that the condition $V'\left(r\right)=0$ features two independent solutions, corresponding to the previously mentioned pair of LRs: an unstable LR at a radius $r_{LR}=3M$, corresponding to a local minimum of $V\left(r\right)$, and a stable LR at a radius $r=\bar r_{LR}<3M$, corresponding to a local maximum of $V\left(r\right)$, where the value of $\bar r_{LR}$ depends on the values of $r_\Sigma$ and $R$;
    \item If $R<R_b$, one verifies that the condition $V'\left(r\right)=0$ features again a single solution corresponding to a LR at a radius $r_{LR}=3M$, corresponding to a global minimum of $V\left(r\right)$. In this situation, a single unstable circular orbit exists for massless test particles. This does not correspond to a degenerate case as the background spacetime is no longer regular (a singularity appears from the violation of the Buchdahl limit, as it is clear from the change in the behavior of $V\left(r_{LR}\right)$ from a saddle point (degenerate case) to a global minimum (non-degenerate case).
\end{enumerate}
For the case 3 enumerated above, the radius $\bar r_{LR}$ of the stable LR depends on either $R$ or $r_\Sigma$ in different regimes. Indeed, if $r_\Sigma$ is below some critical value, say $r_\Sigma < r_c$, then $\bar r_{LR} = r_\Sigma$, whereas if $r_\Sigma > r_c$, then $\bar r_{LR} = r_c$, where the critical value $r_c$ is the radius of the LR of a configuration with the same mass $M$ and radius $R$ but without thin-shell, i.e., with $R=r_\Sigma$. This critical value $r_c$ is given explicitly in terms of $R$ and $M$ as
\begin{equation}\label{eq:radLR}
r_c=R\sqrt{\frac{R}{R_b}}\sqrt{\frac{R-R_b}{R-2M}}.
\end{equation}
Note that when $R=r_\Sigma=3M$ we obtain $r_c=3M$, thus corresponding to the degenerate case $r_{LR}=\bar r_{LR}$ mentioned before. On the other hand, when $R=R_b$, one obtains $r_c=0$, corresponding to the limiting case at which the solution ceases to be regular. The radius $\bar r_{LR}$ is plotted in Fig.\ref{fig:lightrings} as a function of $R$ and $r_\Sigma$ in this parameter region. Note that we do not plot the radius of the unstable LR, i.e., $r_{LR}=3M$, since its value is independent of these parameters. 

Recent works \cite{Cunha:2022gde,Cunha:2023xrt} suggest that the existence of a second stable LR might lead to spacetime instabilities caused by the trapping and accumulation of photons at $r=\bar r_{LR}$. Thus, to avoid potentially unstable configurations, in what follows we restrict our analysis to configurations with at most a single degenerate LR, i.e., $r_\Sigma \geq 3M$.

\subsubsection{Ingoing and outgoing null geodesics}

In the upcoming sections, we recur to a Mathematica-based ray-tracing code to solve the geodesic equation. For this purpose, it is useful to rewrite Eq.\eqref{eq:eom_geodesic} in a more convenient way via the use of the chain rule
\begin{equation}\label{eq:chainrule}
\dot r = \frac{dr}{d\lambda}=\frac{dr}{d\phi}\frac{d\phi}{d\lambda}=\dot \phi\left(\frac{d\phi}{dr}\right)^{-1}.
\end{equation}
Introducing the transformation given in Eq. \eqref{eq:chainrule} into Eq. \eqref{eq:eom_geodesic} and solving with respect to $d\phi/dr$, Eqs.\eqref{eq:eom_geodesic} and \eqref{eq:def_potential} take the form
\begin{equation}\label{eq:raytrace}
    \phi'\left(r\right)=\pm \frac{b}{r^2}\frac{\sqrt{-g_{tt}g_{rr}}}{\sqrt{1+g_{tt}\frac{b^2}{r^2}}},
\end{equation}
where $\pm$ represents ingoing $(-)$ and outgoing $(+)$ geodesics, and where we have introduced the definition of the impact parameter $b\equiv L/E$. Equation \eqref{eq:raytrace} is effectively an ordinary differential equation for $\phi\left(r\right)$ that must be numerically integrated. We start by integrating inwards, i.e., starting with Eq.\eqref{eq:raytrace} with a negative sign, from a chosen numerical infinite $r_\infty\equiv 1000M$ down to the radius of closest approach at which $\phi'\left(r\right)$ diverges. At this point, the first integration stops, the sign of Eq.\eqref{eq:raytrace} is inverted, and a second integration outwards starts. The full integration is completed when the outgoing geodesic hits numerical infinity $r_\infty$. This process is repeated for an appropriate range of the impact parameter $b$ from the peripheral region of the accretion disk down to $b=0$.

\section{Shadows}\label{sec:shadows}

\subsection{Accretion disk models and intensity profiles}

Several different stationary models for the accretion disk and corresponding luminosity profiles can be constructed according to the geodesic structure and orbital stability analysis of the background spacetime, as outlined in the previous section. Indeed, one expects the luminosity profile of the accretion disk to be directly affected by the existence and respective stability of circular orbits, along which the massive particles constituting the accretion disk move. As such, we motivate the choice of the following models for the luminosity profiles:
\begin{enumerate}
    \item The ISCO model: since circular orbits with $r=r_{ISCO}$ are marginally stable, and circular orbits with $r\lesssim r_{ISCO}$ are unstable, i.e., the orbital stability switches its nature at the ISCO, one may argue that the stationary accretion disk should only exist in the region where its constituent particles follow stable orbits. Thus, in this model we consider that the luminosity profile of the accretion disk increases monotonically from infinity down to the region adjacent to the ISCO, where it peaks, and abruptly decays for $r_o<r_{ISCO}$;
    \item The LR model: for the solutions featuring a LR, one can argue that circular orbits in the region $r_{LR}<r_o<r_{ISCO}$ exist, even though they are unstable, and thus one should consider the possibility of extending the inner-edge of the accretion disk all the way down to $r_{LR}$. Thus, in this model we consider that the luminosity profile of the accretion disks increases monotonically from infinity down to a region adjacent to the LR, where it peaks, and abruptly decays for $r_o<r_{LR}$.
    \item The Centre model: since the background spacetimes under consideration are regular, i.e., they feature neither singularities nor event horizons, one can argue that photons being emitted by infalling matter in interior regions of the spacetime, where orbital motion is either unstable (due to friction effects or the concavity of the effective potential) or impossible (as it is the case for solutions with LRs), should still be visible by an asymptotic observer. Thus, in this model we consider that the luminosity profile of the accretion disk increases monotonically from infinity all the way down to $r=0$, where it peaks. 
    \item The EH model: although the background spacetimes under consideration do not have event horizons, an alternative to the Centre disk model must be provided to allow for the comparison between the Schwarzschild solution and the relativistic fluid configurations, as in a black-hole spacetime one can not have any flux of radiation from inside the EH. Thus, in this model we consider that the luminosity profile of the accretion disk increases monotonically from infinity all the way down to a region nearby $r=2M$, where it peaks, and then rapidly decays when approaching the EH.
\end{enumerate}
Note that not all of the models described above are suitable for every possible combination of $r_\Sigma$ and $R$. For example, for models with $r_\Sigma>3M$, which do not feature light-rings, the LR model is not a well-motivated description of the intensity profile of the accretion disk. Nevertheless, to allow for a direct comparison between the models and to clarify behavioral changes, in what follows we shall produce the shadow images for all the models and for all of the chosen combinations of parameters.

\begin{figure}
    \centering
    \includegraphics[scale=1]{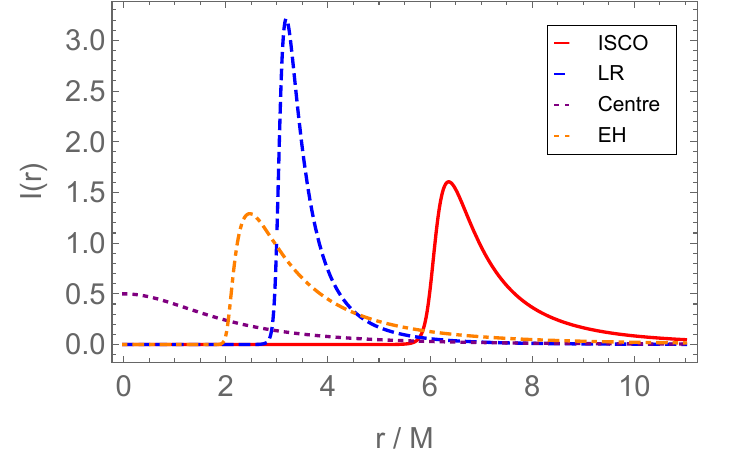}
    \caption{Intensity profiles of the GLM model given in Eq.\eqref{eq:GLMmodel} for the three models for the accretion disk models considered.}
    \label{fig:disks}
\end{figure}

To model the intensity profiles described above, we adopt the GLM model published recently in Ref.\cite{Gralla:2020srx}. This model was shown to be in a close agreement with the observational predictions for the intensity profiles of astrophysical accretion disks obtained via general-relativistic magneto-hydrodynamics \cite{Vincent:2022fwj}. The intensity profile of the GLM model is described by
\begin{equation}\label{eq:GLMmodel}
    I\left(r,\gamma,\mu,\sigma\right)=\frac{\exp\left\{-\frac{1}{2}\left[\gamma+\text{arcsinh}\left(\frac{r-\mu}{\sigma}\right)\right]^2\right\}}{\sqrt{\left(r-\mu\right)^2+\sigma^2}},
\end{equation}
where the constants $\gamma$, $\mu$, and $\sigma$ are free parameters that control the shape of the intensity profile $I\left(r\right)$: $\gamma$ controls the rate of increase of the intensity profile from infinity down to the peak; $\mu$ performs a translation of the whole intensity profile as to shift the peak to a desired radial position; and $\sigma$ controls the dilation of the intensity profile as a whole. For the three models described before, the values chosen for the parameters $\gamma$, $\mu$ and $\sigma$ are given in Table \ref{tab:diskparameters}, and the respective intensity profiles are plotted in Fig. \ref{fig:disks}. These intensity profiles correspond to the intensity profiles in the reference frame of the emitter, i.e., the accretion disk, where photons are emitted with a frequency $\nu_e$, where the subscript $\ _e$ denotes "emitter". In the reference frame of the observer, the observed frequency $\nu_o$, where the subscript $\ _o$ denotes the observer, will be redshifted with respect to the emitted counterpart, as $\nu_o=\sqrt{g_{tt}}\nu_e$. The associated specific intensity $i\left(\nu\right)$ thus scales as $i\left(\nu_o\right)=\left(\nu_o/\nu_e\right)^3i\left(\nu_e\right)=g_{tt}^{3/2}i\left(\nu_e\right)$, which implies that the intensity $I=\int i\left(\nu\right)d\nu$ in the reference frame of the observer takes the form
\begin{equation}\label{eq:intensity}
    I_o\left(r\right)=g_{tt}^2\left(r\right)I_e\left(r\right).
\end{equation}
In our Mathematica-based code, once the integration of the geodesic equation in Eq.\eqref{eq:raytrace} is performed, the radius at which the integrated geodesic intersects the accretion disk is computed and Eq.\eqref{eq:intensity} is taken into consideration to perform the gravitational redshift of the intensity profile observed in the reference frame of the observer.
\begin{table}
    \centering
    \begin{tabular}{c|c c c}
         &  $\gamma$ & $\mu$ & $\sigma$ \\ \hline
        ISCO & $-2$ & $6M$ & $M/4$ \\
        LR & $-2$ & $3M$ & $M/8$ \\
        Center & $0$ & $0$ & $2M$ \\
        EH & $-3$ & $2M$ & $M/8$
    \end{tabular}
    \caption{Values of the parameters $\gamma$, $\mu$ and $\sigma$ of the GLM model given in Eq.\eqref{eq:GLMmodel} for the four different accretion disk models considered. Note that the EH model is only used for a comparison with the Schwarzschild spacetime.}
    \label{tab:diskparameters}
\end{table}

\subsection{Observed shadows and intensity profiles}

To analyze the effects of the presence of the thin-shell and the LR in the overall observational properties of relativistic fluid stars, a total of six configurations with different combinations of parameters $r_\Sigma$ and $R$ were implemented in the mentioned ray-tracing code. These solutions are labeled as $S_{ij}$, where $i$ denotes $R/M$, and $j$ denotes $r_\Sigma /M$. A summary of the configurations considered and their respective properties is provided in Table \ref{tab:configurations}. The combinations of parameters were chosen in such a way to obtain: (i) solutions without thin-shells with different compacticities, $S_{55}$, $S_{44}$ and $S_{33}$; (ii) solutions with the same compacticity without LRs, with and without thin-shells, $S_{54}$ and $S_{44}$; and (iii) solutions with the same compacticity with LRs, with and without thin-shells, $S_{53}$, $S_{43}$, and $S_{33}$. This allows for a detailed study of the effects of both the thin-shell and the LR in the overall appearance of the solutions. For each of the configurations, the accretion disk is placed on the equatorial plane, i.e., $\theta=\pi/2$, whereas the observer is placed on the axial axis $\theta=0$ at a radius of $r=r_\infty$, dubbed the numerical infinite, a distance large enough for the observer to be well-approximated by an asymptotic observer, i.e., the light-rays reaching this distance are approximately parallel to the vertical axis. 

\begin{table}
    \centering
    \begin{tabular}{c|c c c c}
         & $R$ & $r_\Sigma$ & TS & LR \\ \hline
      $S_{55}$ & $5M$ & $5M$ & no & no \\
      $S_{54}$ & $5M$ & $4M$ & yes & no \\
      $S_{53}$ & $5M$ & $3M$ & yes & yes \\ \hline
      $S_{44}$ & $4M$ & $4M$ & no & no \\
      $S_{43}$ & $4M$ & $3M$ & yes & yes \\ \hline
      $S_{33}$ & $3M$ & $3M$ & no & yes \\
    \end{tabular}
    \caption{Configurations $S_{ij}$ chosen for ray-tracing, respective values of the quantities $R$ and $r_\Sigma$, and reference to the presence of a thin-shell (TS) or a light-ring (LR) in the spacetime.}
    \label{tab:configurations}
\end{table}

\begin{figure*}
    \centering
    \includegraphics[scale=0.5]{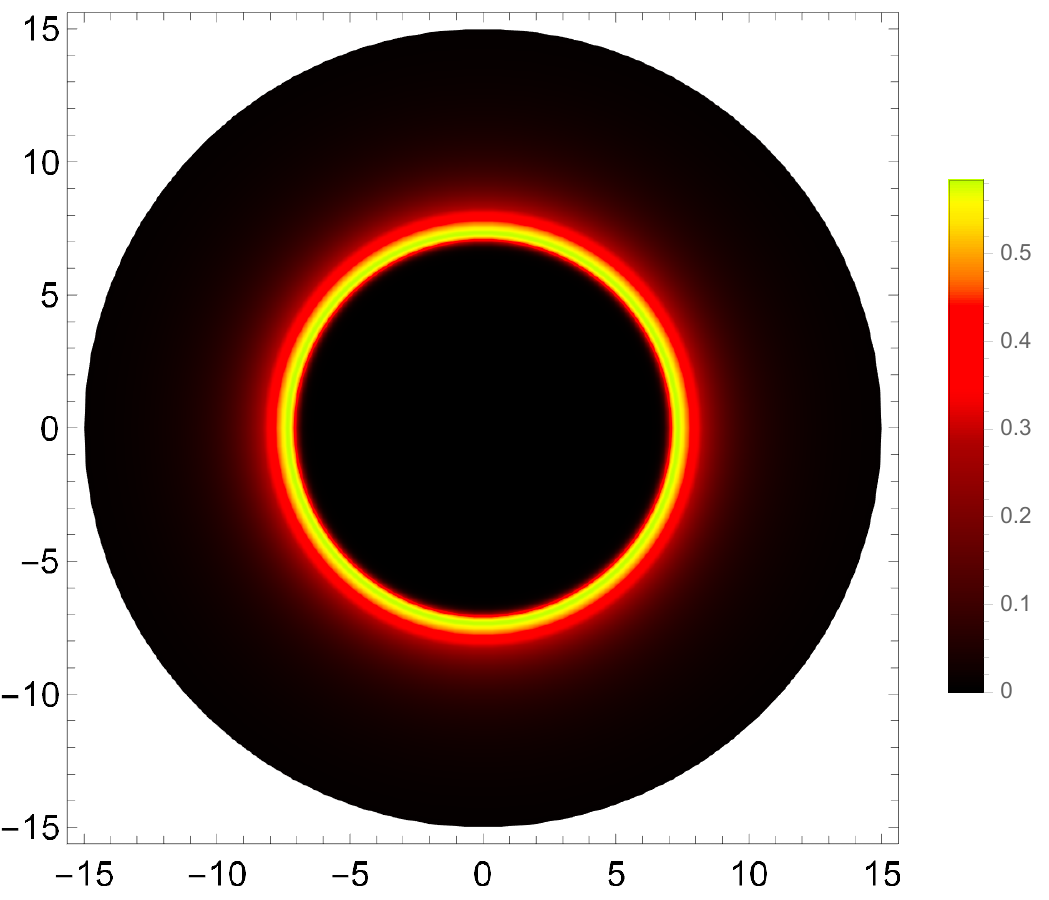}\\
    \includegraphics[scale=0.5]{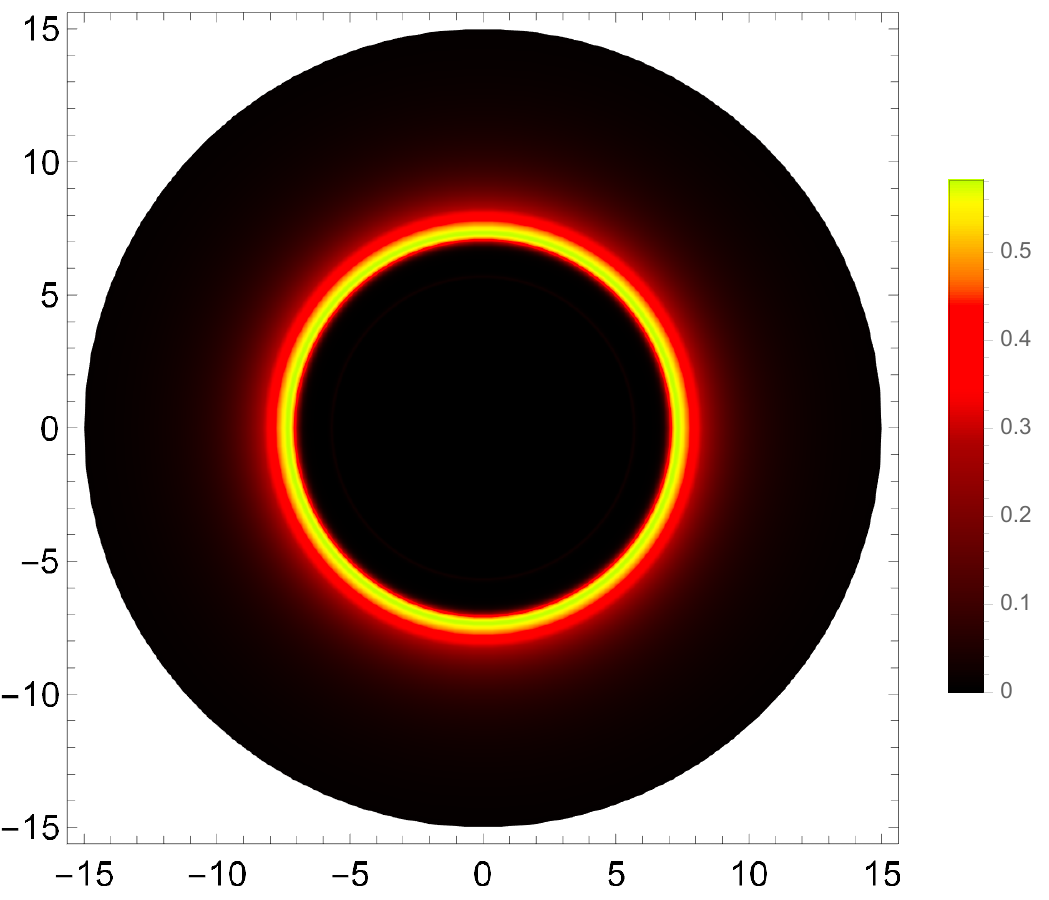}\qquad
    \includegraphics[scale=0.5]{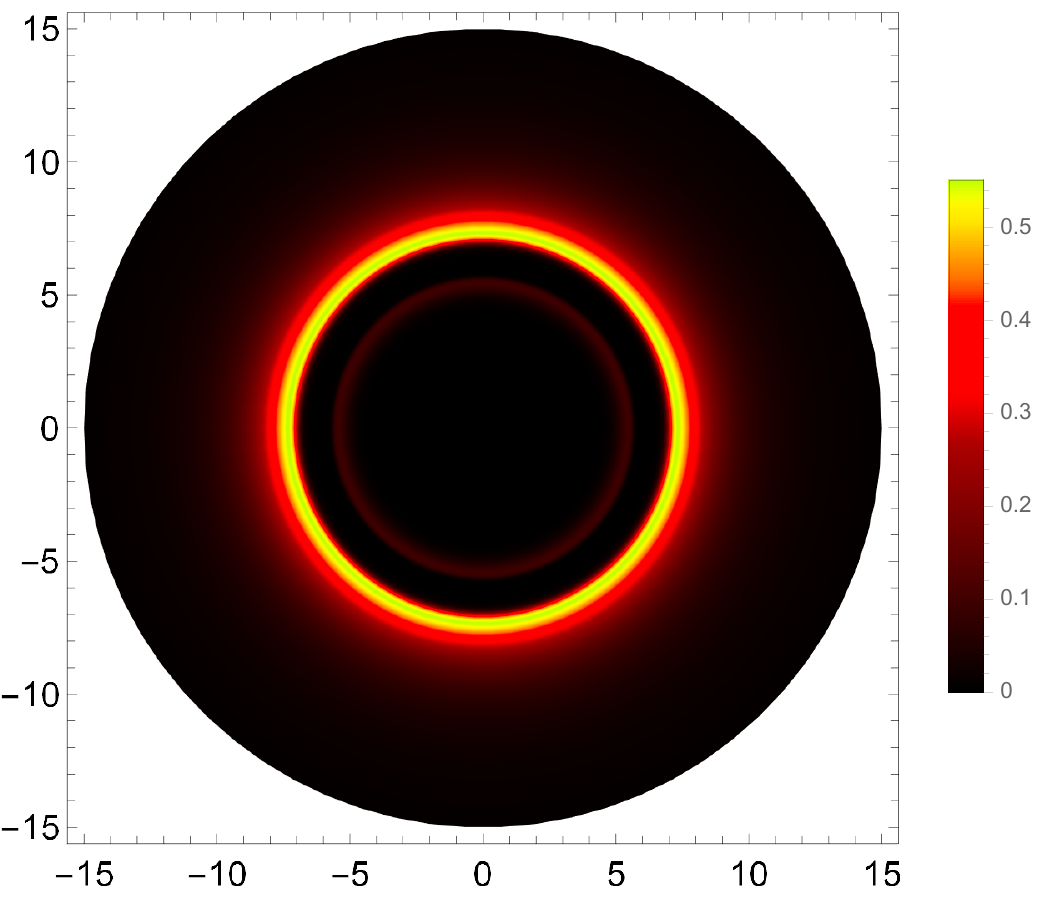}\\
    \includegraphics[scale=0.5]{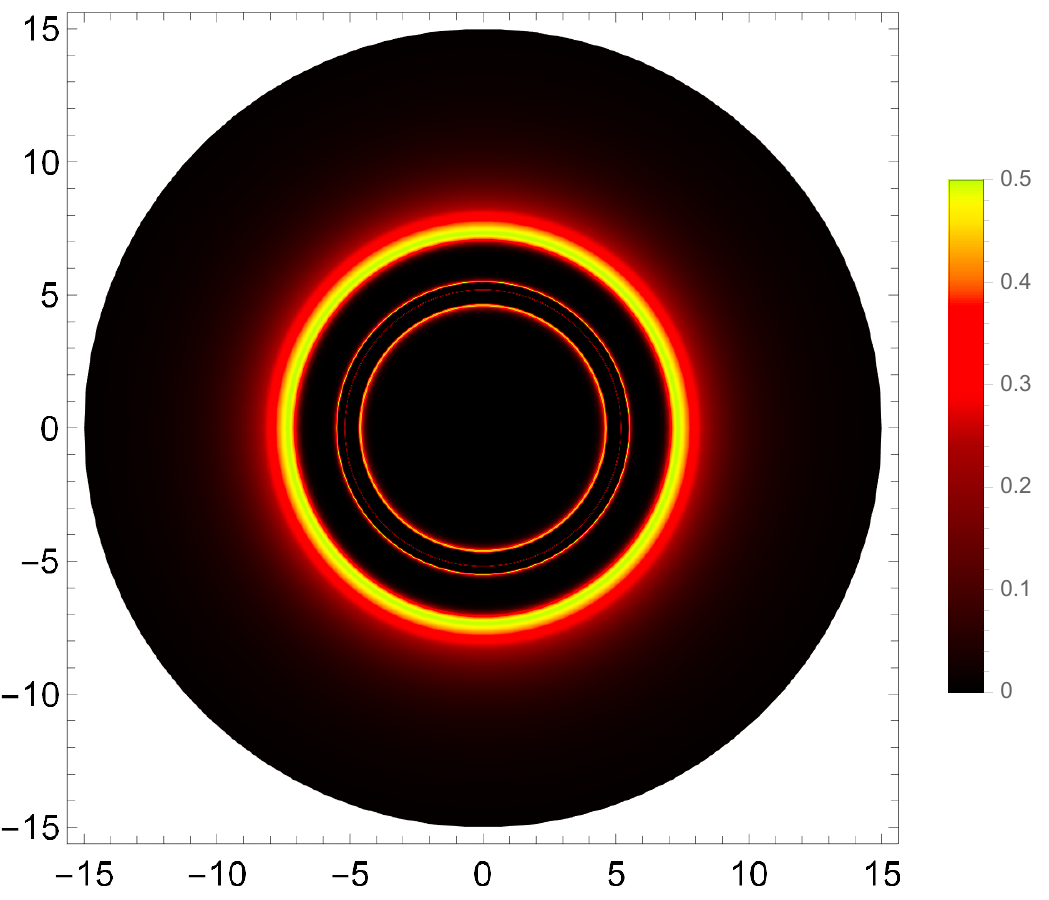}\qquad
    \includegraphics[scale=0.5]{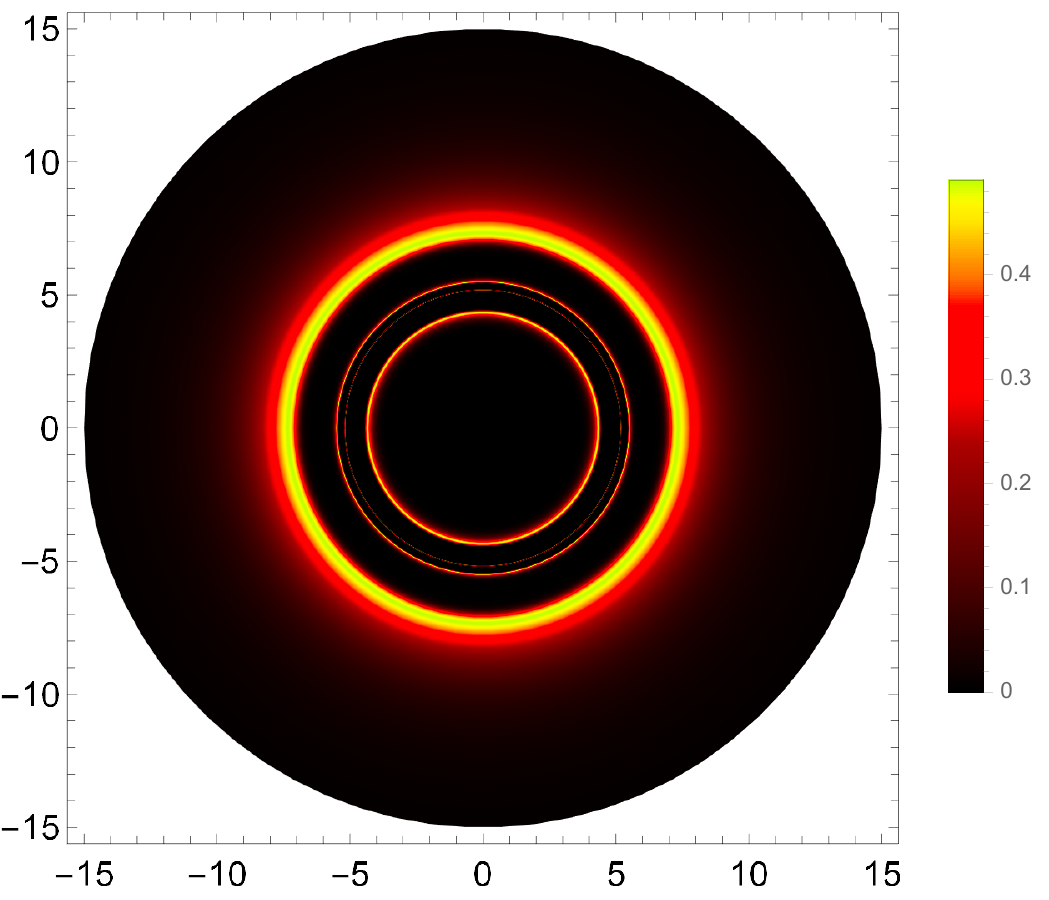}
    \includegraphics[scale=0.5]{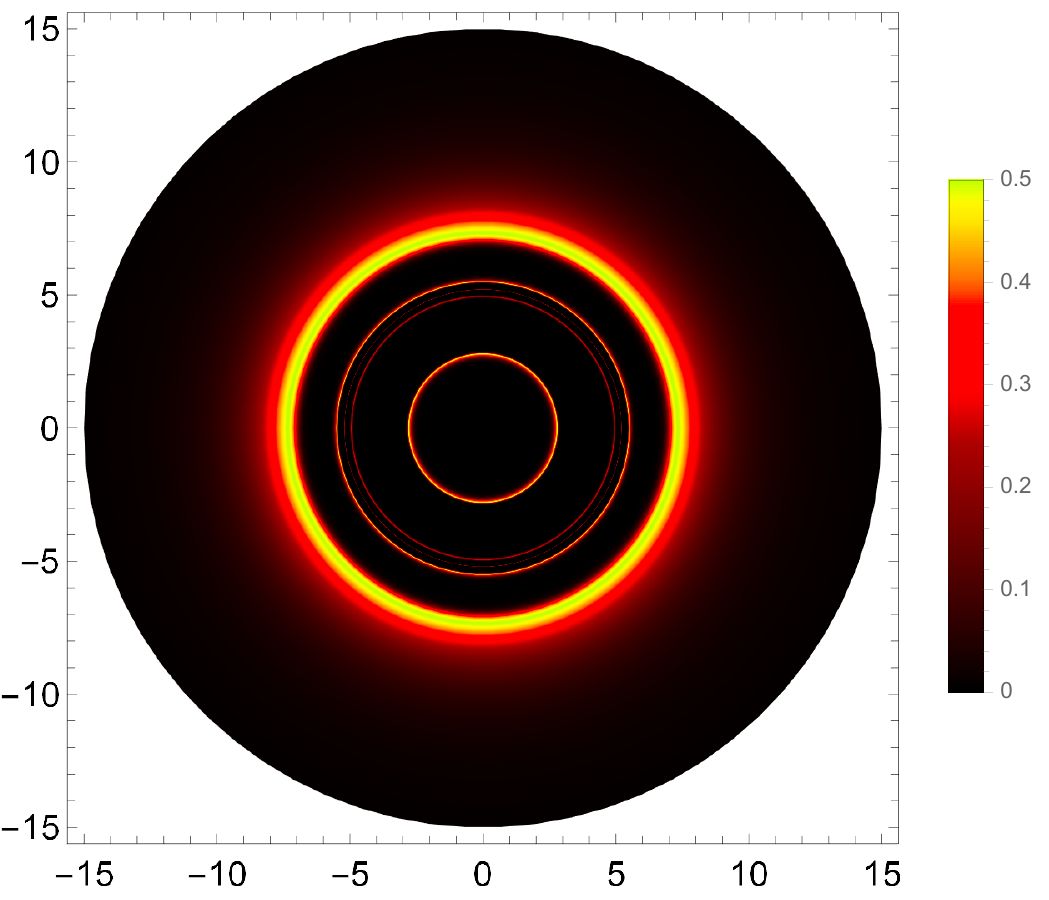}
    \caption{Shadows images with the ISCO accretion disk model (see Fig. \ref{fig:disks}) for the six configurations summarized in Table \ref{tab:configurations}, i.e., $S_{55}$ (top row), $S_{54}$ (middle row left), $S_{44}$ (middle row right), $S_{53}$ (bottom row left), $S_{43}$ (bottom row center), and $S_{33}$ (bottom row right).}
    \label{fig:shadows_ISCO}
\end{figure*}

\begin{figure*}
    \centering
    \includegraphics[scale=0.63]{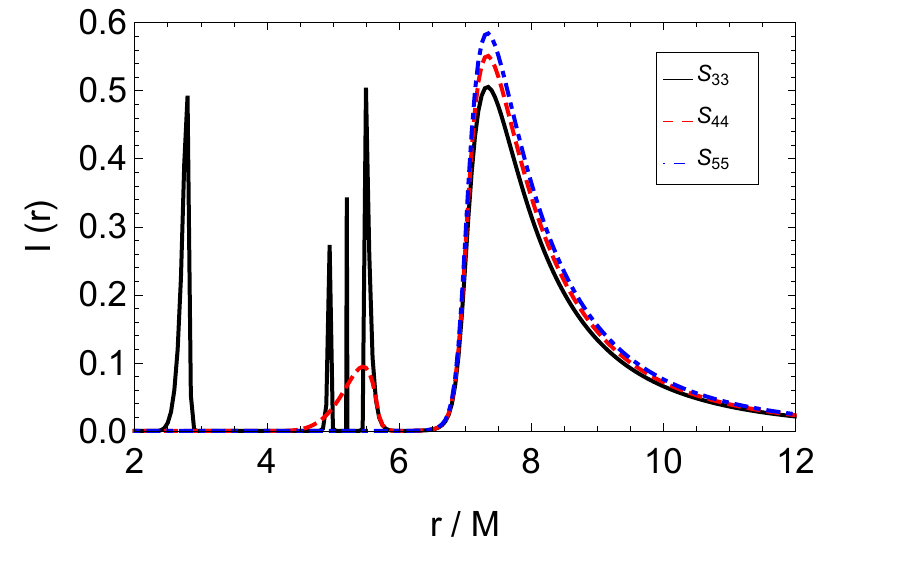}
    \includegraphics[scale=0.63]{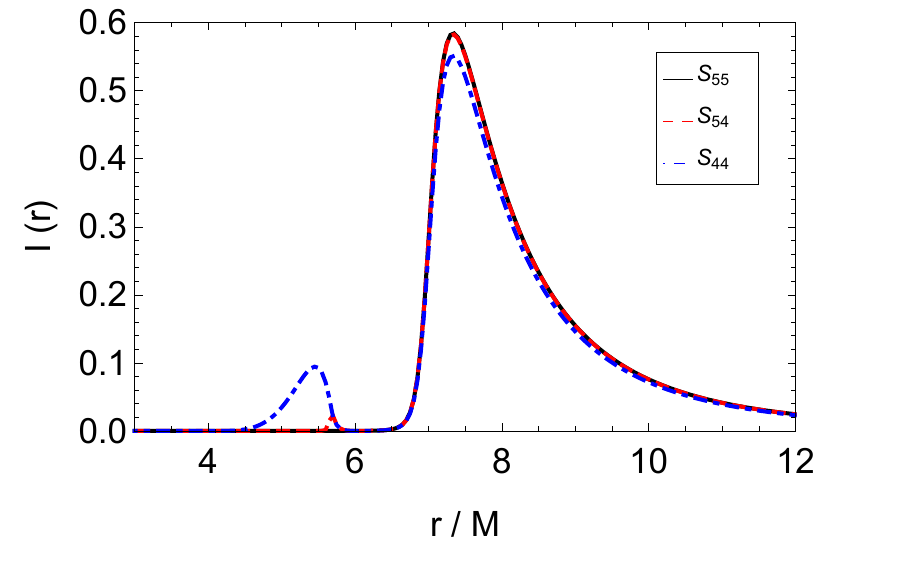}
    \includegraphics[scale=0.63]{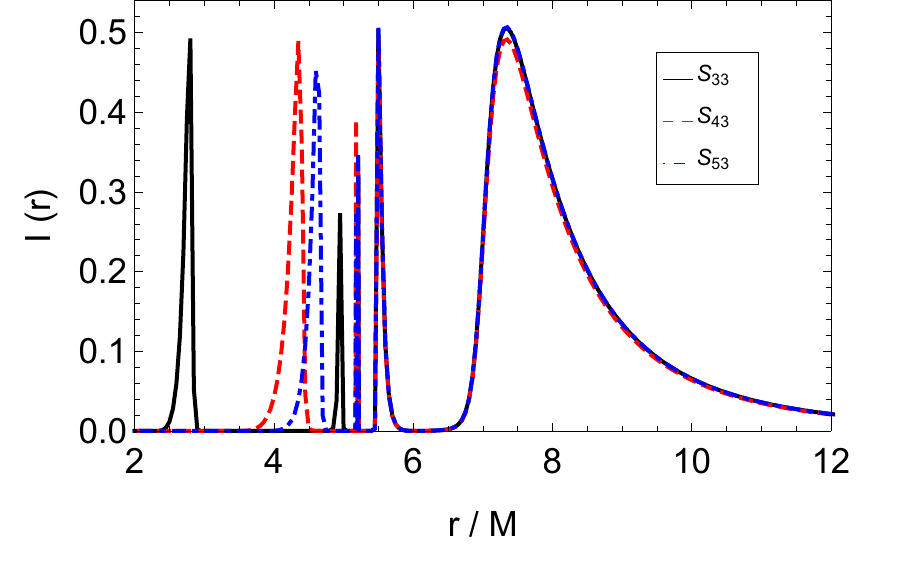}
    \caption{Observed intensity profiles $I_o$ as a function of the normalized radial coordinate $r/M$ with the ISCO accretion disk model for the six configurations summarized in Table \ref{tab:configurations}. We compare configurations without thin-shells (left panel), configurations without LRs (middle panel), and configurations with LRs (right panel).}
    \label{fig:intensity_ISCO}
\end{figure*}

\begin{figure*}
    \centering
    \includegraphics[scale=0.5]{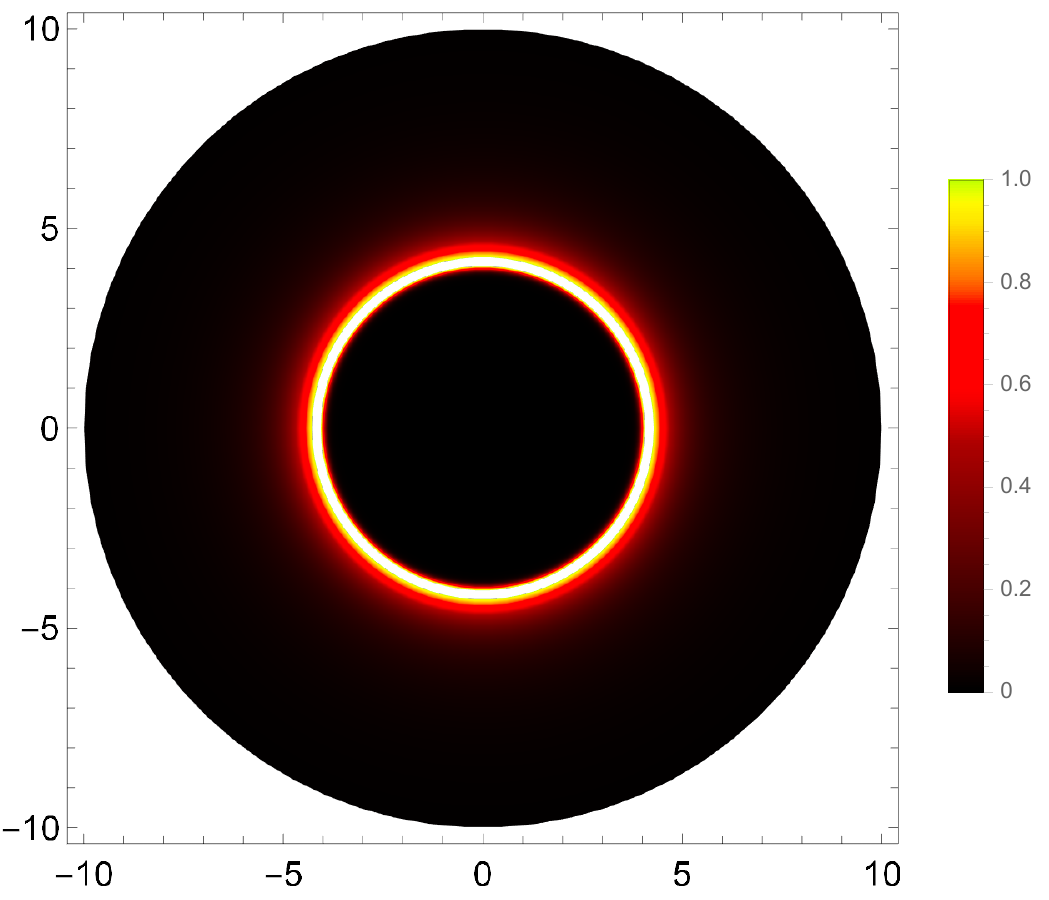}\\
    \includegraphics[scale=0.5]{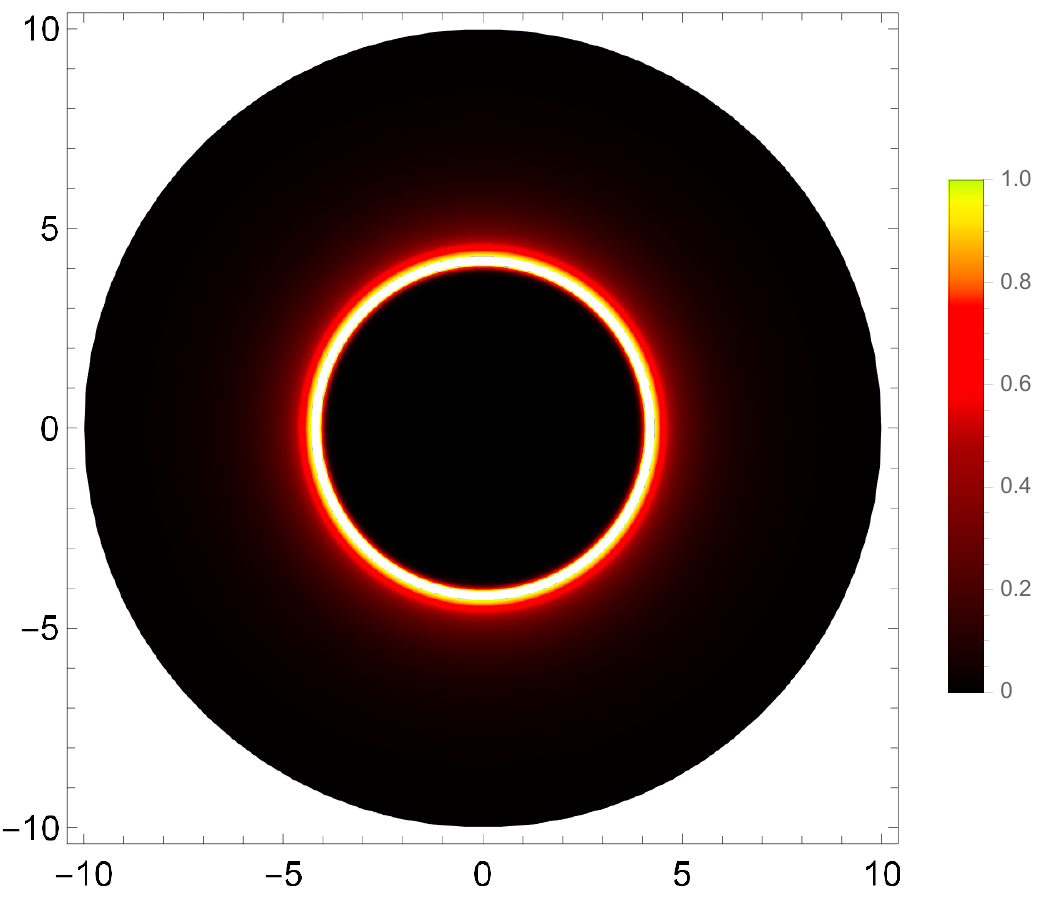}\qquad
    \includegraphics[scale=0.5]{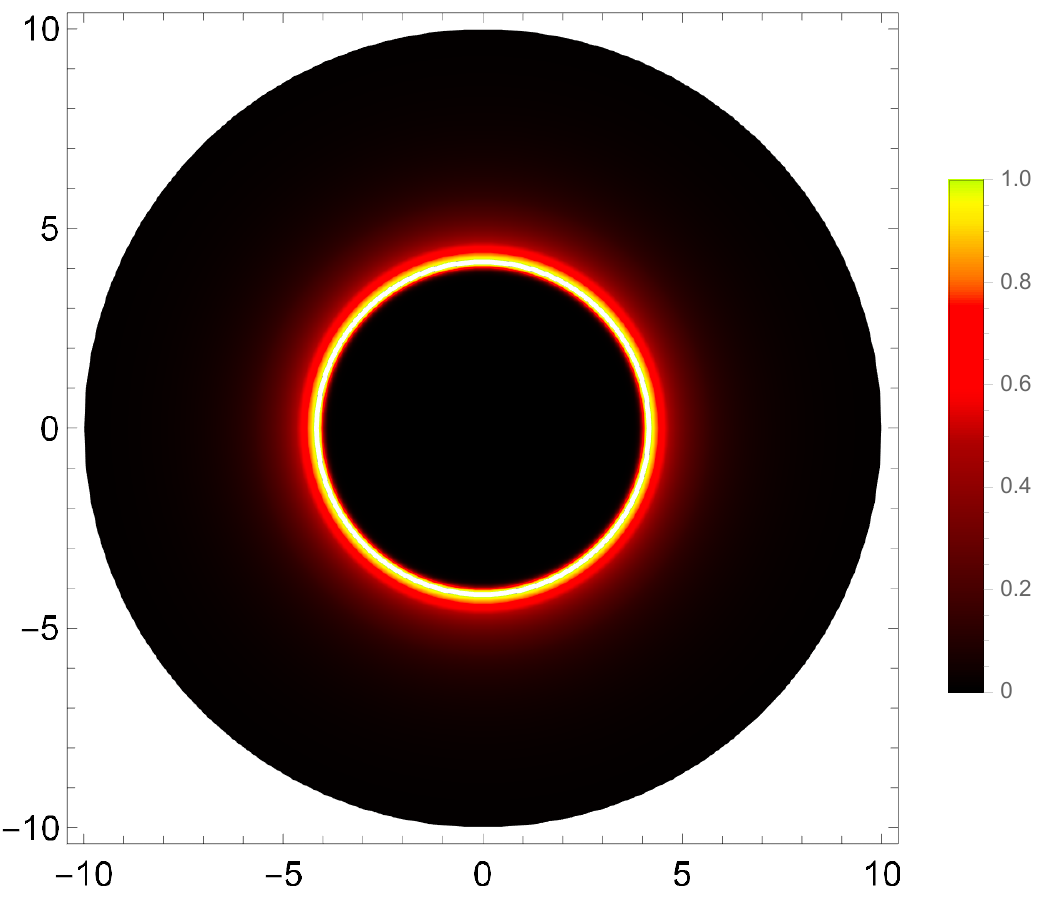}\\
    \includegraphics[scale=0.5]{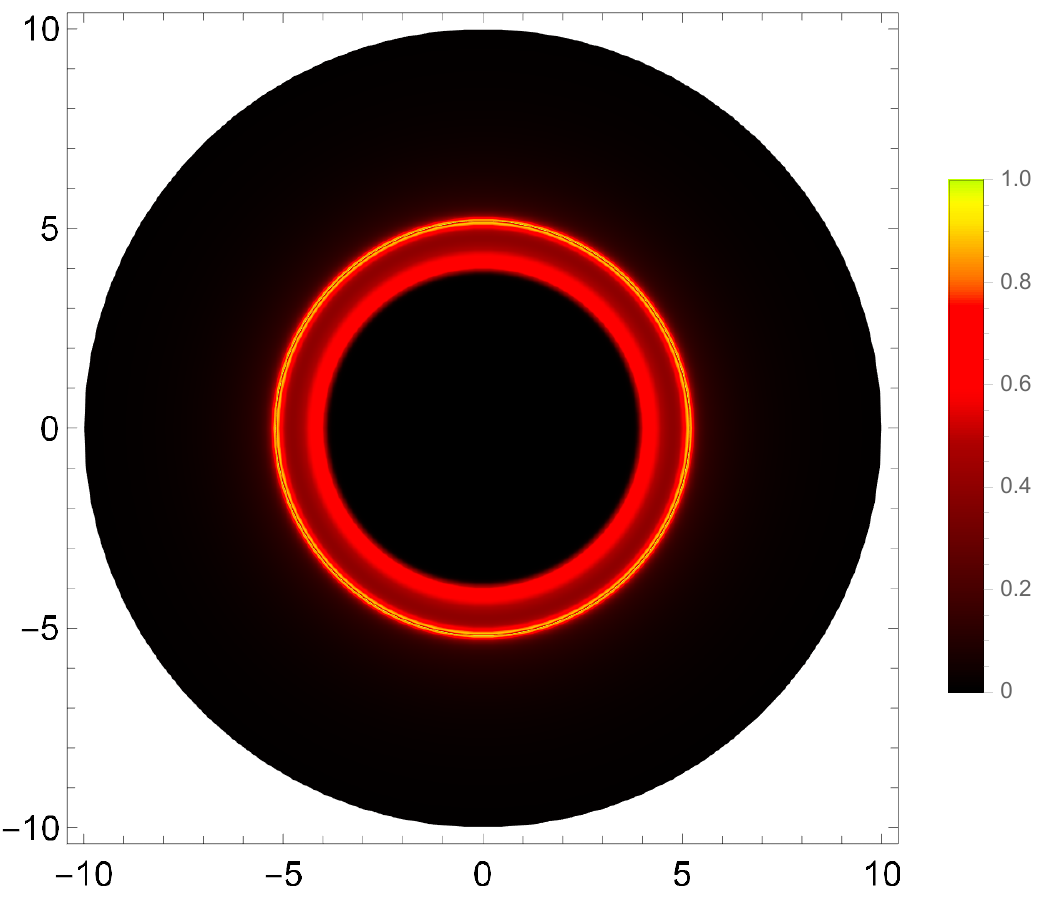}\qquad
    \includegraphics[scale=0.5]{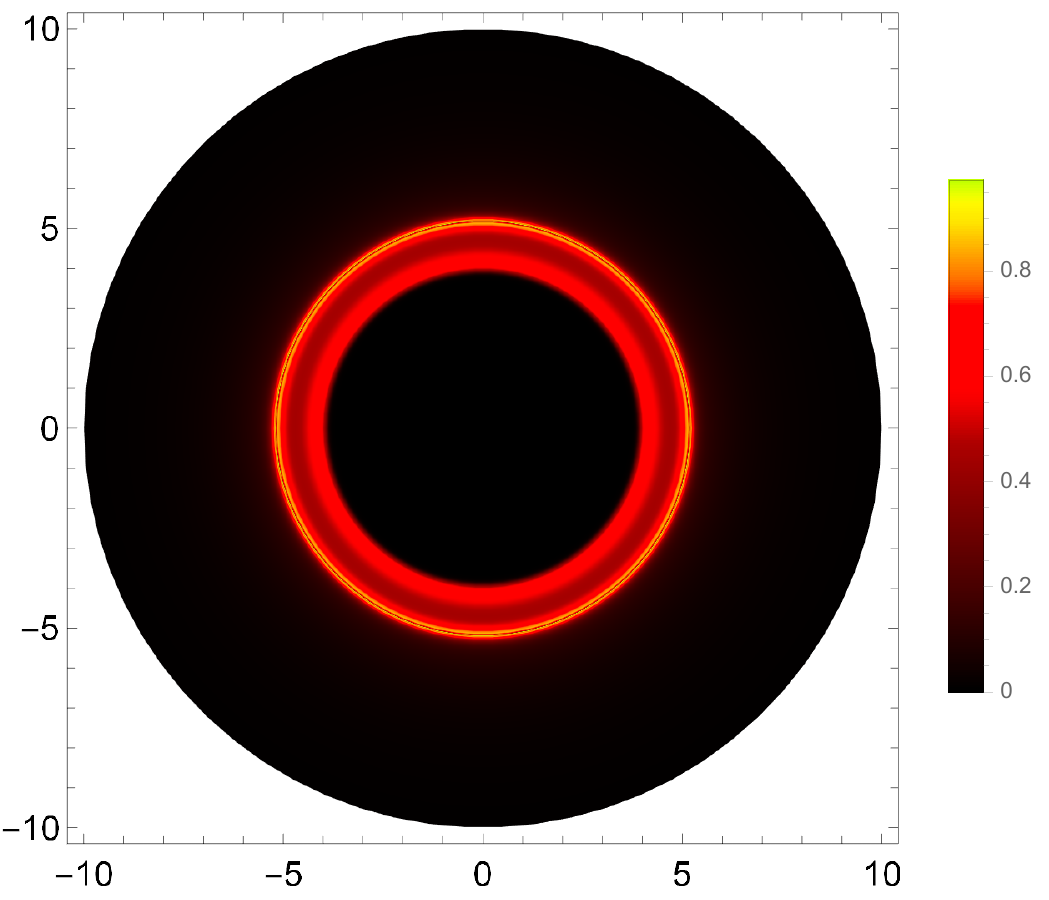}
    \includegraphics[scale=0.5]{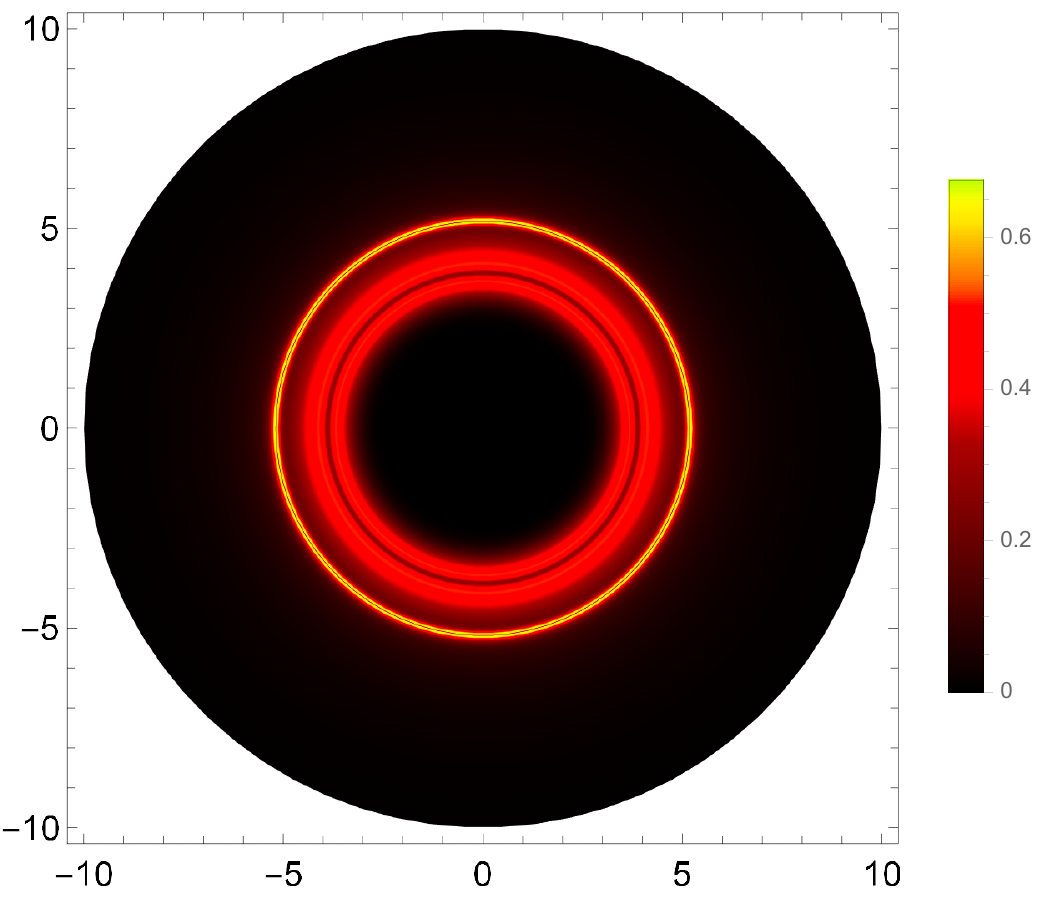}
    \caption{Shadows images with the LR accretion disk model (see Fig. \ref{fig:disks}) for the six configurations summarized in Table \ref{tab:configurations}, i.e., $S_{55}$ (top row), $S_{54}$ (middle row left), $S_{44}$ (middle row right), $S_{53}$ (bottom row left), $S_{43}$ (bottom row center), and $S_{33}$ (bottom row right).}
    \label{fig:shadows_LR}
\end{figure*}

\begin{figure*}
    \centering
    \includegraphics[scale=0.63]{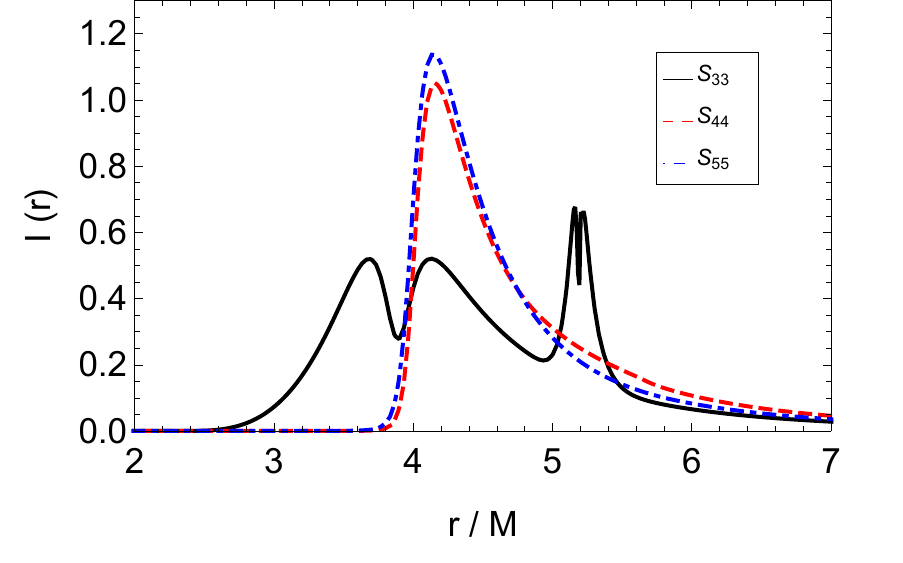}
    \includegraphics[scale=0.63]{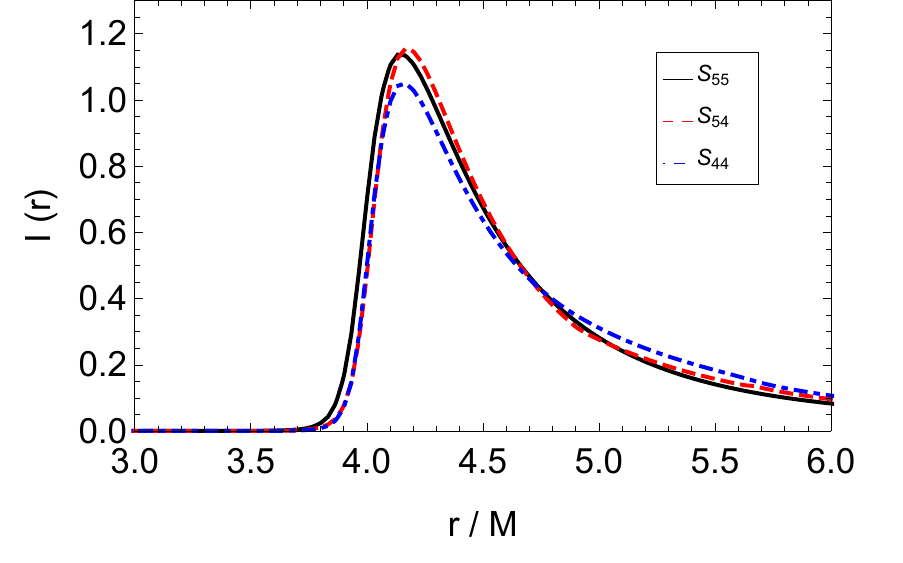}
    \includegraphics[scale=0.63]{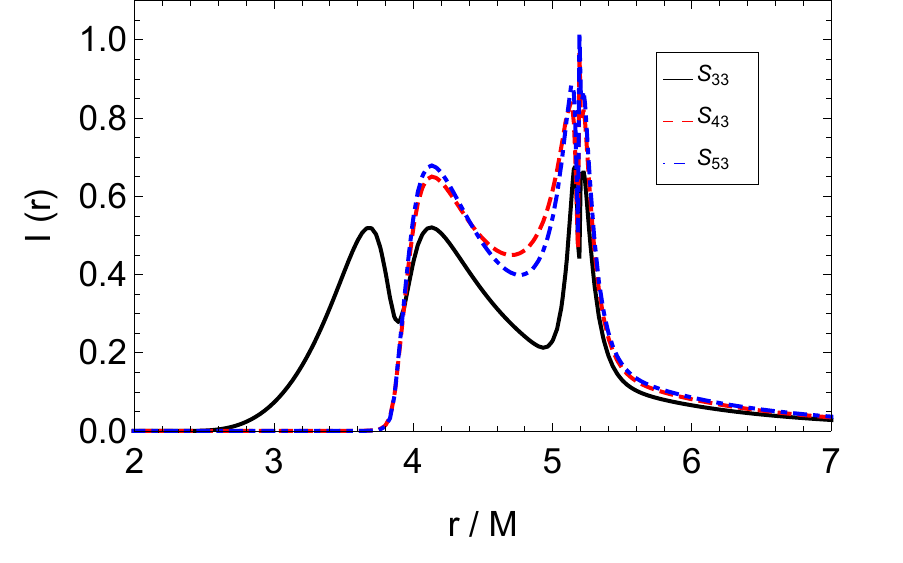}
    \caption{Observed intensity profiles $I_o$ as a function of the normalized radial coordinate $r/M$ with the LR accretion disk model for the six configurations summarized in Table \ref{tab:configurations}. We compare configurations without thin-shells (left panel), configurations without LRs (middle panel), and configurations with LRs (right panel).}
    \label{fig:intensity_LR}
\end{figure*}

\begin{figure*}
    \centering
    \includegraphics[scale=0.5]{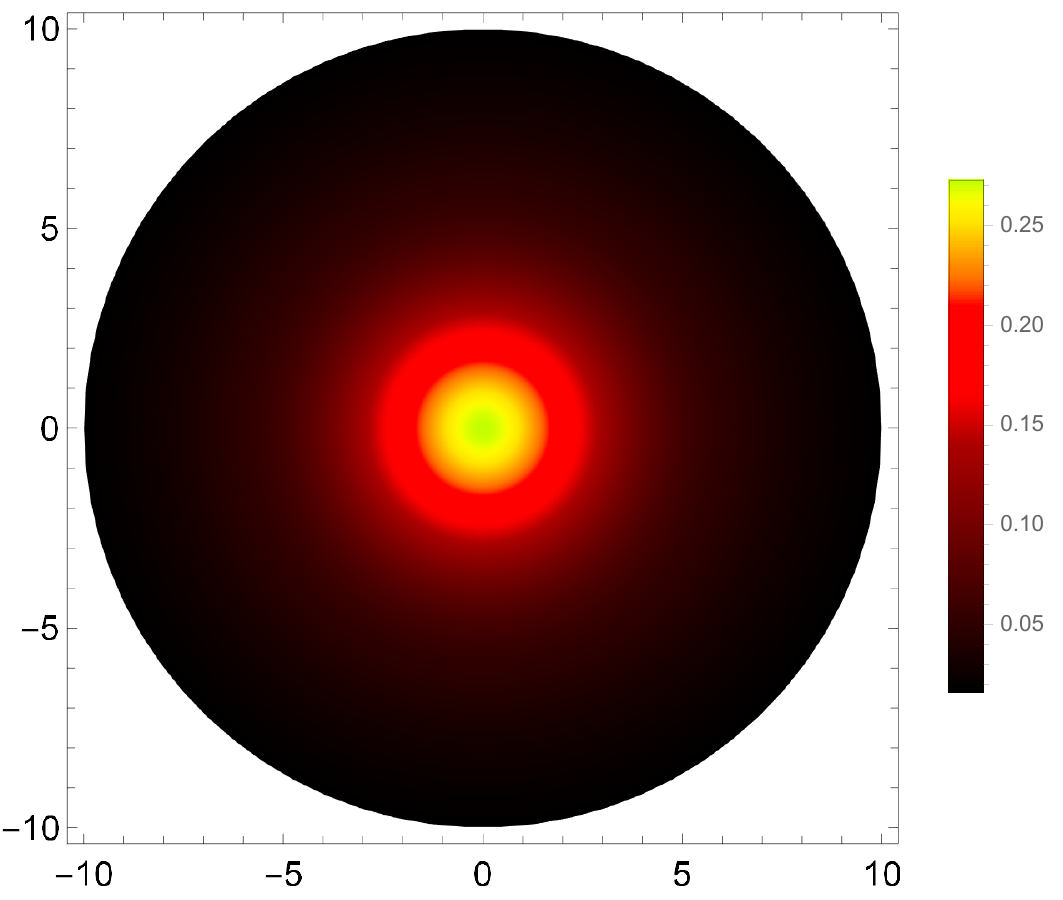}\\
    \includegraphics[scale=0.5]{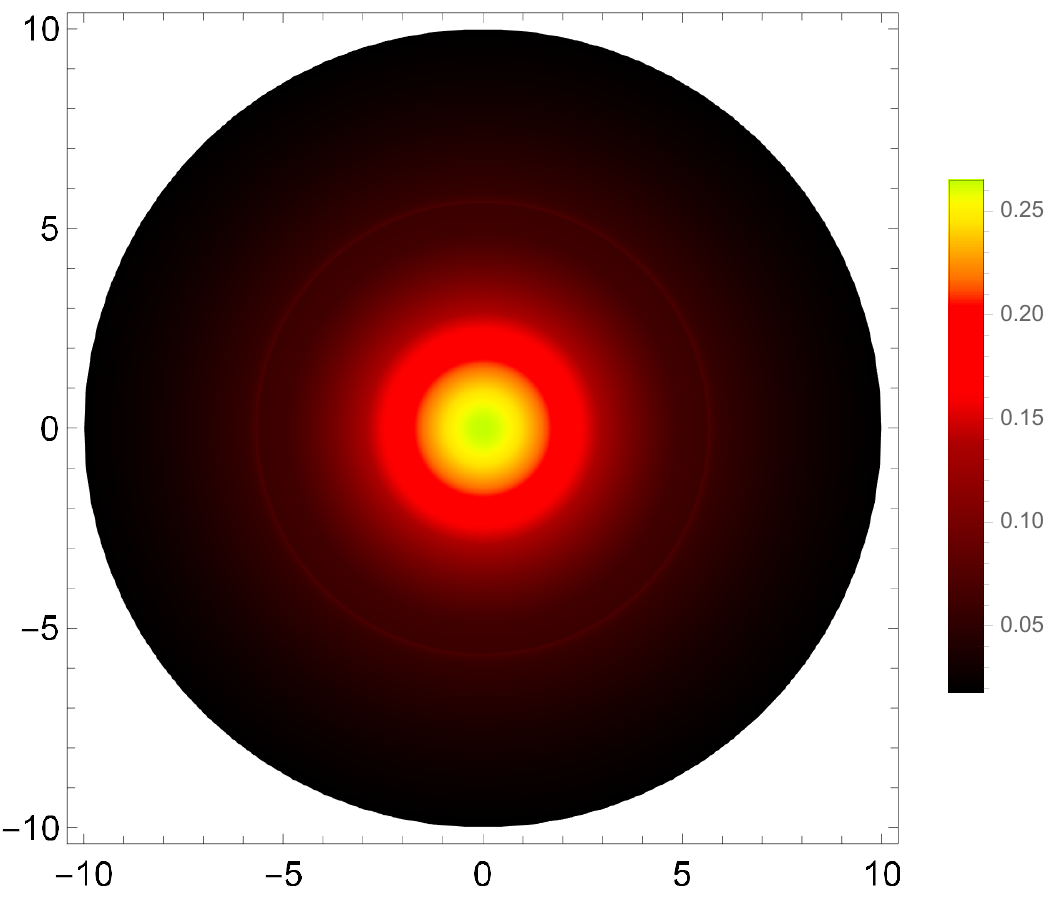}\qquad
    \includegraphics[scale=0.5]{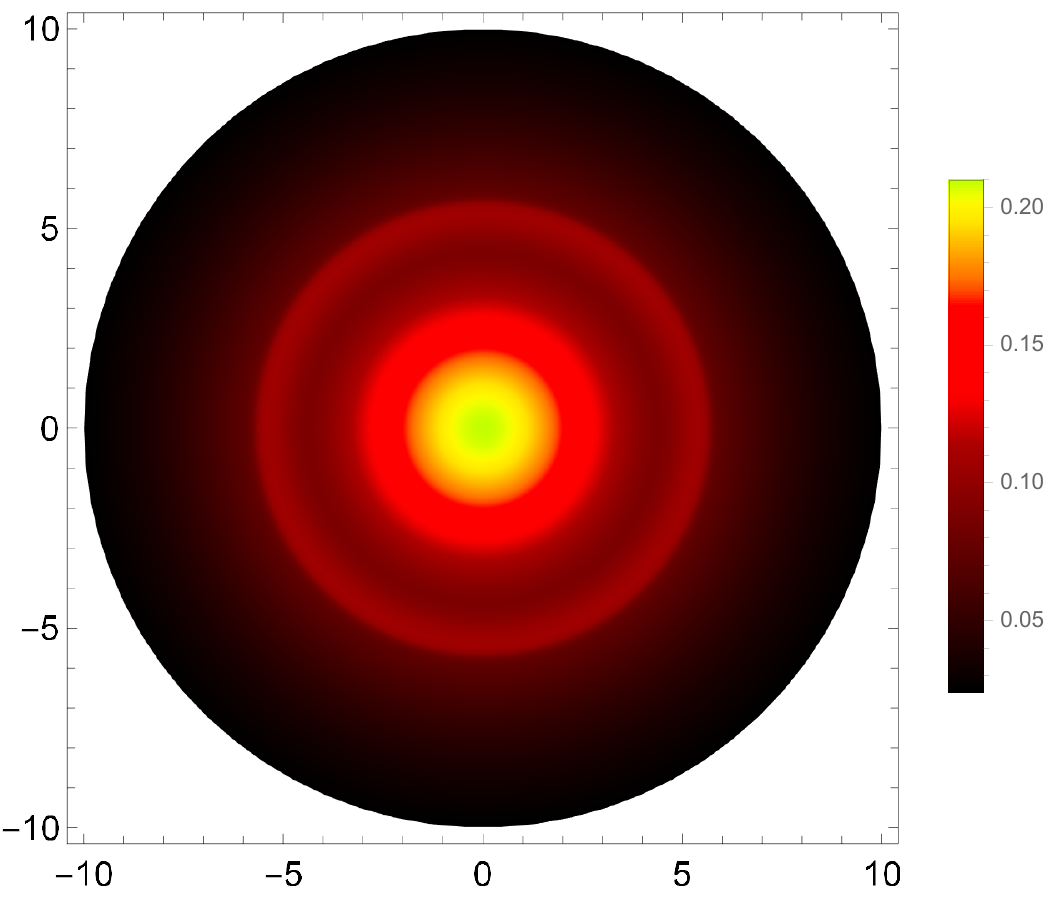}\\
    \includegraphics[scale=0.5]{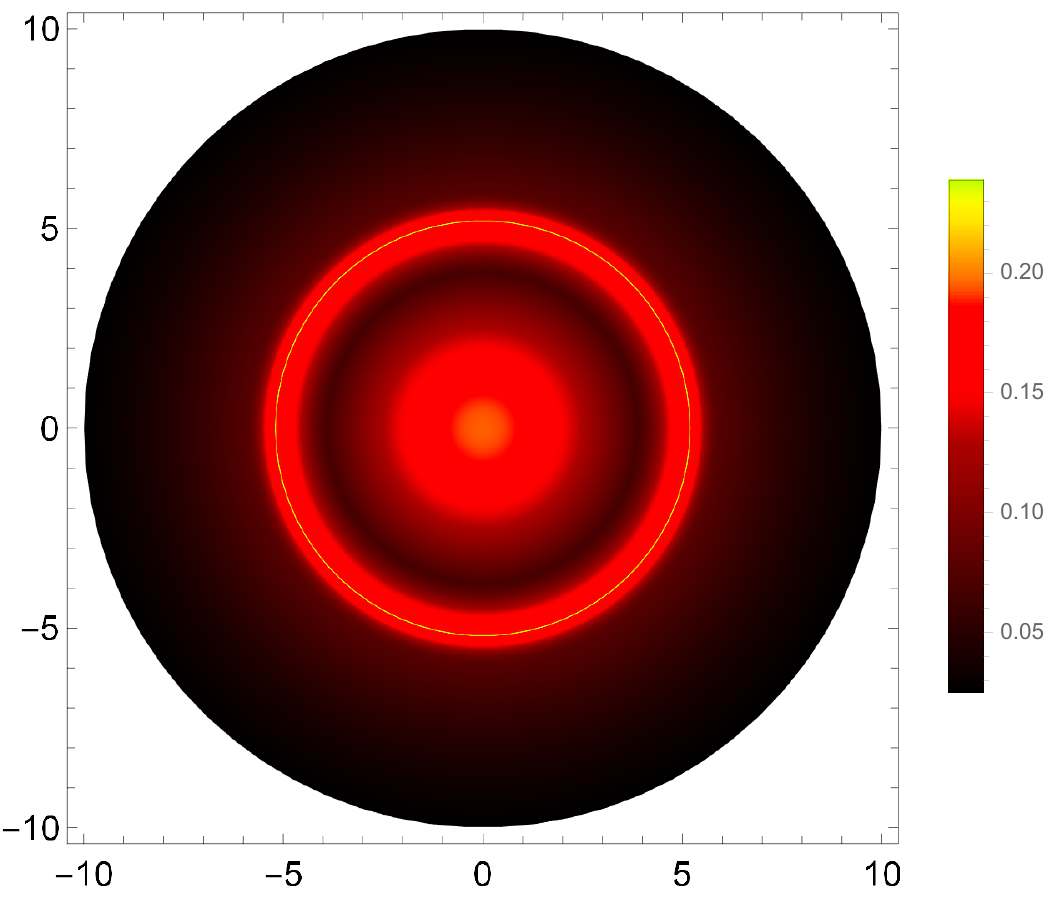}\qquad
    \includegraphics[scale=0.5]{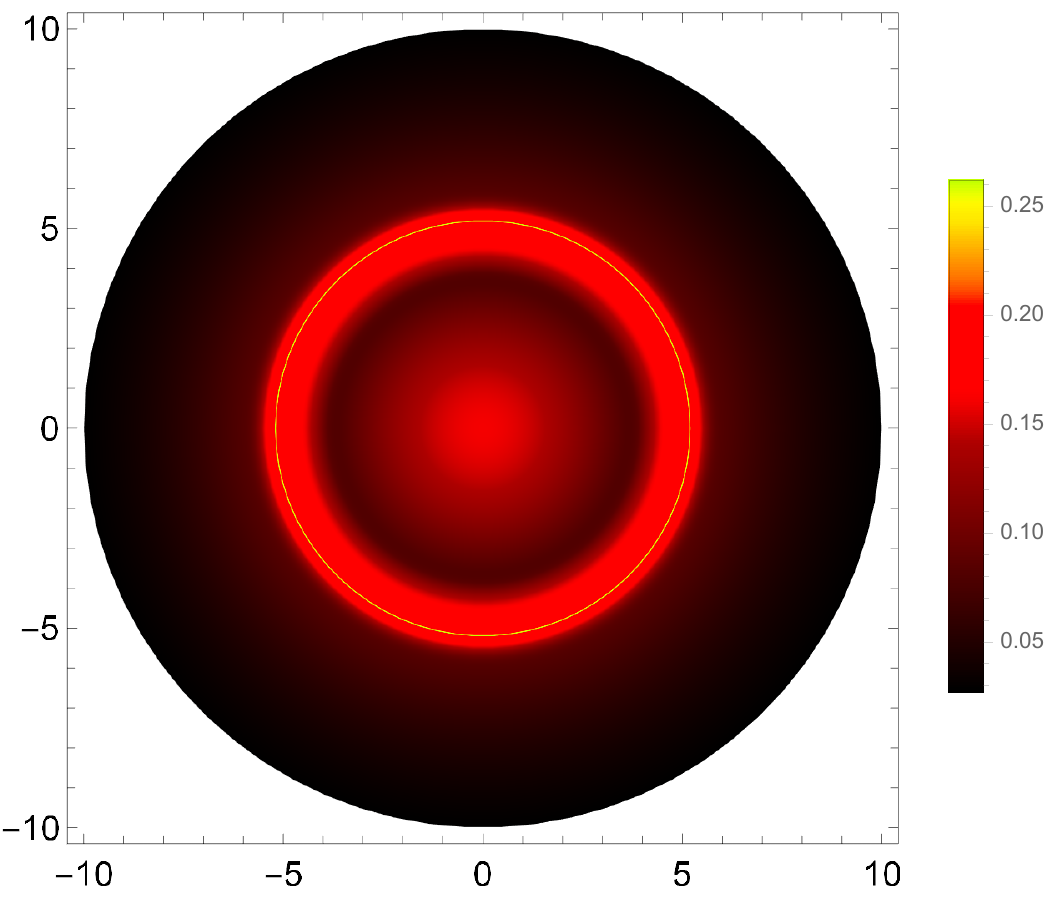}
    \includegraphics[scale=0.5]{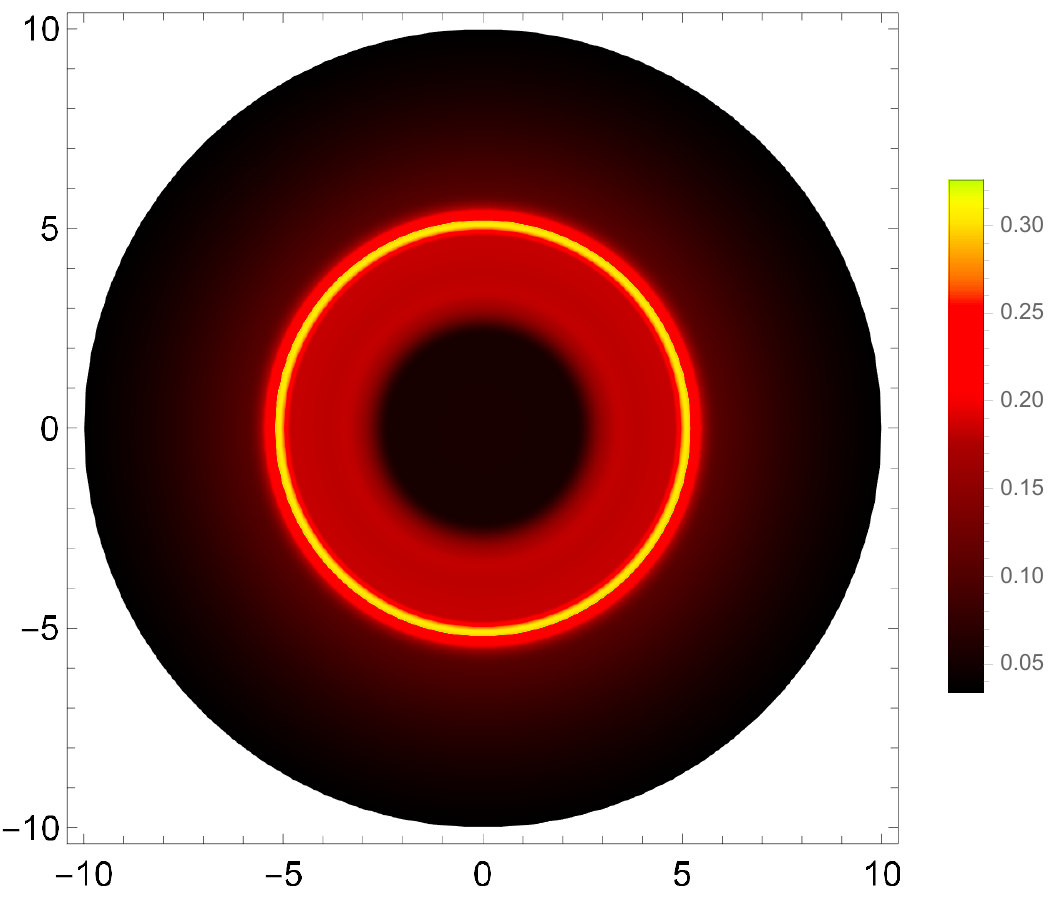}
    \caption{Shadows images with the Centre accretion disk model (see Fig. \ref{fig:disks}) for the six configurations summarized in Table \ref{tab:configurations}, i.e., $S_{55}$ (top row), $S_{54}$ (middle row left), $S_{44}$ (middle row right), $S_{53}$ (bottom row left), $S_{43}$ (bottom row center), and $S_{33}$ (bottom row right).}
    \label{fig:shadows_C}
\end{figure*}

\begin{figure*}
    \centering
    \includegraphics[scale=0.63]{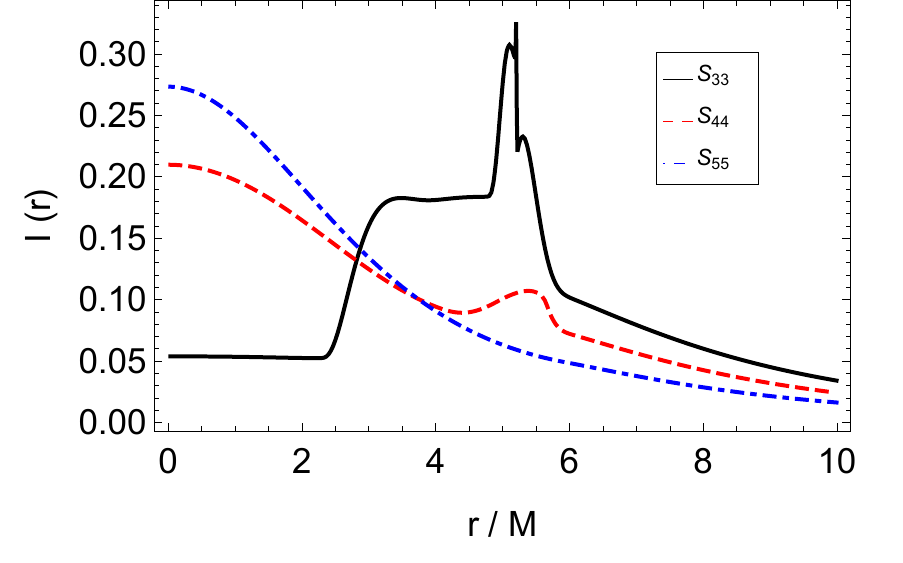}
    \includegraphics[scale=0.63]{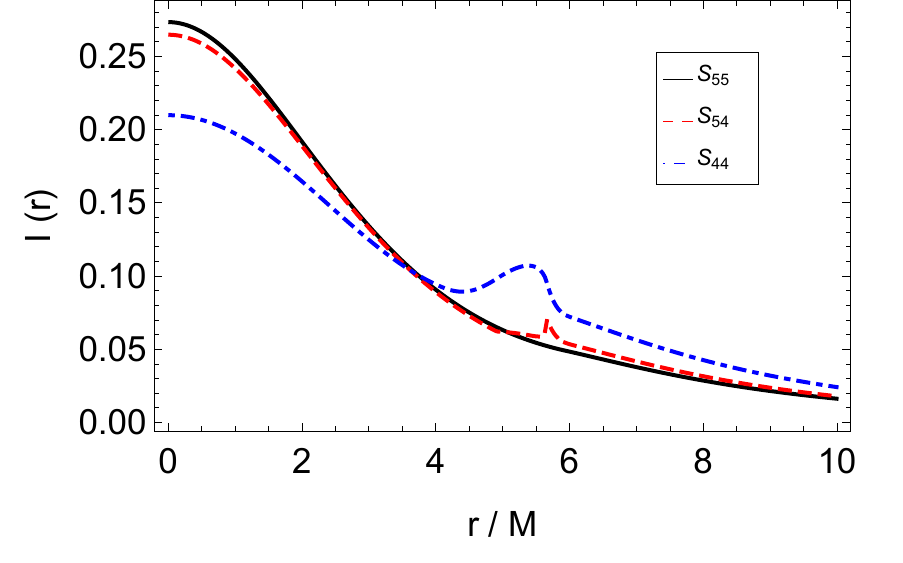}
    \includegraphics[scale=0.63]{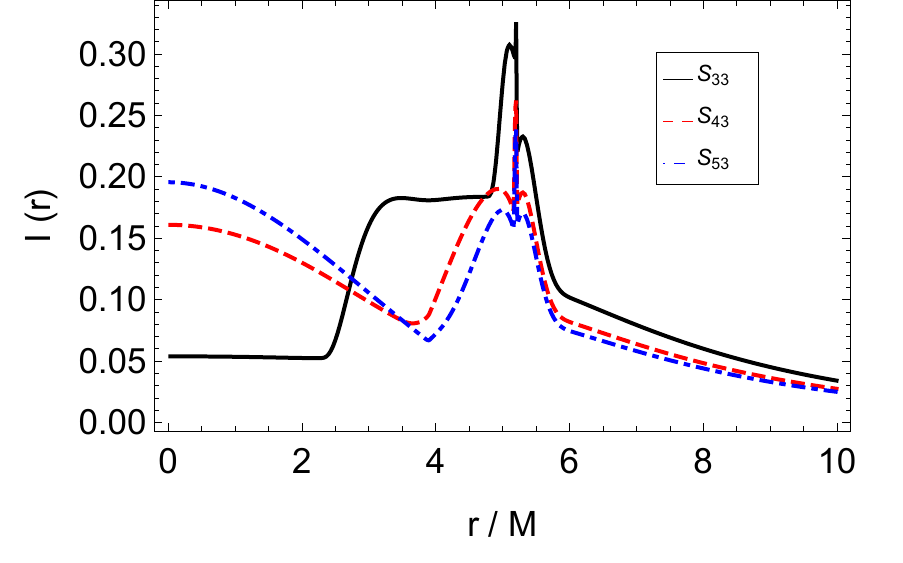}
    \caption{Observed intensity profiles $I_o$ as a function of the normalized radial coordinate $r/M$ with the Centre accretion disk model for the six configurations summarized in Table \ref{tab:configurations}. We compare configurations without thin-shells (left panel), configurations without LRs (middle panel), and configurations with LRs (right panel).}
    \label{fig:intensity_C}
\end{figure*}

The shadow images for the configurations summarized in Table \ref{tab:configurations} are given in Fig. \ref{fig:shadows_ISCO} for the ISCO disk model, Fig. \ref{fig:shadows_LR} for the LR disk model, and Fig. \ref{fig:shadows_C} for the Central disk model. These images are organized in a triangular shape where each row denotes a different compacticity, from $r_\Sigma=5M$ (top row) to $r_\Sigma=3M$ (bottom row), in steps of $M$, and in the same row one decreases the radius $R$ of the star from the left to the right, from $R=5M$ (leftmost row) to $R=r_\Sigma$ (rightmost row), also in steps of $M$. Configurations without LRs are represented in the top and middle rows, whereas configurations with LRs are represented in the bottom row. Also, configurations with thin-shells are represented in the bottom-left triangle of three configurations, whereas solutions without thin-shells are represented in the right edge of three configurations. The observed intensity profiles are given in Fig. \ref{fig:intensity_ISCO} for the ISCO disk model, Fig. \ref{fig:intensity_LR} for the LR disk model, and Fig. \ref{fig:intensity_C} for the Central disk model. These figures are divided into three sub-plots where we compare configurations without thin-shells (left panel), configurations without LRs (middle panel), and configurations with LRs (right panel). 

Considering the intensity profiles, one verifies that, unlike the emitted intensity profiles given in Fig. \ref{fig:disks}, where a single peak of intensity is present, the observed intensity profiles, as well as the produced images, may feature several intensity peaks, depending on the compacticity of the configuration. These multiple peaks of intensity are caused by photons that reach the observer after orbiting around the central object a different number of half-orbits. We can thus identify three main components of the observed intensity profiles:
\begin{enumerate}
    \item Direct component: also known as the primary image, this component corresponds to the photons emitted directly from the accretion disk at the equatorial plane, i.e., at $\theta_e=\pi/2$, to the observer at $\theta_o=0$, having thus orbited a total angular distance of $\Delta\theta\equiv\theta_e-\theta_o=\pi/2$ around the central object. This is the dominant and widest component of the intensity profiles, and it is present in all configurations independently of their compacticity;
    \item Lensed component: also known as the secondary image, this component corresponds to the photons that are emitted from the accretion disk in the direction opposite to where the observer is, but have been lensed around the compact object once, thus orbitting a total angular distance of $\Delta\theta=3\pi/2$ around the central object. This is the second largest component of the intensity profiles and it is only present in the configurations that are compact enough to produce a lensing effect of this magnitude, including all of those that feature a LR;
    \item Light-Ring component: this component corresponds to the photons that have orbitted at least one full orbit around the central compact object, close to the LR (whenever it exists), before reaching the observer. This component presents an infinite structure of sub-rings \cite{Gralla:2020srx,Wielgus:2021peu,Johnson:2019ljv}, which correspond to the photons that have orbitted a total angular distance of $\Delta\theta=5\pi/2+n\pi$, with $n\geq 0$ an integer. This is the narrowest and least dominant component of the intensity profiles and it is only present in the configurations that are compact enough to develop a LR.
\end{enumerate}
For the ISCO disk model, when more than one of the components of the observed intensity profile described above are present, they can be clearly identified as separate peaks of intensity. However, for the LR and Centre disk models, the different components appear superimposed in the observed intensity profile, when they are present. 

An analysis of the images produced and the corresponding intensity profiles allows one to trace several interesting remarks. First, it appears that the presence of the thin-shell affects the results only negligibly when a LR is absent. Indeed, for the three accretion disk models, one verifies that the top three images in Figs. \ref{fig:shadows_ISCO}, \ref{fig:shadows_LR}, and \ref{fig:shadows_C} are qualitatively similar. This similarity is also noticeable in the middle panel of Figs. \ref{fig:intensity_ISCO}, \ref{fig:intensity_LR}, and \ref{fig:intensity_C}, where all intensity profiles within this panel present a similar behavior. The only noticeable qualitative differences in these images and intensity profiles is a variation in the intensity of the secondary image, which is absent for the configuration $S_{55}$ and more noticeable for the configuration $S_{44}$, with $S_{54}$ representing an intermediate step between the two. Note that for the LR disk model the primary and secondary peaks are superimposed, and thus this feature is not as clear, although it must be present for consistency. 

The situation changes drastically when the configurations are compact enough to develop a LR. Indeed, the bottom rows of Figs. \ref{fig:shadows_ISCO}, \ref{fig:shadows_LR}, and \ref{fig:shadows_C} shows three images that are qualitatively different from the three images on the top and middle rows. For these configurations, the contribution of the secondary image to the produced image increases, with its respective peak attaining the same order of magnitude as the peak of the primary image, and the light-ring component is visible. These features of the intensity profiles are also visible in the right panels of Figs. \ref{fig:intensity_ISCO}, \ref{fig:intensity_LR}, and \ref{fig:intensity_C}, and the qualitative behavioural transition from configurations without LRs to configurations with LRs is also visible in the left panels of the same images. These results suggest that, unlike the thin-shell whose presence has a minor effect in the observational properties of these configurations when the LR is absent, the LR itself drastically affects the results and calls for a more detailed analysis regarding this transition. More details on this transition can be found in Appendix \ref{sec:appendixA}.

Finally, it is worth mentioning that, even though the thin-shells seems to have a negligible effect when the LR is absent, the same is not true regarding the configurations that feature a LR. Indeed, comparing the three images in the bottom row of Figs. \ref{fig:shadows_ISCO}, \ref{fig:shadows_LR}, and \ref{fig:shadows_C}, one verifies that the rightmost image, for which the thin-shell is absent, is qualitatively different from the two leftmost images. For the ISCO and LR disk models, one verifies that the configuration without a thin-shell features an extra circular structure in comparison to the configurations with a thin-shell, whereas for the Centre model one verifies that the configuration without a thin-shell features a strong central dimming of intensity which is not present in the configurations with a thin-shell. These differences are also visible in the right panels of Figs. \ref{fig:intensity_ISCO}, \ref{fig:intensity_LR}, and \ref{fig:intensity_C}, where the black solid curve corresponding to $S_{33}$ presents a qualitatively different behavior in comparison to $S_{53}$ and $S_{43}$. Since the presence of the thin-shell seems to affect the observational properties of the configurations when a LR is present, a more detailed analysis of this transition is also well motivated and can be found in Appendix \ref{sec:appendixA}.

\subsection{Comparison with the black-hole scenario}

From all of the configurations analyzed in this section, the configuration $S_{33}$ is the one that resembles the most the Schwarzschild spacetime. Indeed, both Schwarzschild and the configuration $S_{33}$ feature an ISCO and a LR in the same radial locations, and share the same circular orbital stability properties for massive test particles, i.e., circular orbits are stable for $r_o>6M$, unstable for $3M<r_o<6M$, and nonexistent for $r<3M$. There are, however, fundamental differences between the two spacetimes that have a strong impact on their observational properties, namely the regularity of $S_{33}$ and absence of an event horizon, in comparison with the Schwarzschild spacetime. Given the higher relevance of $S_{33}$ in comparison with the other configurations considered for astrophysical and observational purposes, we provide a comparison of the observational properties of $S_{33}$ and the Schwarzschild spacetime. We have also produced three images, one for each of the disk models considered, considering an observer with a certain inclination with respect to the vertical axis, in this case $\theta=80^\circ=4\pi/9$. We have not produced these images for the remaining configurations due to the large necessity of computational power. 

\begin{figure*}
    \centering
    \includegraphics[scale=0.5]{plot_r3_rs3_ISCO.pdf}
    \includegraphics[scale=0.5]{plot_r3_rs3_LR.pdf}
    \includegraphics[scale=0.5]{plot_r3_rs3_C.pdf}\\
    \includegraphics[scale=0.5]{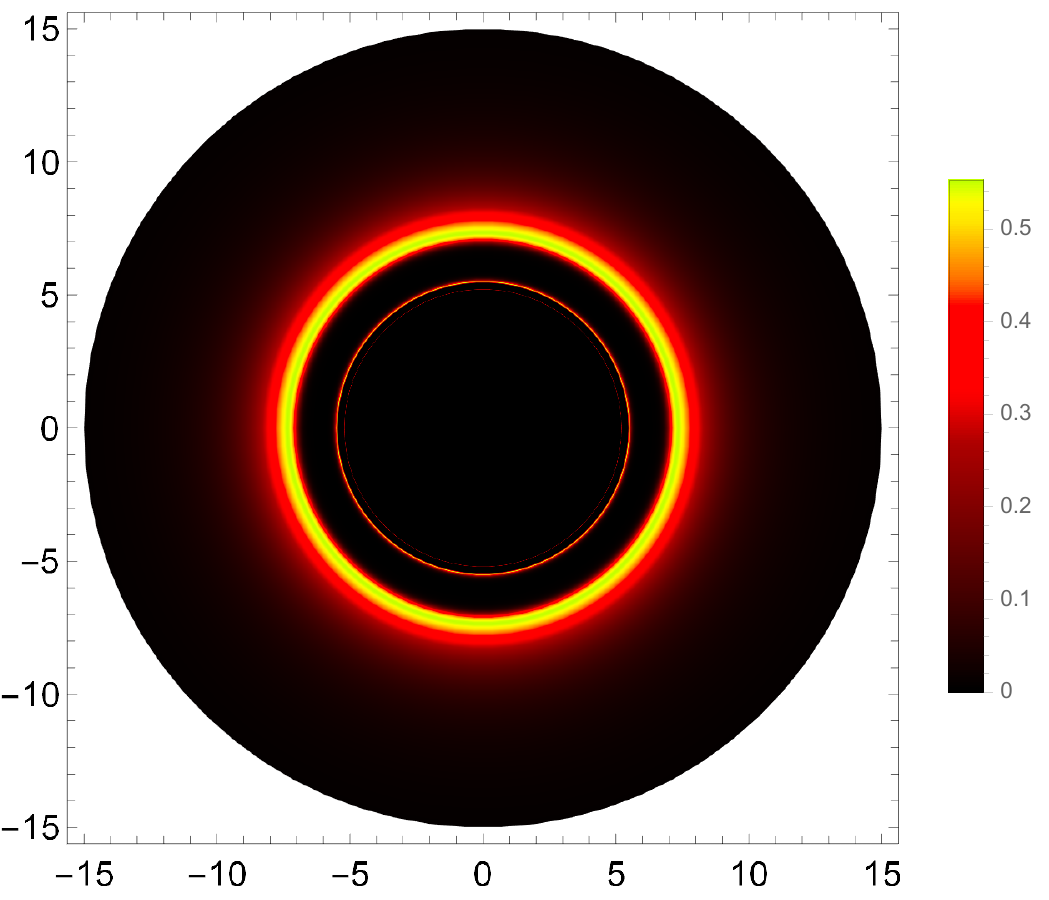}
    \includegraphics[scale=0.5]{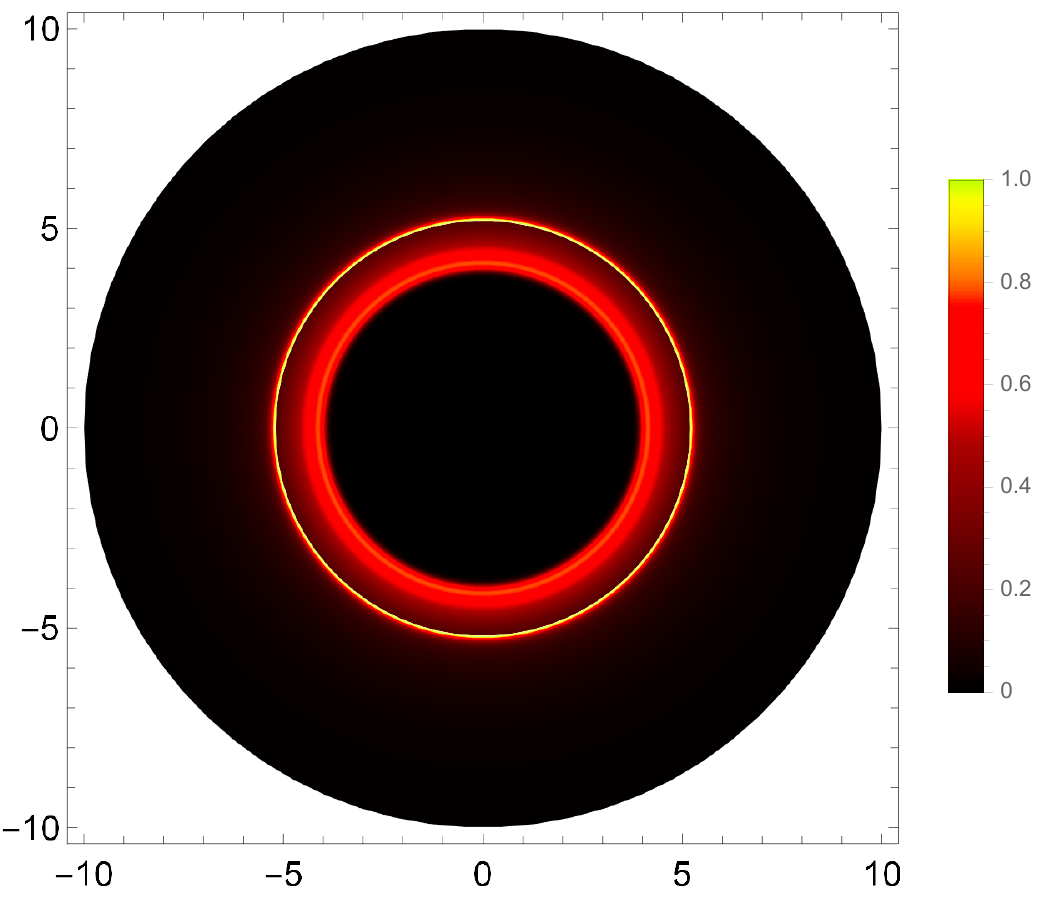}
    \includegraphics[scale=0.5]{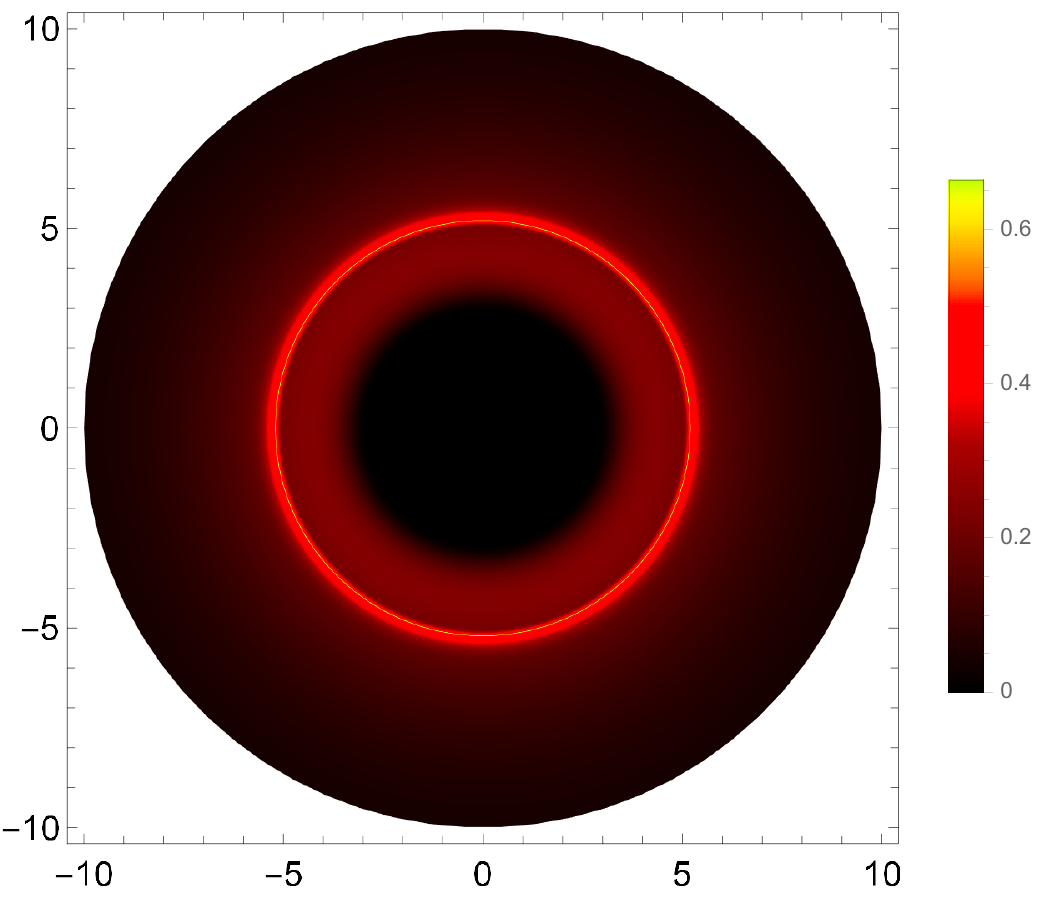}
    \caption{Comparison between the images produced for the $S_{33}$ configuration (top row) and the Schwarzschild spacetime (bottom row) for the ISCO disk model (left column), the LR disk model (middle column), and either the Centre (top right) or the EH (bottom right) disk models for an observer in the vertical axis $\theta=0$. Note that we have used the EH disk model instead of the Centre disk model in the Schwarzschild spacetime due to the existence of an event horizon.}
    \label{fig:axial_comp}
\end{figure*}

\begin{figure*}
    \centering
    \includegraphics[scale=0.5]{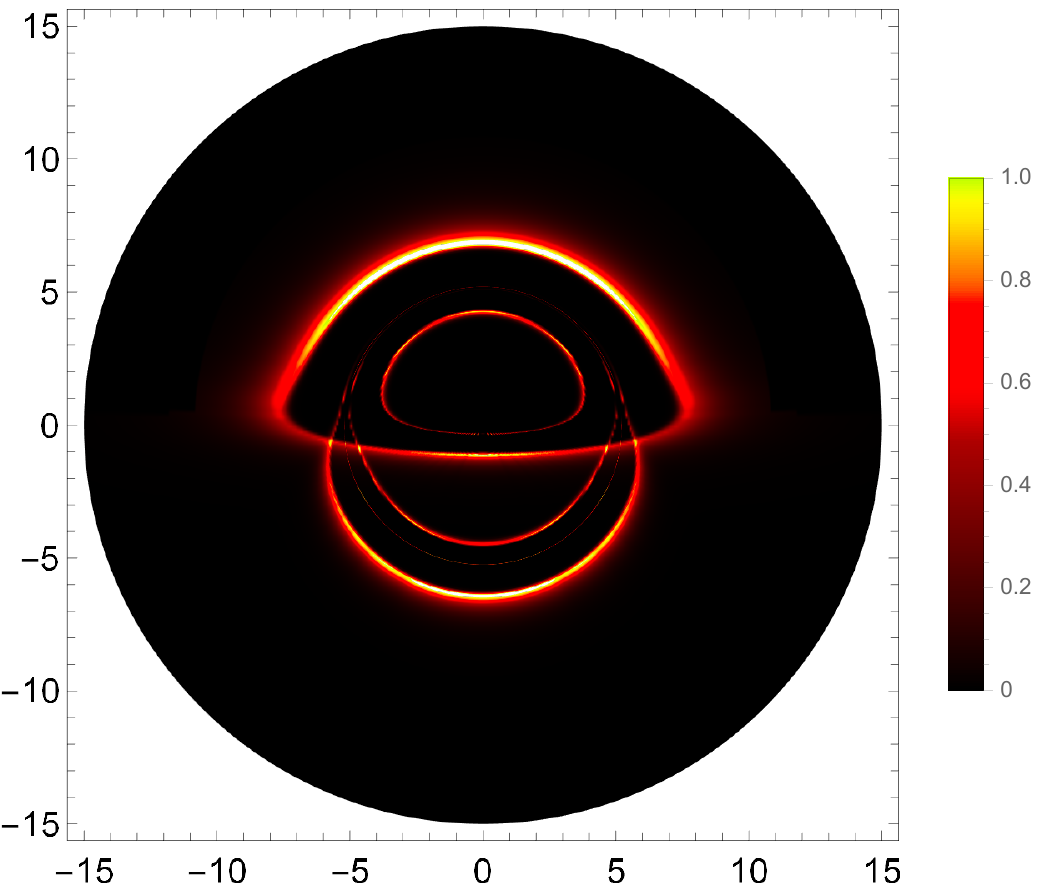}
    \includegraphics[scale=0.5]{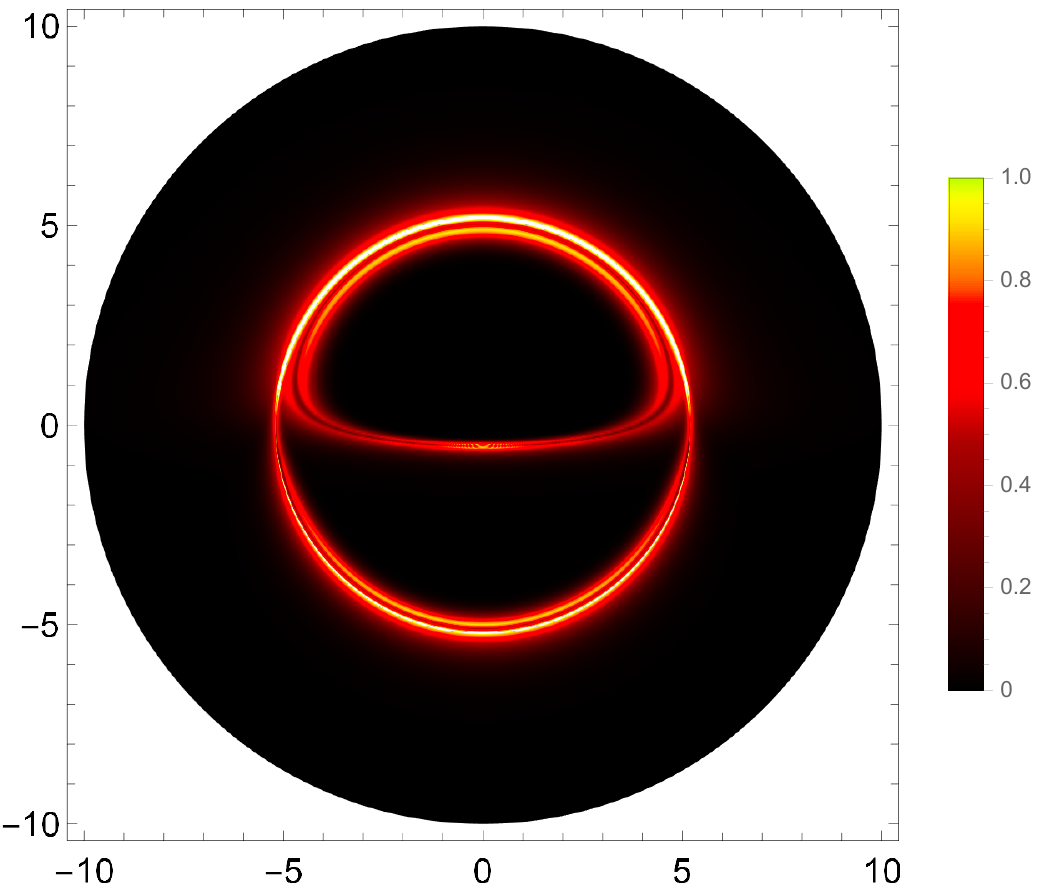}
    \includegraphics[scale=0.5]{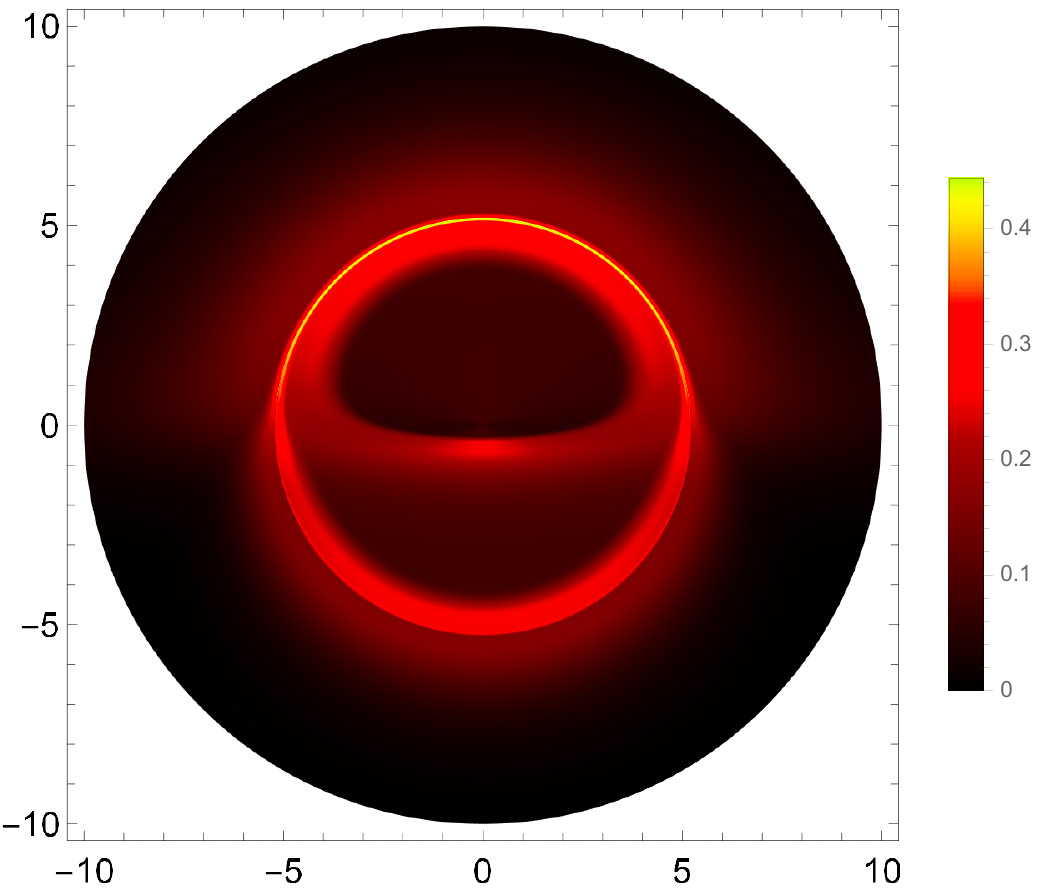}\\
    \includegraphics[scale=0.5]{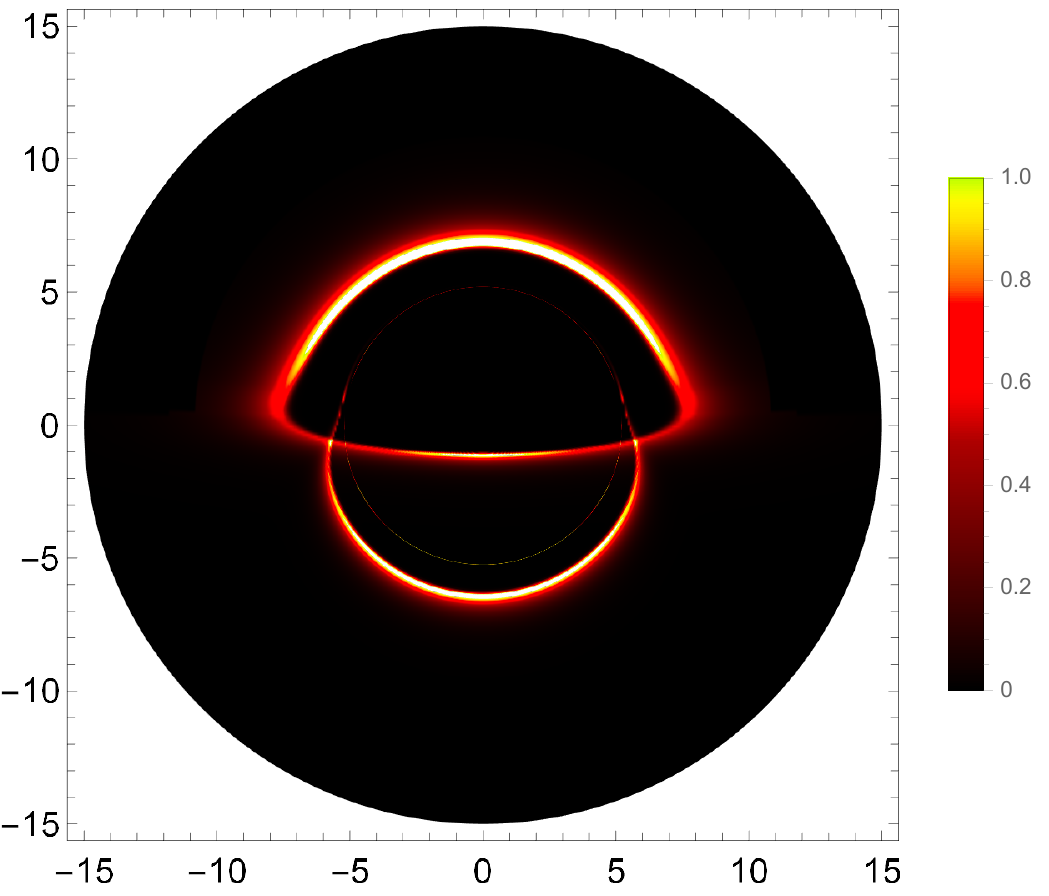}
    \includegraphics[scale=0.5]{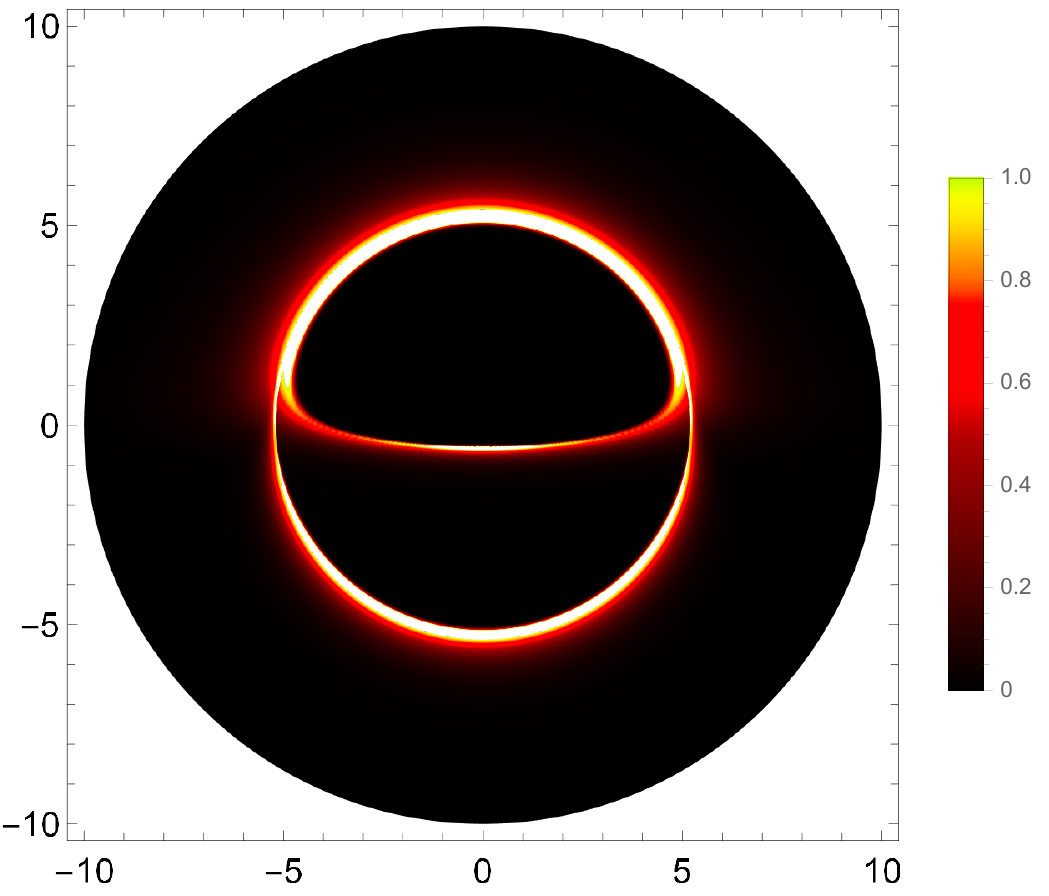}
    \includegraphics[scale=0.5]{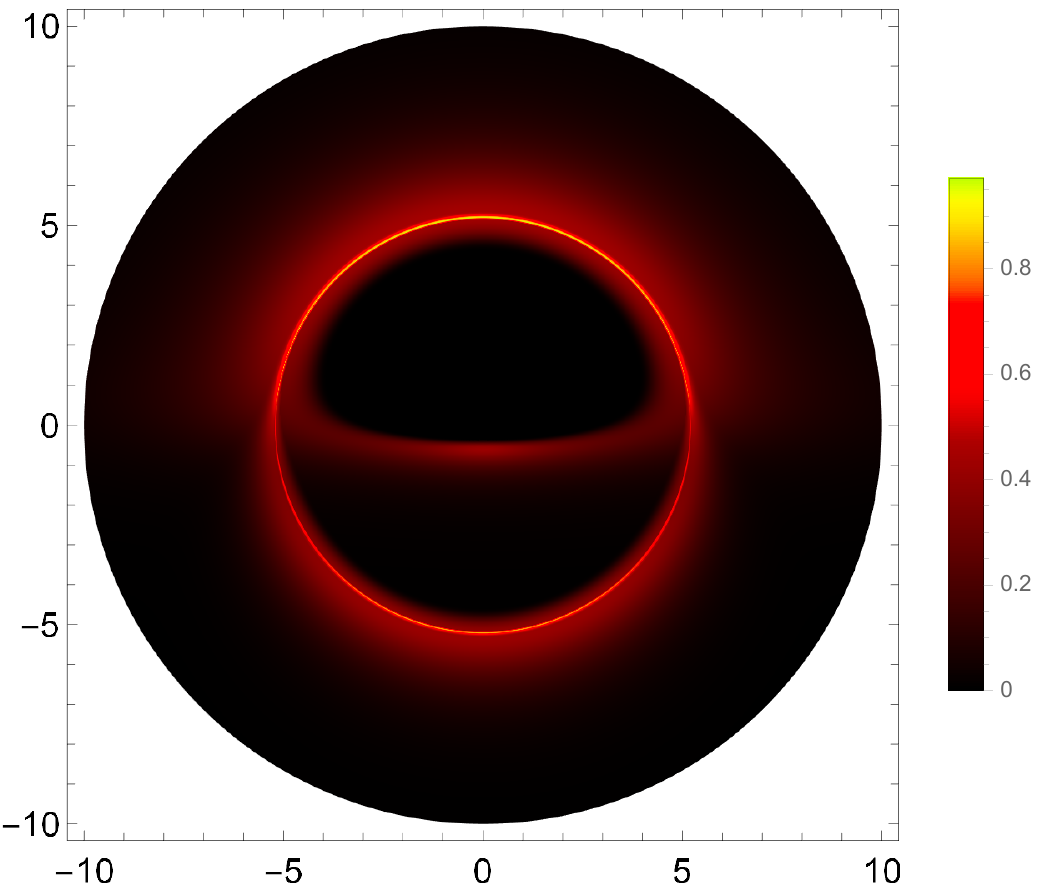}
    \caption{Comparison between the images produced for the $S_{33}$ configuration (top row) and the Schwarzschild spacetime (bottom row) for the ISCO disk model (left column), the LR disk model (middle column), and either the Centre (top right) or the EH (bottom right) disk models for an observer with an observation inclination of $\theta=4\pi/9$. Note that we have used the EH disk model instead of the Centre disk model in the Schwarzschild spacetime due to the existence of an event horizon.}
    \label{fig:angular_comp}
\end{figure*}

\begin{figure*}
    \centering
    \includegraphics[scale=0.63]{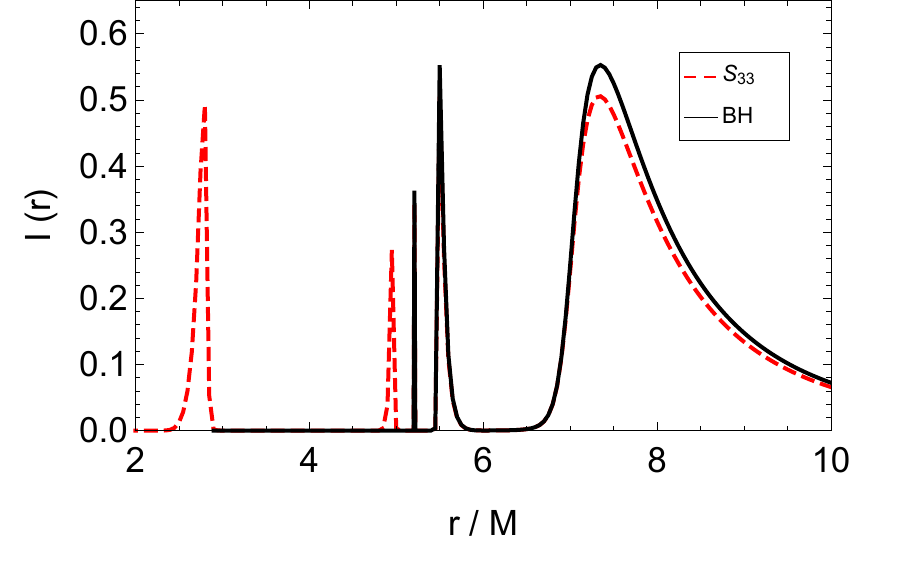}
    \includegraphics[scale=0.63]{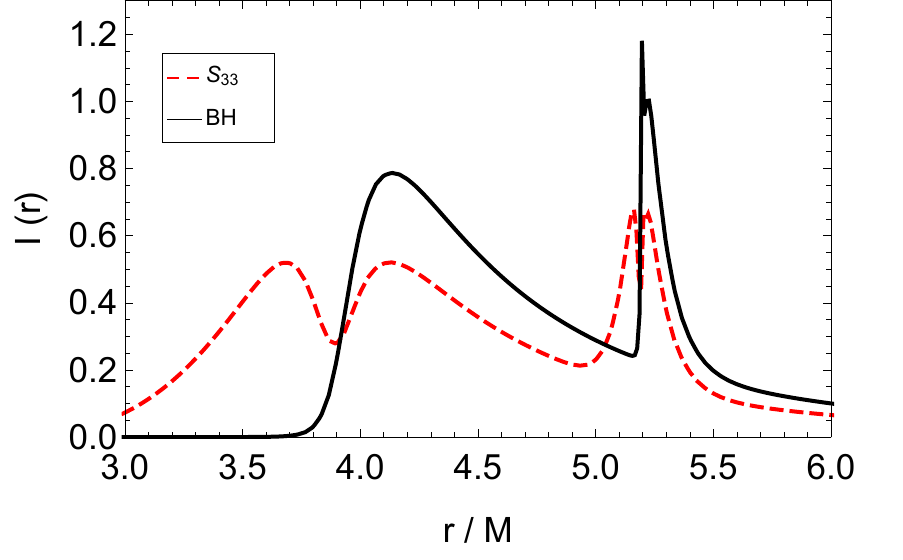}
    \includegraphics[scale=0.63]{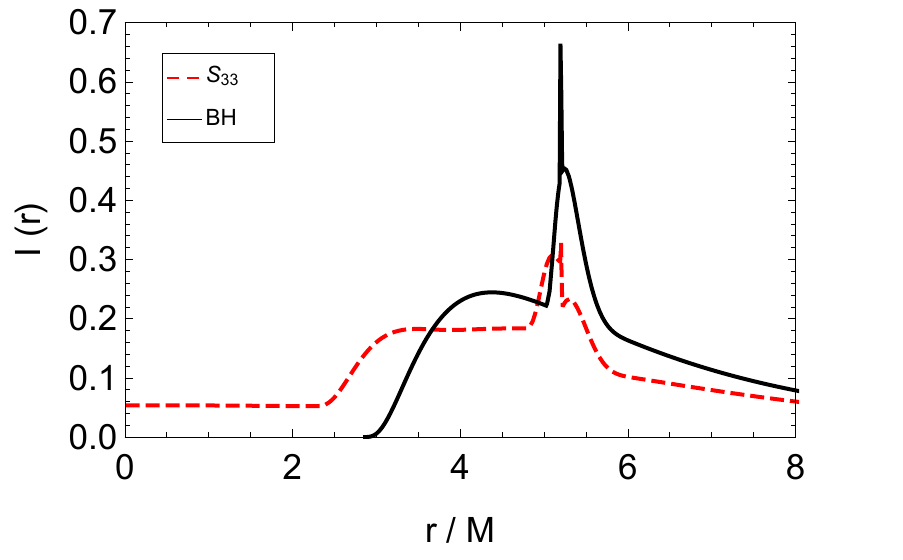}
    \caption{Comparison between the observed intensity profiles for the $S_{33}$ configuration (red dashed curve) and the Schwarzschild spacetime (black solid curve) for the ISCO disk model (left column), the LR disk model (middle column), and either the Centre or the EH disk models (right column). Note that we have used the EH disk model instead of the Centre disk model in the Schwarzschild spacetime due to the existence of an event horizon.}
    \label{fig:intensity_comp}
\end{figure*}

The comparison between the axial images of $S_{33}$ and the Schwarzschild spacetime is provided in Fig. \ref{fig:axial_comp}, inclined images are provided in Fig. \ref{fig:angular_comp}, and the comparison between the observed intensity profiles is given in Fig. \ref{fig:intensity_comp}. Note that since the Centre disk model is not physically adequate for the Schwarzschild spacetime, we have instead used the EH disk model in this comparison. For the three comparison setups, the main differences between the $S_{33}$ configuration and the Schwarzschild spacetime are the appearance of extra secondary components in the former. Indeed, for the ISCO and the LR disk models, one can clearly observe two extra secondary images in the $S_{33}$ that are absent from the Schwarzschild case, one of which close to the inner edge of the LR component (in the LR disk model, this is only perceivable in the observed intensity profiles), and another closer to the center of the image. These components are absent from the images of the Schwarzschild spacetime because their respective photons have an impact parameter smaller than the critical impact parameter $b_c=3\sqrt{3}M$ of the Schwarzschild spacetime, and thus they are absorbed by the event horizon, which is absent from the $S_{33}$ configuration. Thus, although the two spacetimes produce similar observational properties, e.g. they both feature a shadow-like dimming of radiation in the center and feature light-ring contributions, there are qualitative differences e.g. extra images and different shadow sizes, allowing one to tell them apart.

Although for the ISCO disk model the extra components of the secondary image are barely noticeable from an observational point of view due to limitations in the current interferometry capabilities (this situation may change in the future thanks to the next generation of experiments like the ngEHT \cite{Johnson:2019ljv}), for the LR disk model the extra inner component effectively decreases the radius of the observed shadow. A similar effect is visible for the Centre disk model in the $S_{33}$ configuration in comparison with the EH disk model in the Schwarzschild spacetime, for which the effects of gravitational redshift of $S_{33}$ produce a shadow-like dimming in the interior region of the image, but the radius of this feature is again smaller than the radius of the shadow in the Schwarzschild spacetime. Note however that these two effects, although similar in an observational context, have completely different origins: while in the LR disk model the shadow of $S_{33}$ is caused by the fact that the accretion disk is truncated at a finite radius, in the Centre disk model the shadow is caused by the gravitational redshift of the primary component.


\section{Conclusions}\label{sec:concl}

In this work we have studied the observational properties of a recently proposed family of incompressible relativistic fluid spheres surrounded by optically thin accretion disks, i.e., transparent to their own radiation. In particular, we have studied how the presence of a thin-shell and a LR affect the corresponding observational properties, and we have made a qualitative comparison of the results with the Schwarzschild scenario. 

Our results indicate that when the configurations are not compact enough to develop a LR, the presence of a thin-shell affects only negligibly the observational properties of these spacetimes, and thus should be hard to detect due to the limited resolution of the EHT. On the other hand, the presence of a LR contributes significantly to a qualitative change in the appearance of these configurations, not only providing extra contributions to the observed intensity profile but also in modifying the shape of the contributions already present before the LR develops. Also, when the configurations feature a LR, the otherwise negligible contribution of the thin-shell becomes important, as the ratio between the mass stored at the thin-shell and distributed in the volume of the star also induce qualitative changes to the observed intensity profile and images. 

From all of the configurations analyzed, one turned to be the most physically relevant in comparison with the black-hole scenario, namely $S_{33}$, a configuration with a radius $R=3M$ and without a thin-shell. For all of the accretion disk models considered, $S_{33}$ produced the observed intensity profiles and images with the greatest resemblance to the ones obtained for the Schwarzschild spacetime. Nevertheless, a few qualitative differences were pin-pointed, namely, the existence of additional secondary image contributions in $S_{33}$ that decrease the overall size of the observed shadow non-negligibly, thus potentially offering a framework to distinguish between these two spacetimes.

One of the most interesting outcomes of this analysis is the fact that, when compact enough e.g. the $S_{33}$ case, these configurations produce a shadow-like dimming in the intensity profiles even if one assumes that the radiation emission peaks at the center of these objects. This dimming is caused by the gravitational redshift and was also previously observed for bosonic star spacetimes. The possibility of shadow-like features emerging in spacetimes without event horizons motivates a further study of such horizonless compact objects as suitable alternatives to the black-hole scenario that could potentially be compatible with the recent and future observations of the EHT.

To conclude, this work aims to motivate the study of fluid stars as possible alternatives to the black-hole scenario and to provide an elementary first step towards a more detailed and physically robust analysis of these configurations.  Indeed, more realistic accretion disk emission profiles taking into account accretion rates and magneto-hydrodynamics could potentially change the observational predictions and lead to a stronger constraint on the configurations considered. On the other hand, other experimental results e.g. the observation of infrared flares near the galactic centre by the GRAVITY collaboration also provide a suitable framework to analyze and constraint the models considered here via the study of the observational properties of hot-spots orbiting these configurations. This analysis is currently being tackled and we hope to report it in the near future.

\begin{acknowledgments}
We thank Diego Rubiera-Garcia for important comments and suggestions. The numerical code used in this work was developed in collaboration with Gonzalo J. Olmo and Diego Rubiera-Garcia. JLR acknowledges the European Regional Development Fund and the programme Mobilitas Pluss for financial support through Project No.~MOBJD647, and project No.~2021/43/P/ST2/02141 co-funded by the Polish National Science Centre and the European Union Framework Programme for Research and Innovation Horizon 2020 under the Marie Sklodowska-Curie grant agreement No. 94533.
\end{acknowledgments}

\pagebreak

\appendix

\section{Phase transitions in observational properties}\label{sec:appendixA}

In Sec. \ref{sec:shadows} we have depicted a few general statements about how the presence of the thin-shell and the LR affect the observational properties of the relativistic fluid stars considered in this work. In particular, we stated that the effects of the thin-shell are negligible when the LR is absent, but that there is a quick qualitative behavioral transition of the intensity profiles and produced images when the LR is present. Thus, in this section we aim to clarify the details of such transitions, by analyzing more closely how small variations of the parameters $r_\Sigma$ and $R$ affect the qualitative behavior of the intensity profiles and images produced close to the parameter region where the transition occurs. In the following subsections, three transitions are analyzed, namely: (i) the development of a LR in the absence of a thin-shell; (ii) the development of a LR in the presence of a thin-shell; and (iii) the development of a thin-shell in the presence of a LR. Note that we do not consider the development of a thin-shell in the absence of a LR since the results in the previous section suggest that such a transition induces negligible qualitative effects.

\subsection{Development of a LR in the absence of a thin-shell}

\begin{figure*}[h]
    \centering
    \includegraphics[scale=0.5]{plot_r4_rs4_ISCO.pdf}
    \includegraphics[scale=0.5]{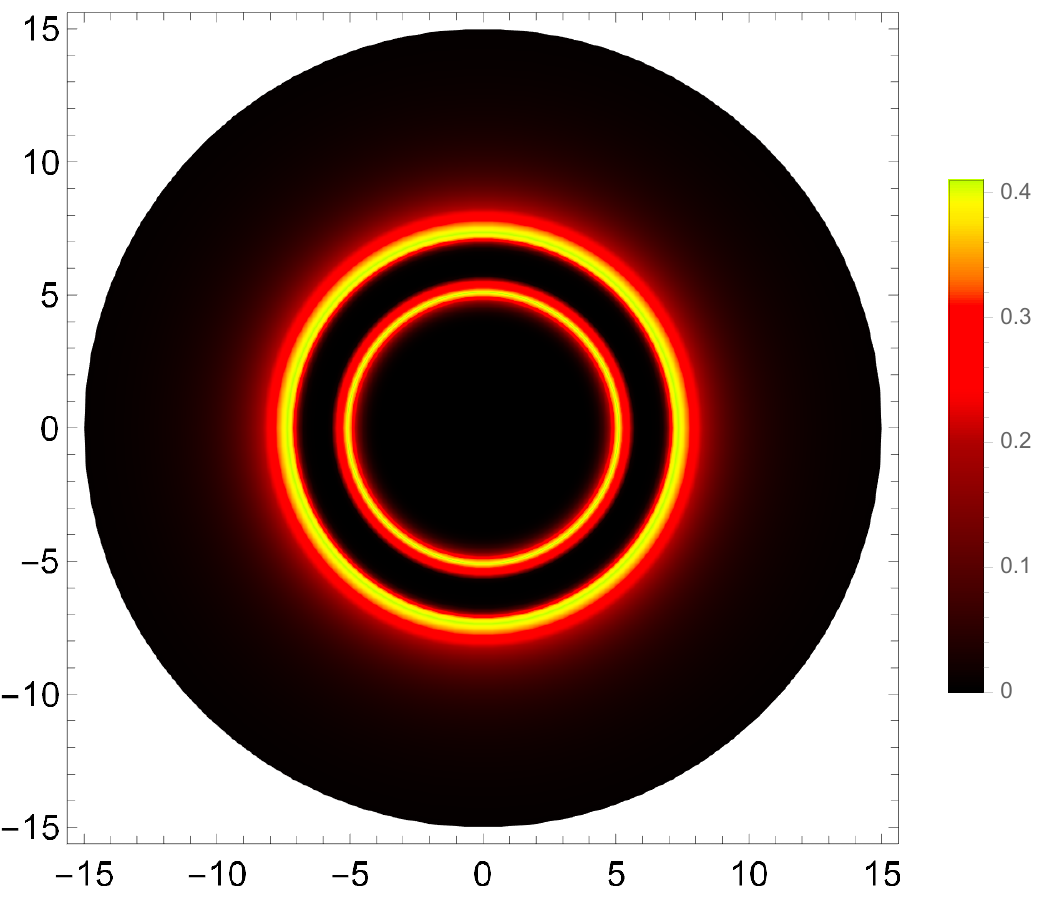}
    \includegraphics[scale=0.5]{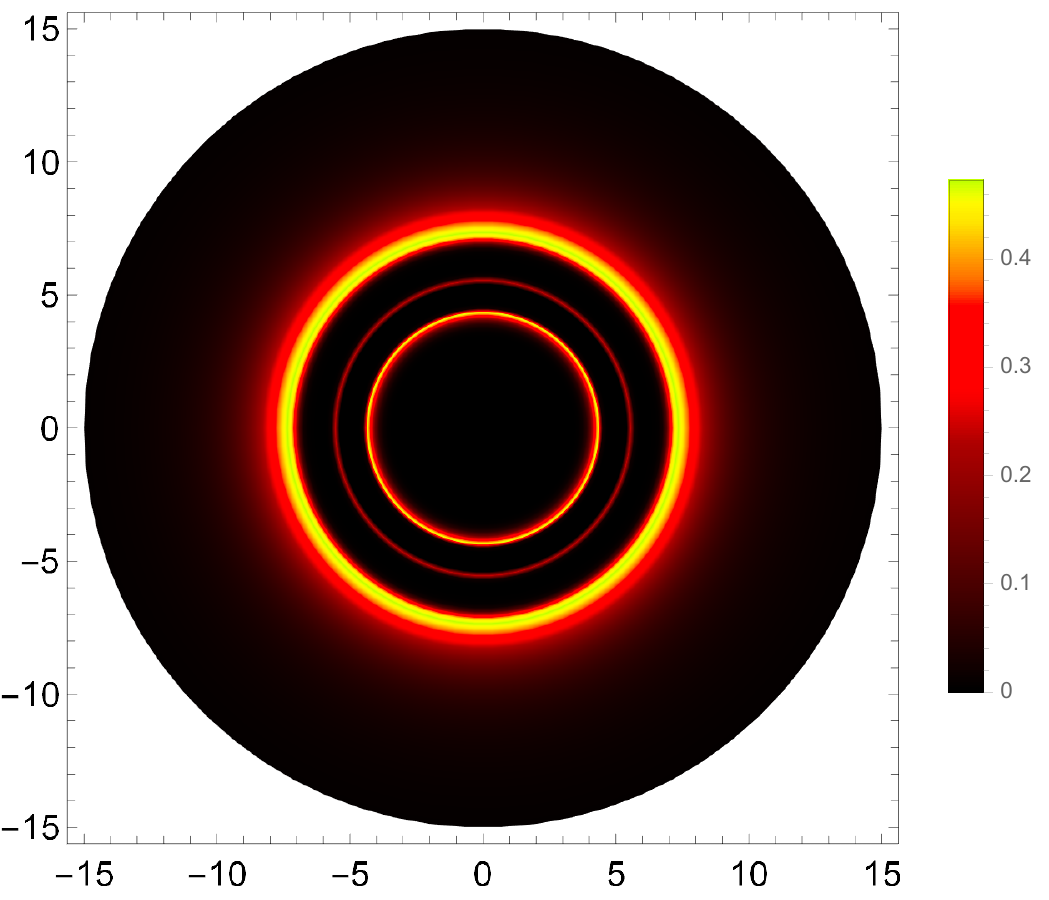}\\
    \includegraphics[scale=0.5]{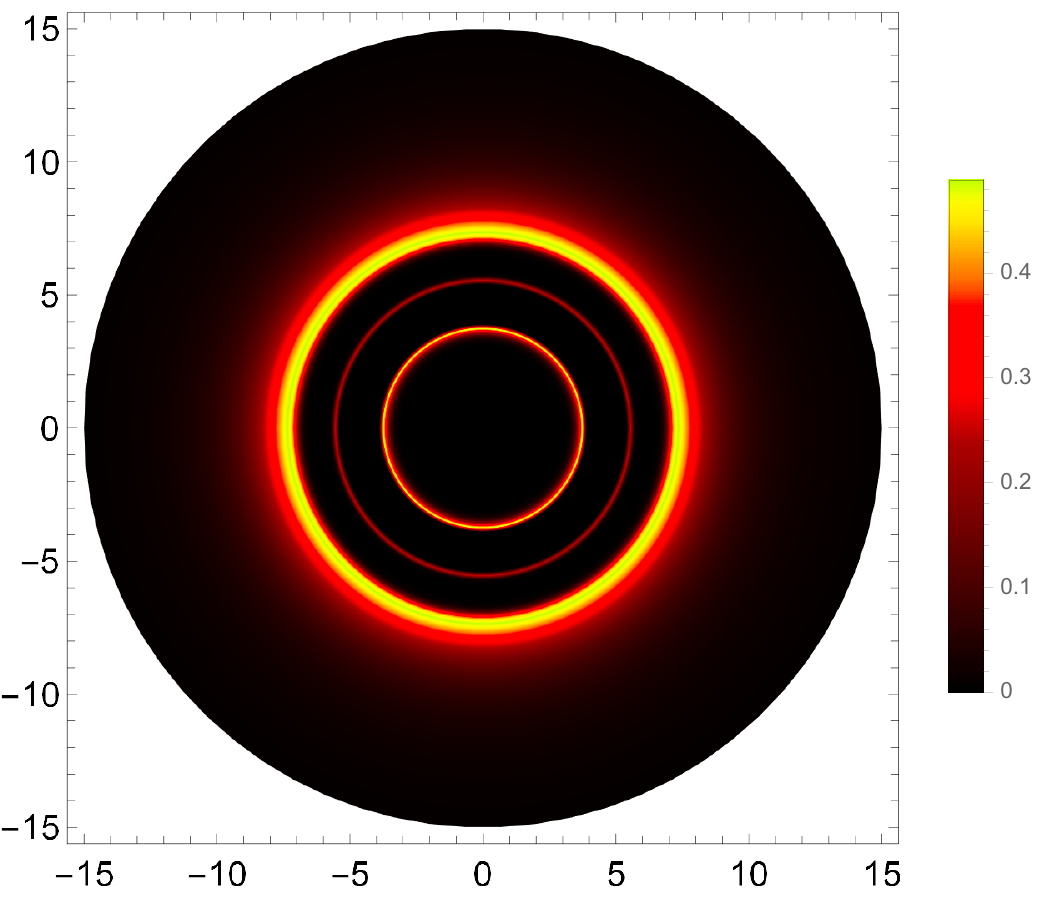}
    \includegraphics[scale=0.5]{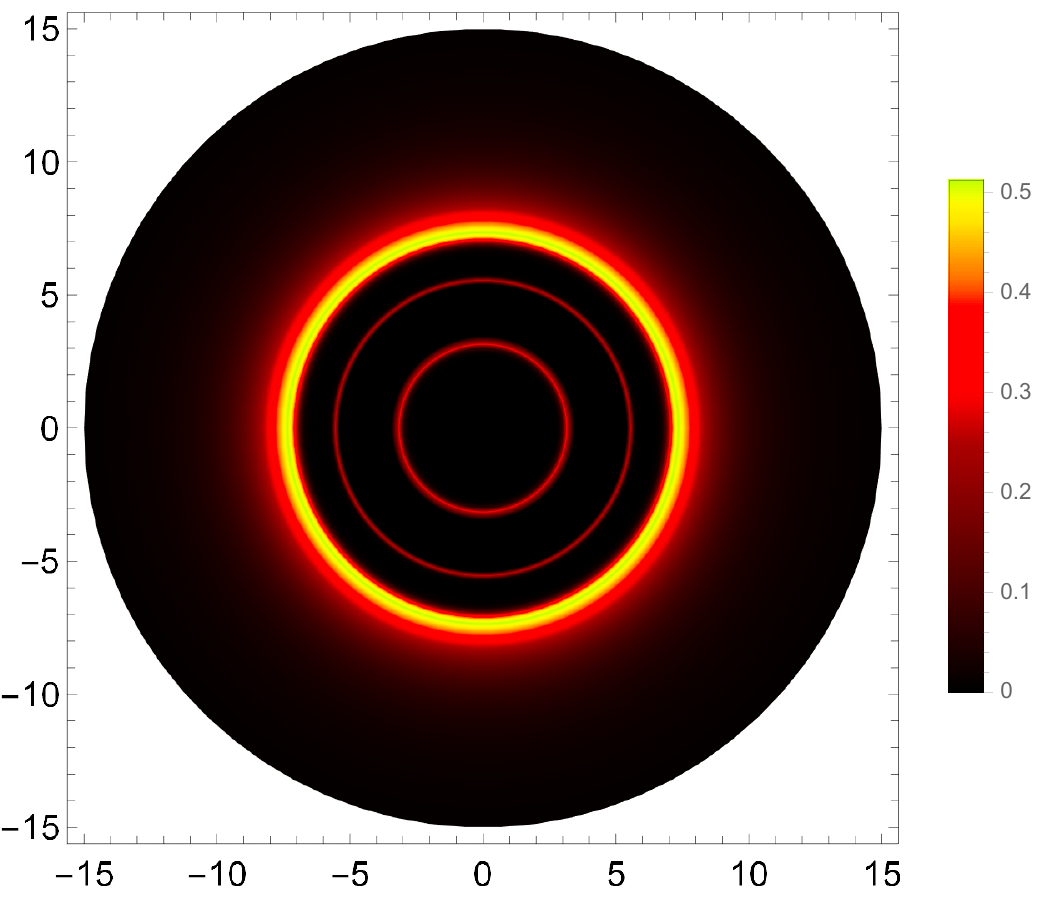}
    \includegraphics[scale=0.5]{plot_r3_rs3_ISCO.pdf}
    \caption{Shadows images with the Centre accretion disk model for configurations with $r_\Sigma=R$ from $R=4M$ to $R=3M$ in steps of $0.2M$, starting in the top left corner down to the bottom right corner.}
    \label{fig:pt1_shadows_ISCO}
\end{figure*}

\begin{figure*}[h]
    \centering
    \includegraphics[scale=0.5]{plot_r4_rs4_LR.pdf}
    \includegraphics[scale=0.5]{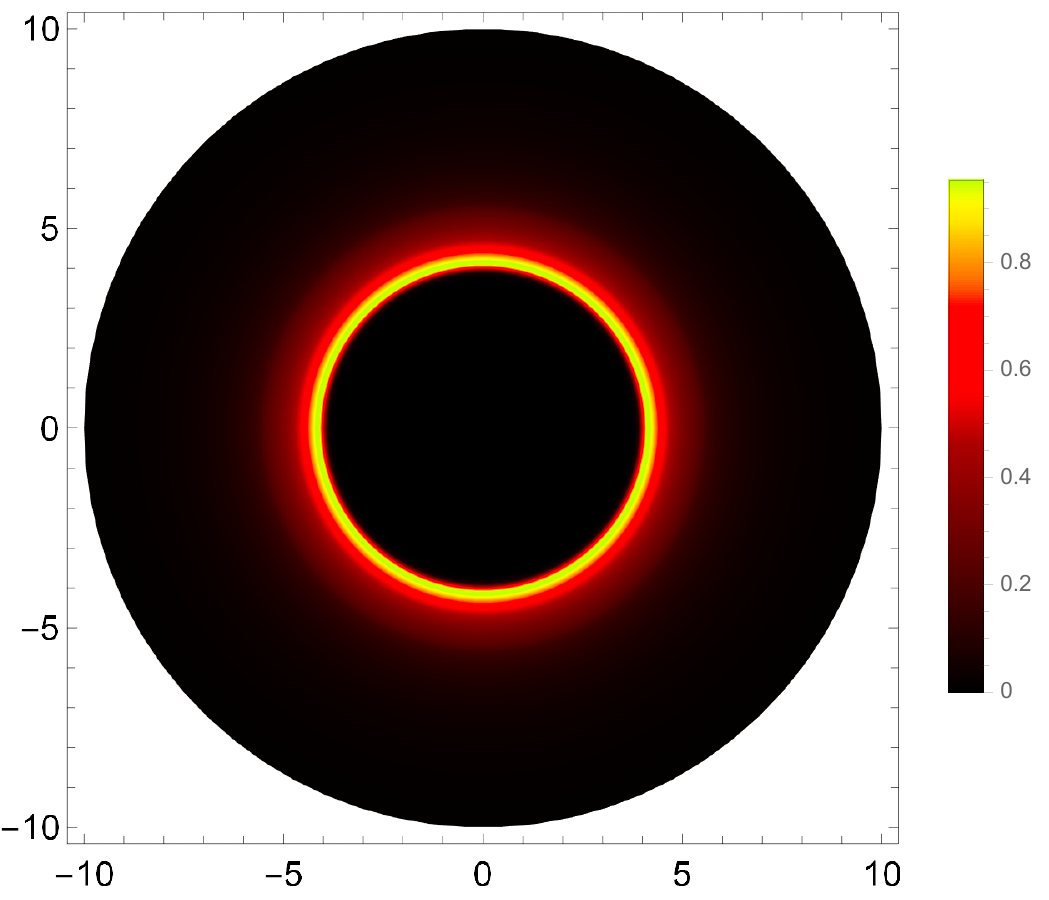}
    \includegraphics[scale=0.5]{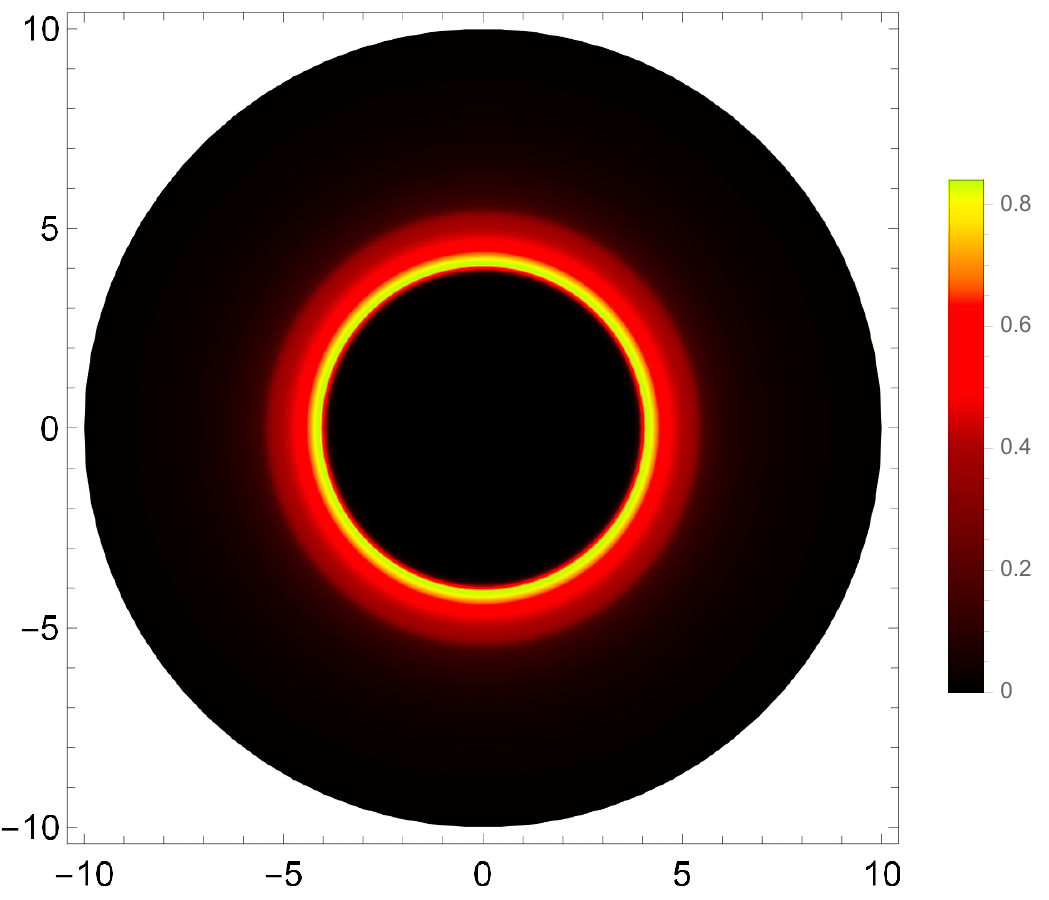}\\
    \includegraphics[scale=0.5]{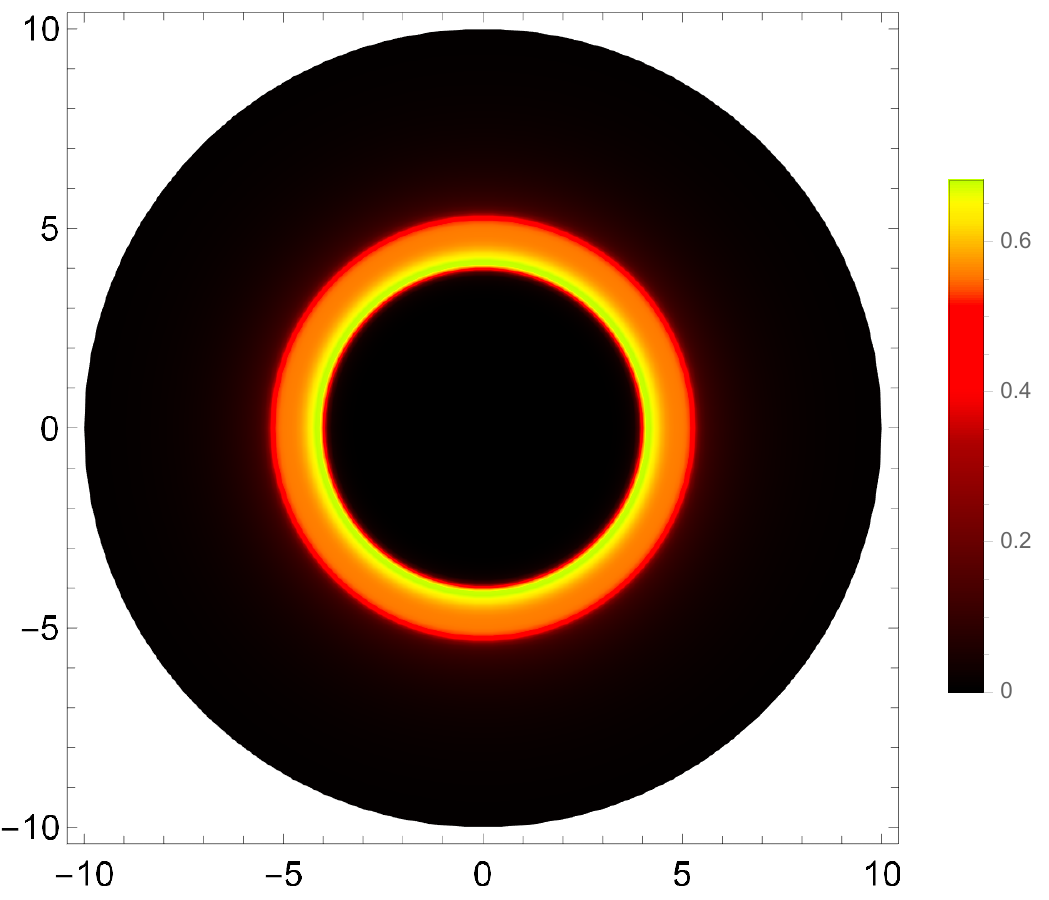}
    \includegraphics[scale=0.5]{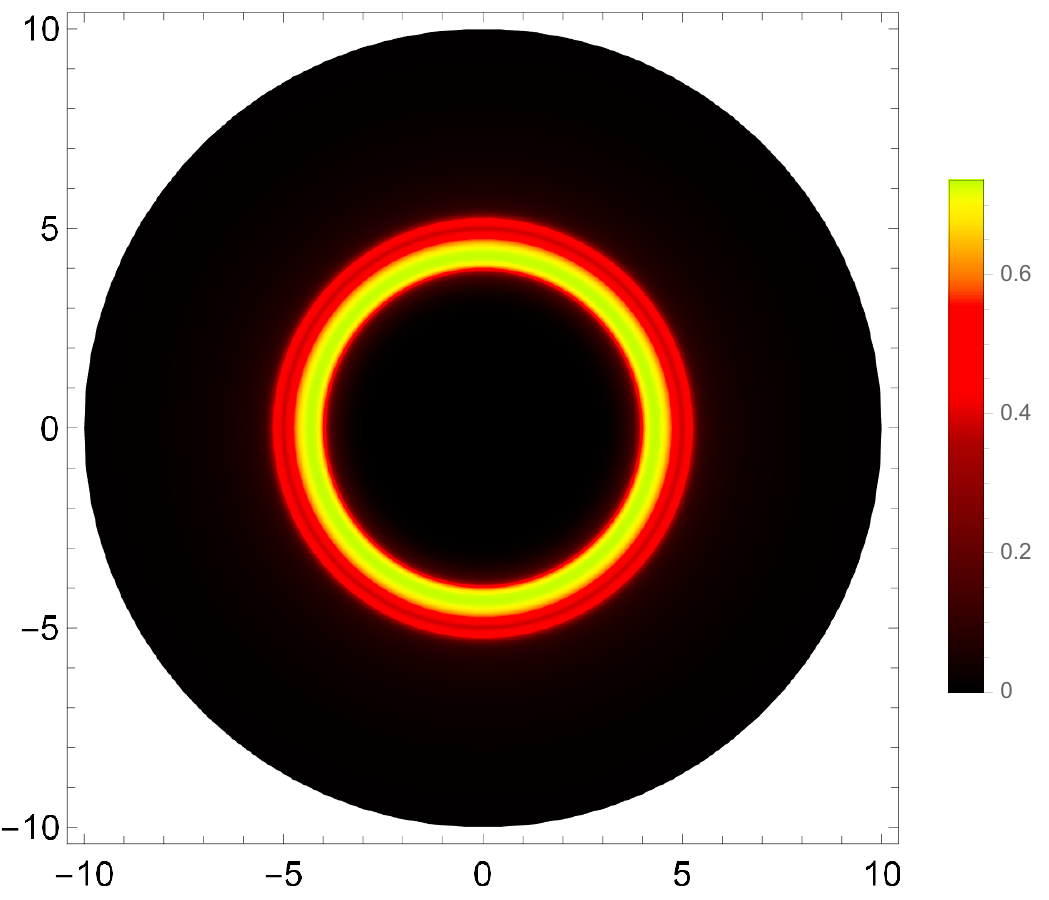}
    \includegraphics[scale=0.5]{plot_r3_rs3_LR.pdf}
    \caption{Shadows images with the Centre accretion disk model for configurations with $r_\Sigma=R$ from $R=4M$ to $R=3M$ in steps of $0.2M$, starting in the top left corner down to the bottom right corner.}
    \label{fig:pt1_shadows_LR}
\end{figure*}

\begin{figure*}[h]
    \centering
    \includegraphics[scale=0.5]{plot_r4_rs4_C.pdf}
    \includegraphics[scale=0.5]{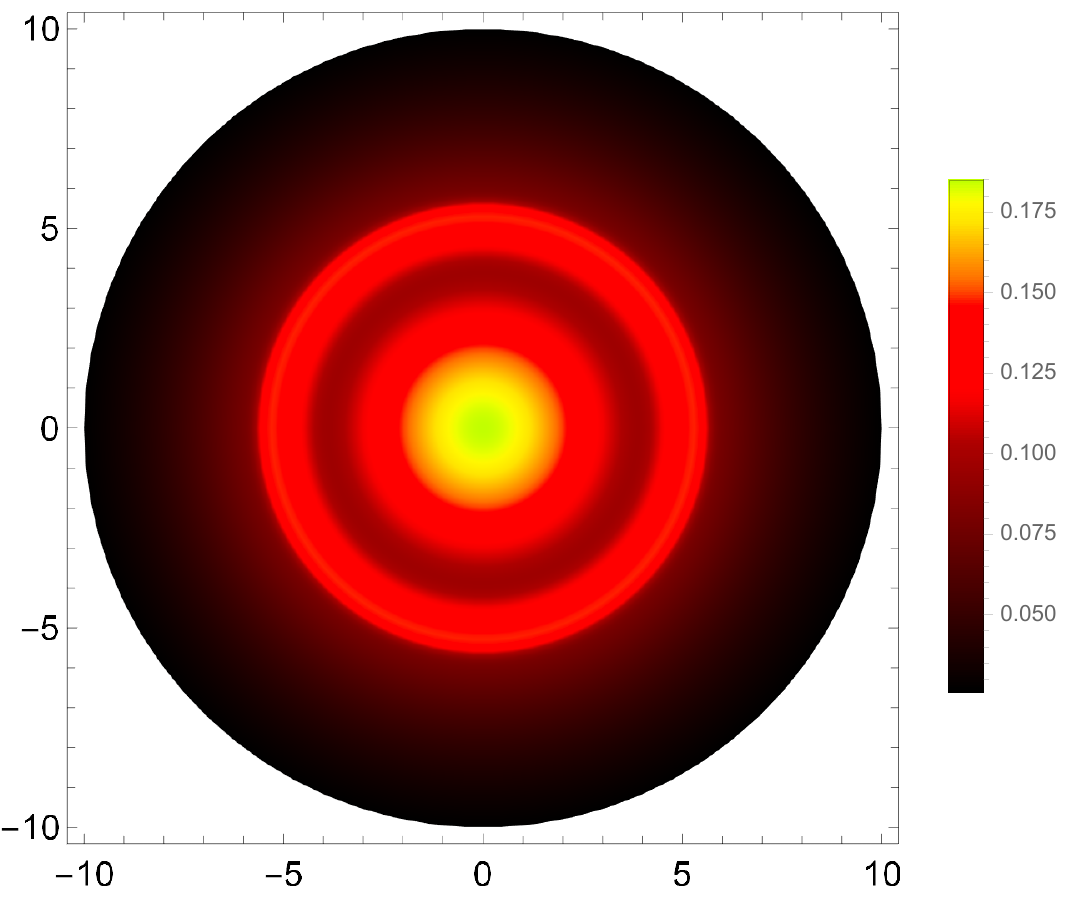}
    \includegraphics[scale=0.5]{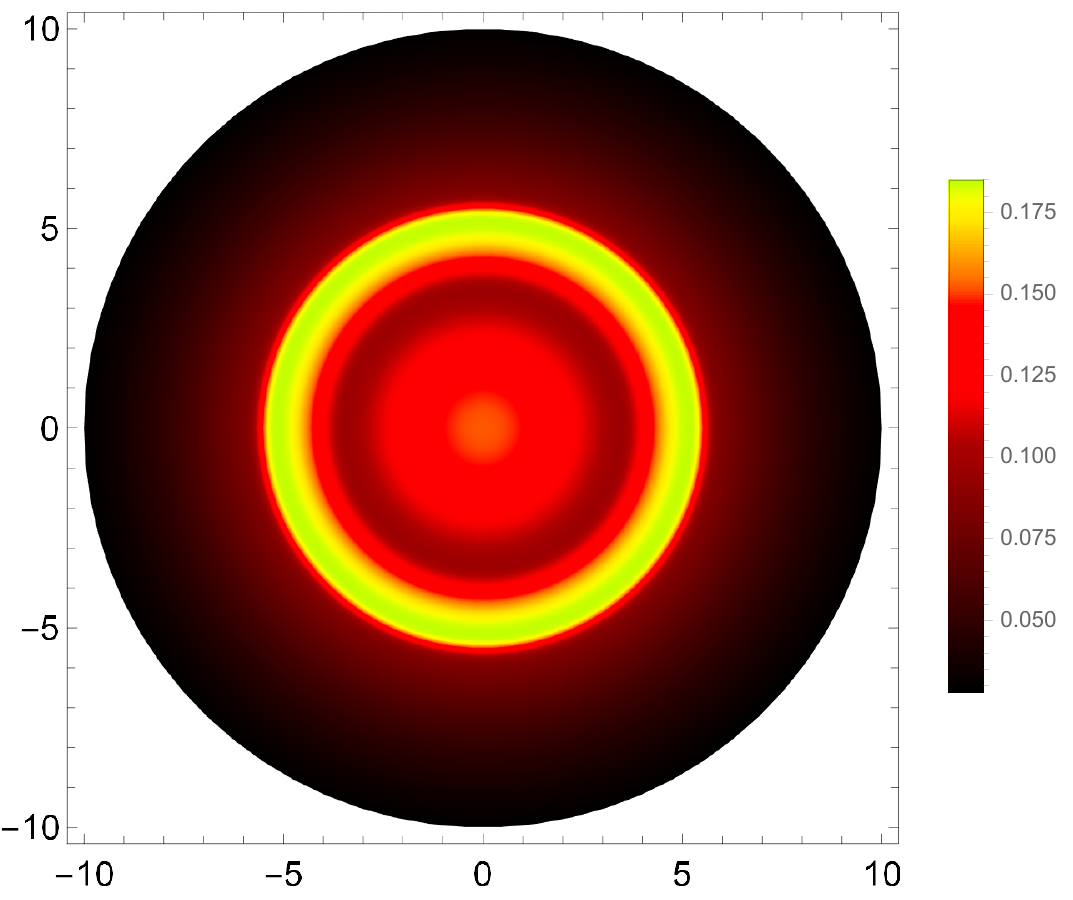}\\
    \includegraphics[scale=0.5]{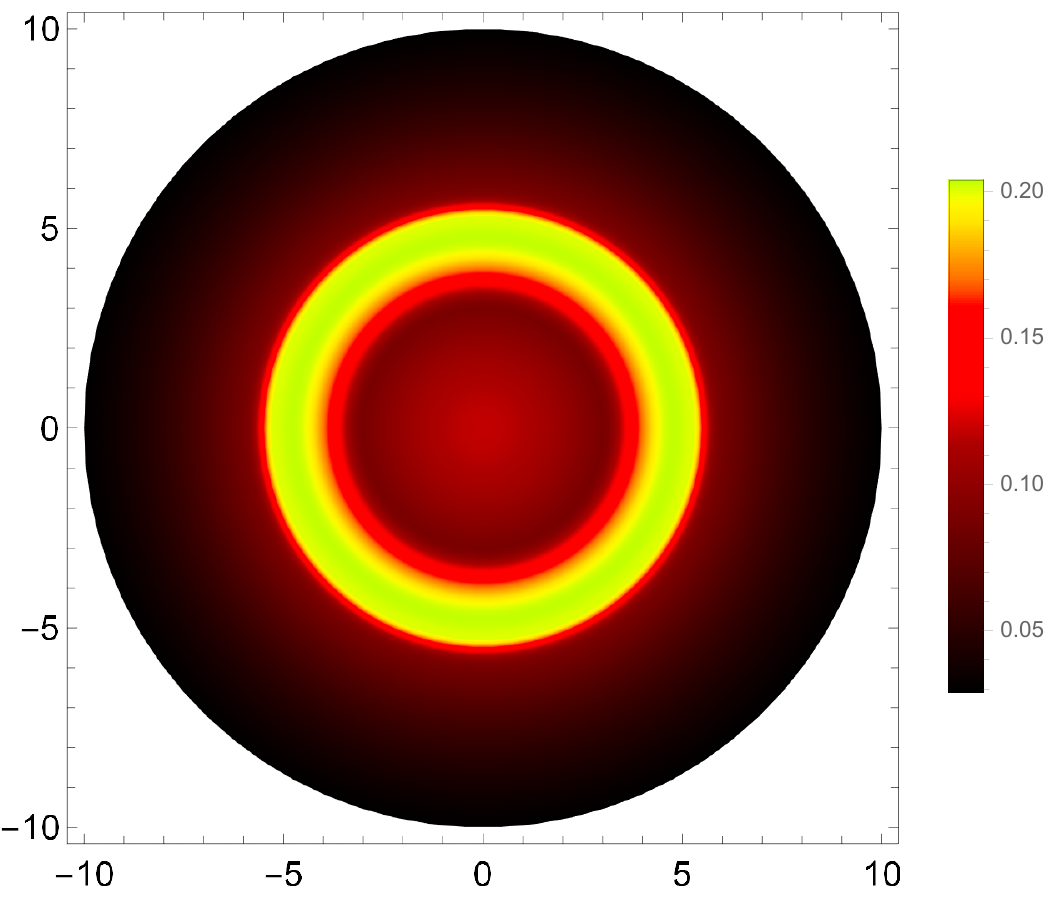}
    \includegraphics[scale=0.5]{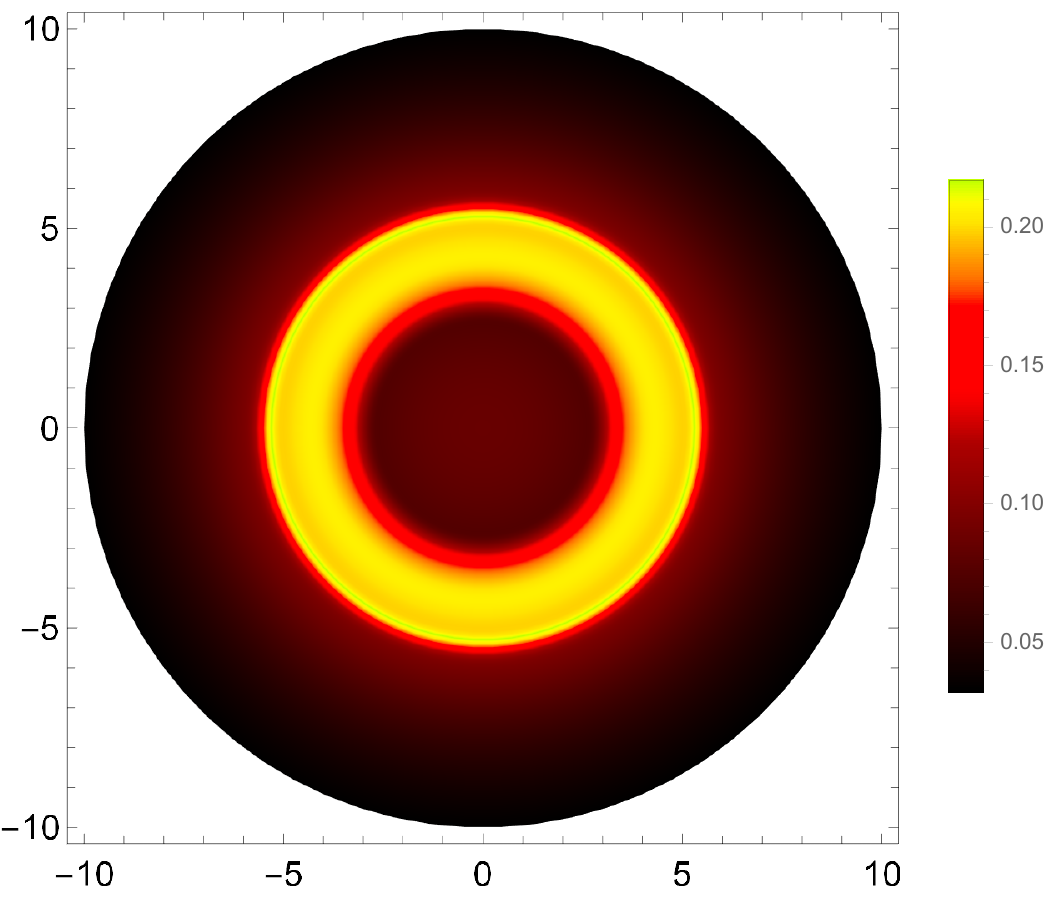}
    \includegraphics[scale=0.5]{plot_r3_rs3_C.pdf}
    \caption{Shadows images with the Centre accretion disk model for configurations with $r_\Sigma=R$ from $R=4M$ to $R=3M$ in steps of $0.2M$, starting in the top left corner down to the bottom right corner.}
    \label{fig:pt1_shadows_C}
\end{figure*}

\begin{figure*}[h]
    \centering
    \includegraphics[scale=0.63]{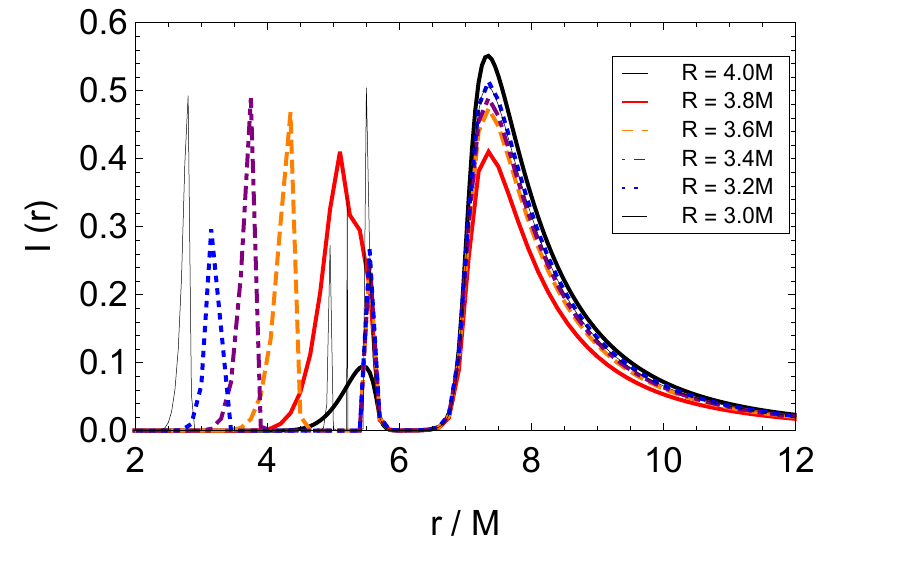}
    \includegraphics[scale=0.63]{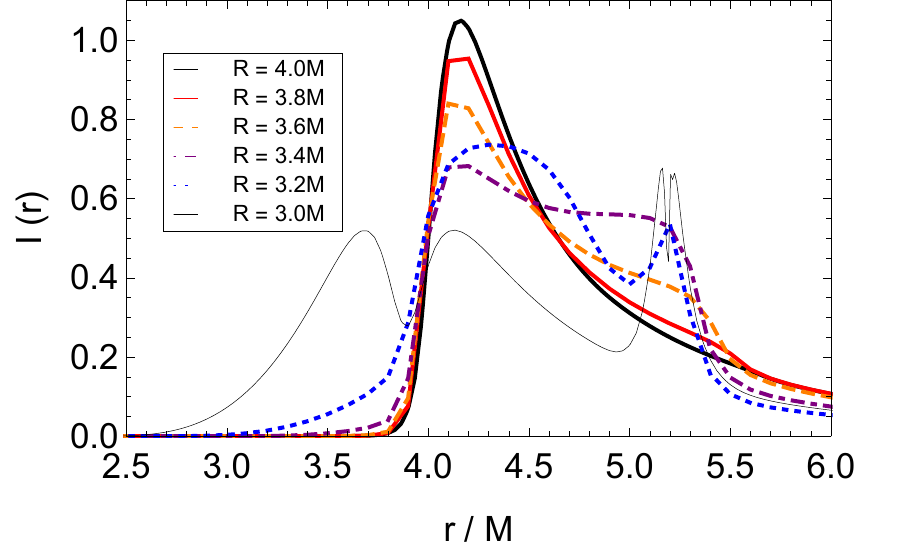}
    \includegraphics[scale=0.63]{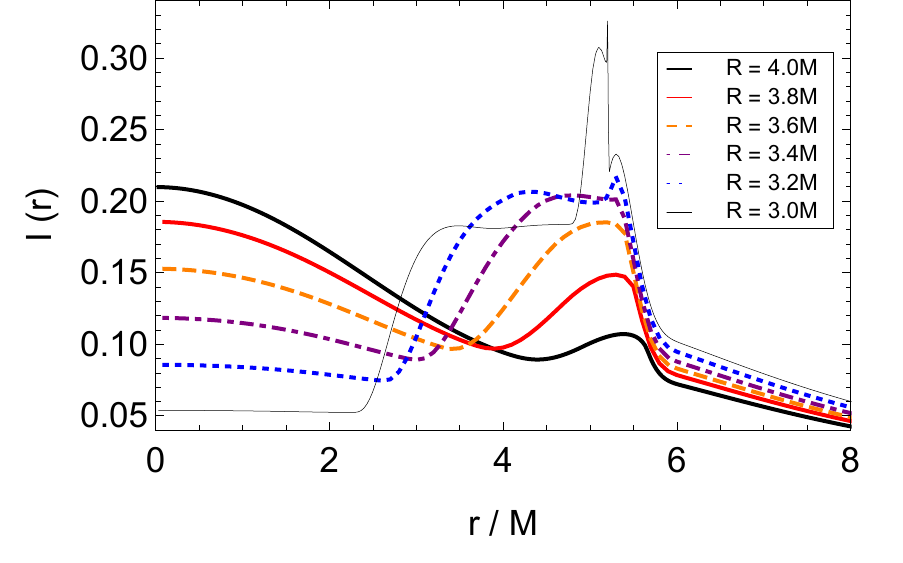}
    \caption{Observed intensity profiles $I_o$ as a function of the normalized radial coordinate $r/M$ with the ISCO (left panel), LR (middle panel), and Centre (right panel) accretion disk models, for configurations with $r_\Sigma=R$ from $R=4M$ to $R=3M$ in steps of $0.2M$.}
    \label{fig:pt1_intensity}
\end{figure*}

To analyze how the appearance of a LR qualitatively affects the observational properties of our configurations in the absence of a thin-shell, i.e., the transition that occurs between the configurations $S_{44}$ and $S_{33}$ in the previous section, we consider four extra configurations with $r_\Sigma=R=\{3.8M;3.6M;3.4M,3.2M\}$, and perform the ray-tracing analysis described in the previous sections. One can think of the analysis in this section as a study of how the increase of the density of the relativistic fluid that these stars are composed of affects their observational properties, while the total mass $M$ is maintained constant. The images produced for these configurations are given in Fig. \ref{fig:pt1_shadows_ISCO} for the ISCO disk model, Fig. \ref{fig:pt1_shadows_LR} for the LR disk model, and Fig. \ref{fig:pt1_shadows_C} for the Centre disk model. The corresponding intensity profiles are given in Fig. \ref{fig:pt1_intensity}.

Figures \ref{fig:pt1_shadows_ISCO} to \ref{fig:pt1_intensity} clarify the stages of the transition between configurations without LRs to configurations with LRs in the absence of a thin-shell. For the ISCO disk model, one verifies that as the compacticity of the star increases, there is an increase in the intensity of the secondary image, until eventually it splits into two separate secondary images, at $R=3.6M$. The outer component remains in the same location independently of the compacticity, whereas the inner component moves radially inwards when the compacticity increases. Finally, when the solutions are compact enough to develop a LR, the outer component of the secondary image splits again giving rise to a third component, alongside with the light-ring component. On the other hand, for the LR disk model one observes a broadening of the dominant image contribution, corresponding to a superposition of the primary and secondary images, as the compacticity increases. It is thus not clear if the secondary image splits into several components at $R=3.6M$ due to this superposition, although we expect such a behavior to be true for consistency with the results for the ISCO disk model. At $R=3.2M$, the split of the secondary image in two components becomes finally visible, and the inner component moves radially inwards when the LR forms, while the outer component suffers a second splitting, consistently with the ISCO model. Finally, for the Centre disk model, one observes an increase of the intensity and broadening of the secondary image the compacticity of the star, followed by a decrease in the intensity of the primary image, the latter caused primarily by an increase in the gravitational redshift effect, until finally the light-ring develops. A first splitting of the secondary image can be observed at $R=3.2M$, followed by a second splitting at $R=3M$, consistently with the LR model. Summarizing, these results show that although the qualitative behavior of the intensity profiles and images differs strongly between the configurations $S_{44}$ and $S_{33}$, the transition between the two behaviors is smooth instead of abrupt.

\subsection{Development of a LR in the presence of a thin-shell}

\begin{figure*}[h]
    \centering
    \includegraphics[scale=0.5]{plot_r4_rs4_ISCO.pdf}
    \includegraphics[scale=0.5]{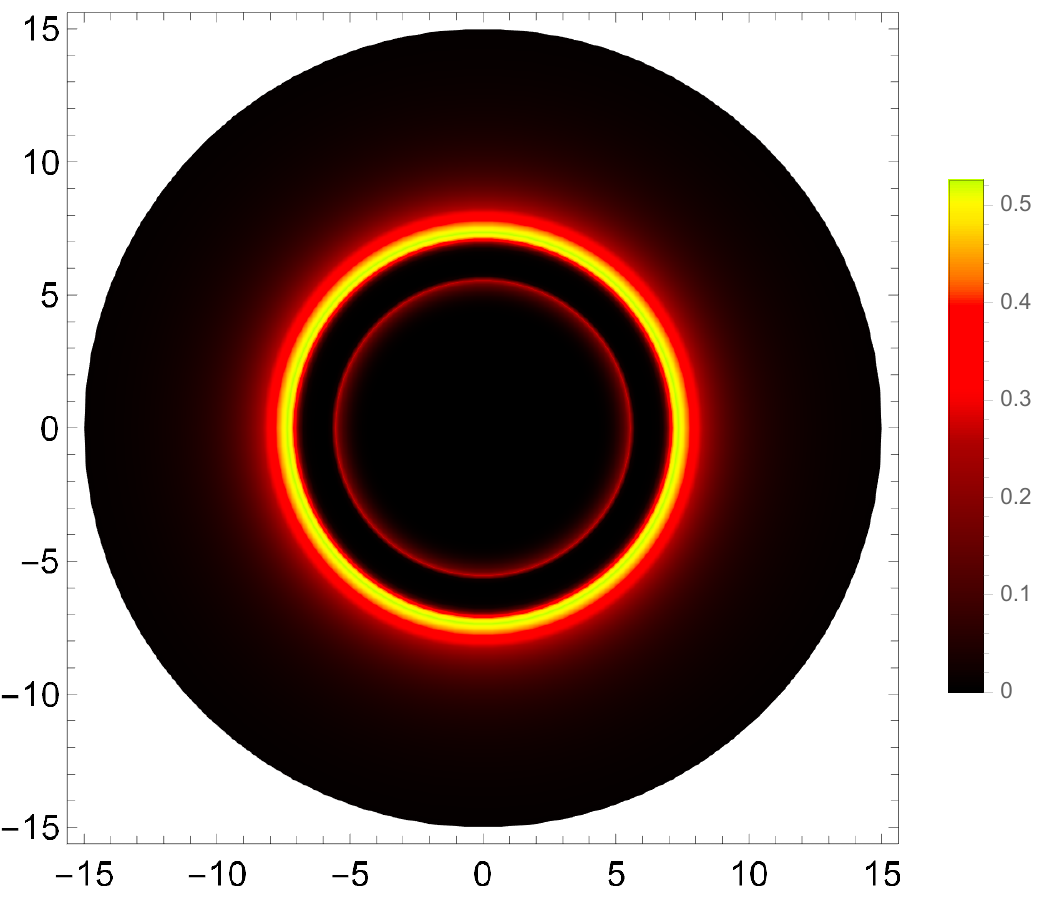}
    \includegraphics[scale=0.5]{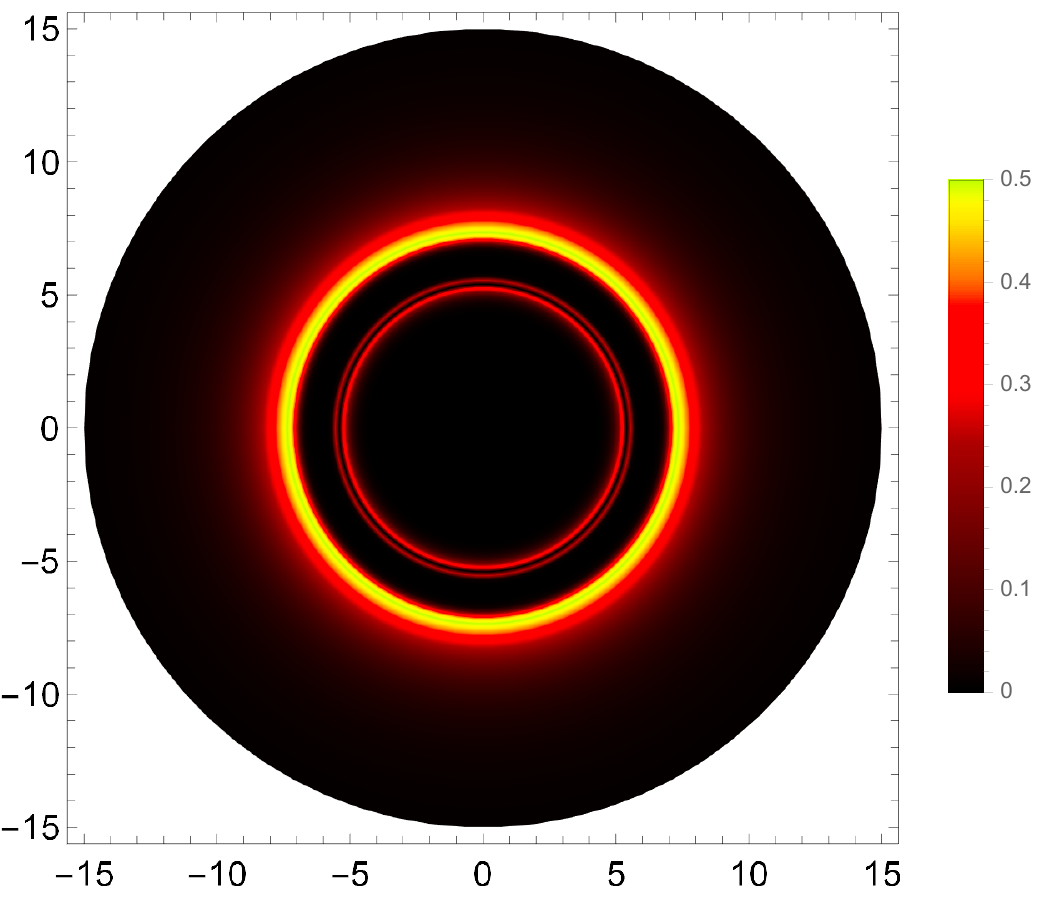}\\
    \includegraphics[scale=0.5]{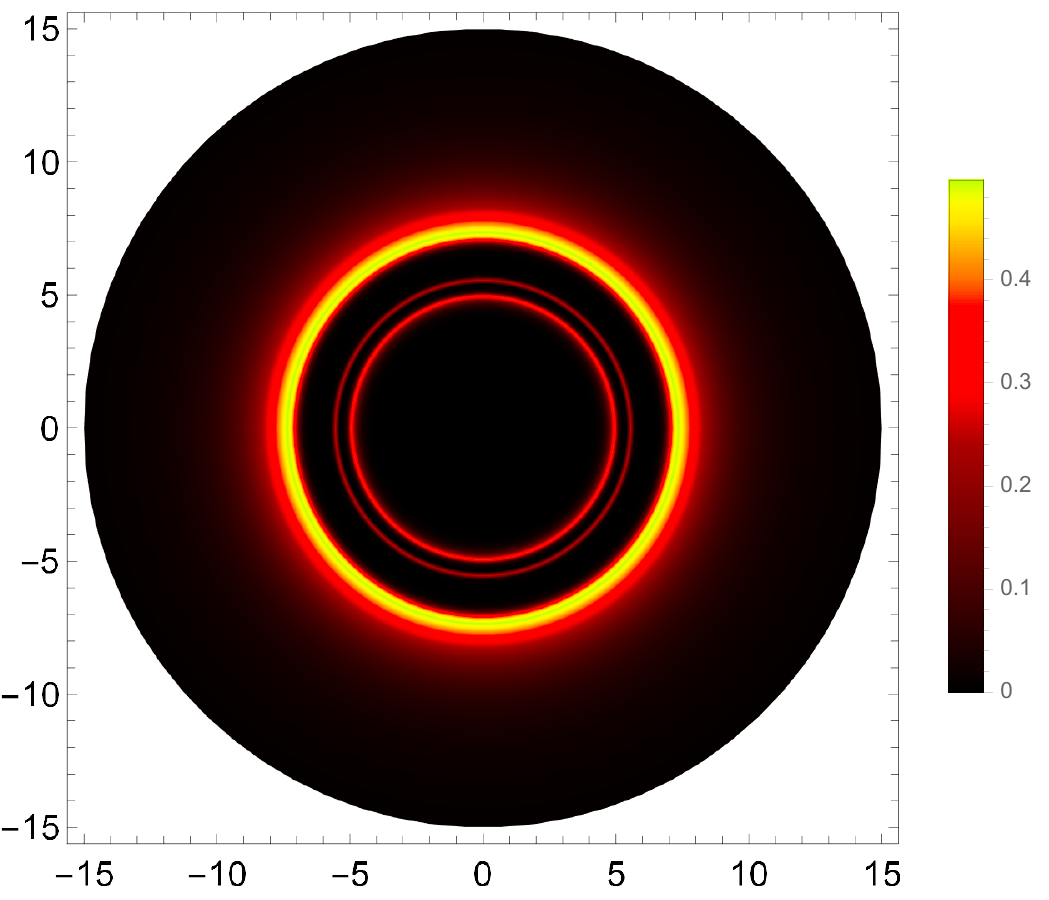}
    \includegraphics[scale=0.5]{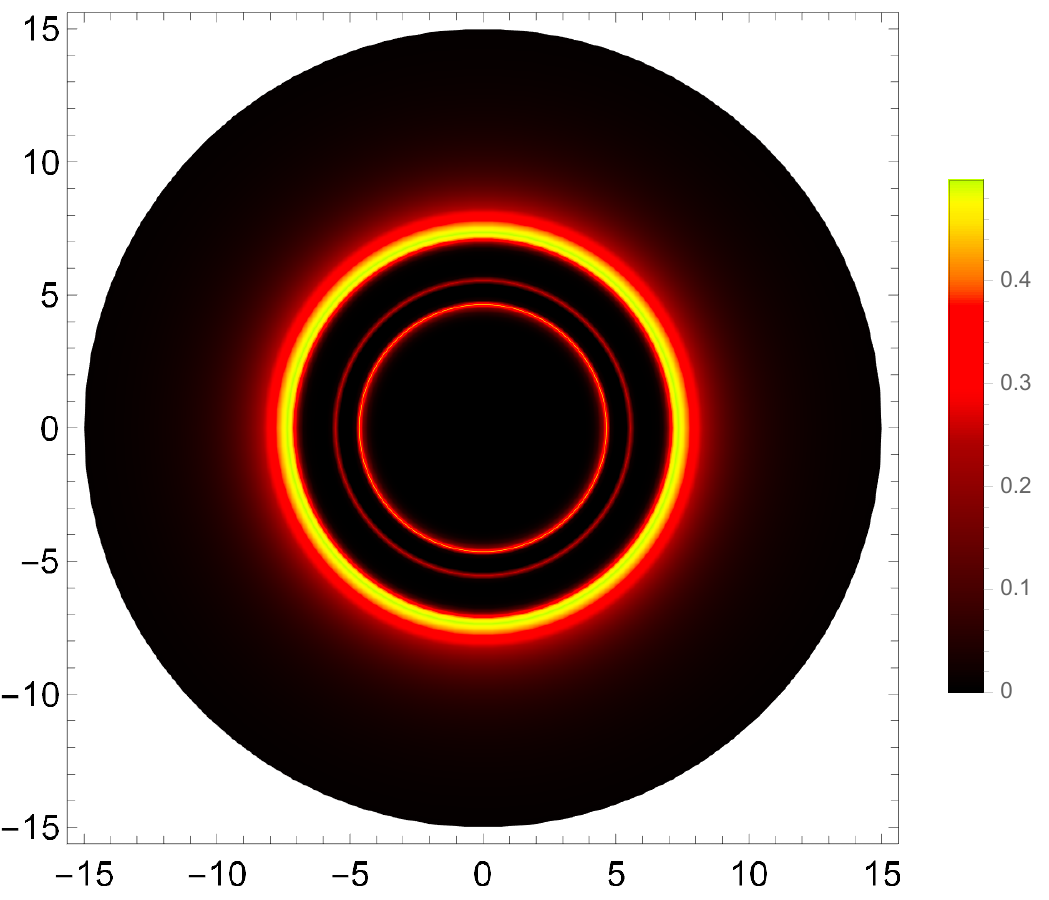}
    \includegraphics[scale=0.5]{plot_r4_rs3_ISCO.pdf}
    \caption{Shadows images with the Centre accretion disk model for configurations with $R=4M$ and $r_\Sigma=4M$ to $r_\Sigma=3M$ in steps of $0.2M$, starting in the top left corner down to the bottom right corner.}
    \label{fig:pt2_shadows_ISCO}
\end{figure*}

\begin{figure*}[h]
    \centering
    \includegraphics[scale=0.5]{plot_r4_rs4_LR.pdf}
    \includegraphics[scale=0.5]{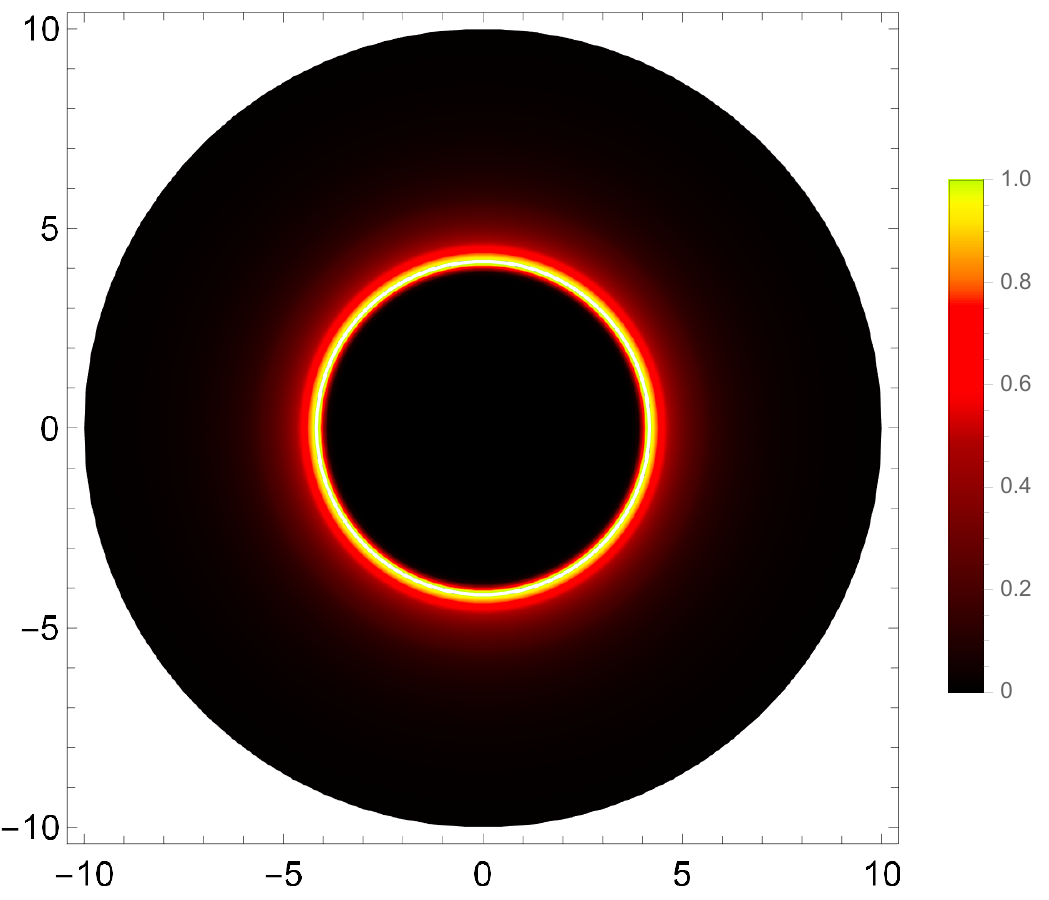}
    \includegraphics[scale=0.5]{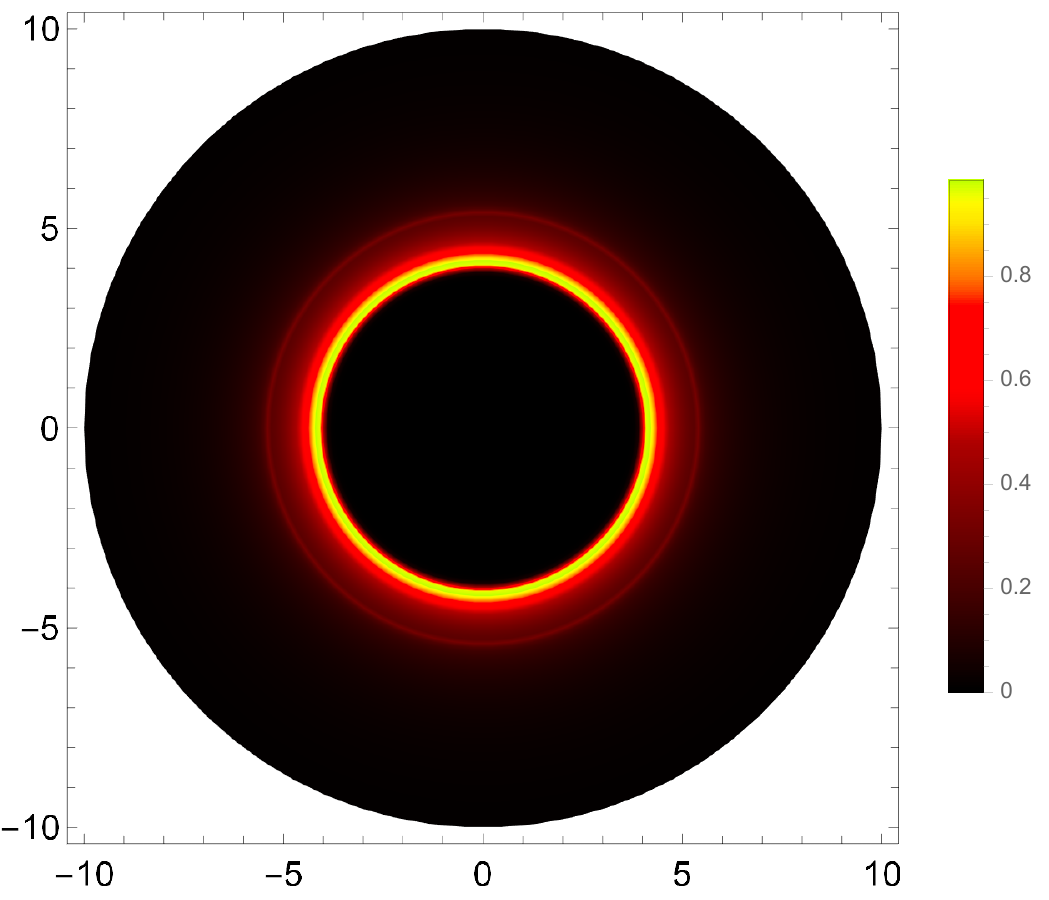}\\
    \includegraphics[scale=0.5]{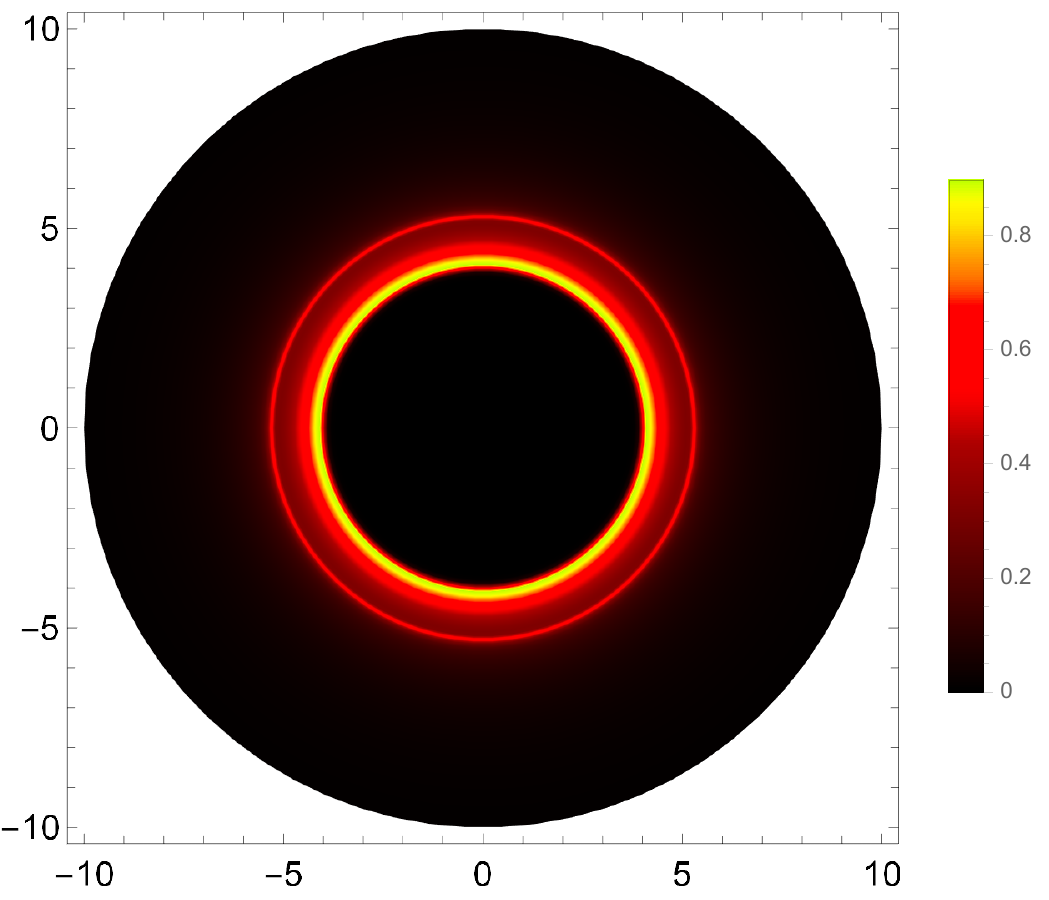}
    \includegraphics[scale=0.5]{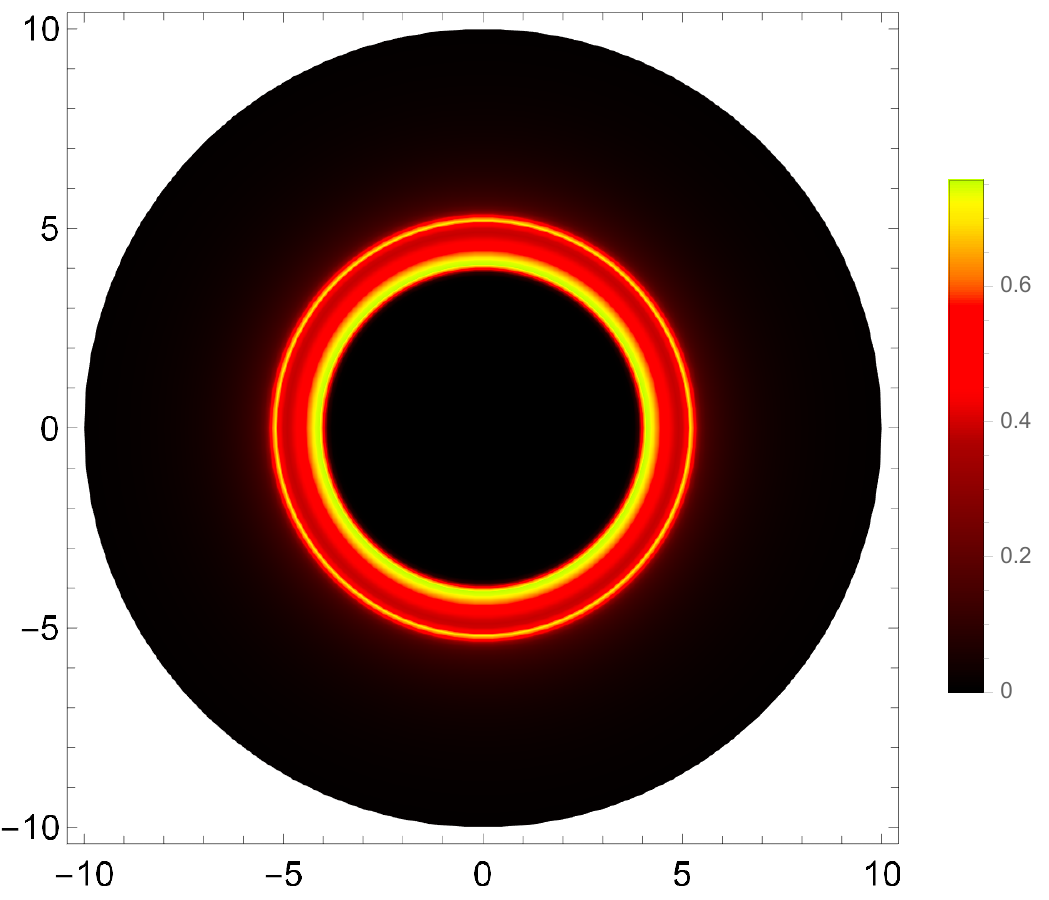}
    \includegraphics[scale=0.5]{plot_r4_rs3_LR.pdf}
    \caption{Shadows images with the Centre accretion disk model for configurations with $R=4M$ and $r_\Sigma=4M$ to $r_\Sigma=3M$ in steps of $0.2M$, starting in the top left corner down to the bottom right corner.}
    \label{fig:pt2_shadows_LR}
\end{figure*}

\begin{figure*}[h]
    \centering
    \includegraphics[scale=0.5]{plot_r4_rs4_C.pdf}
    \includegraphics[scale=0.5]{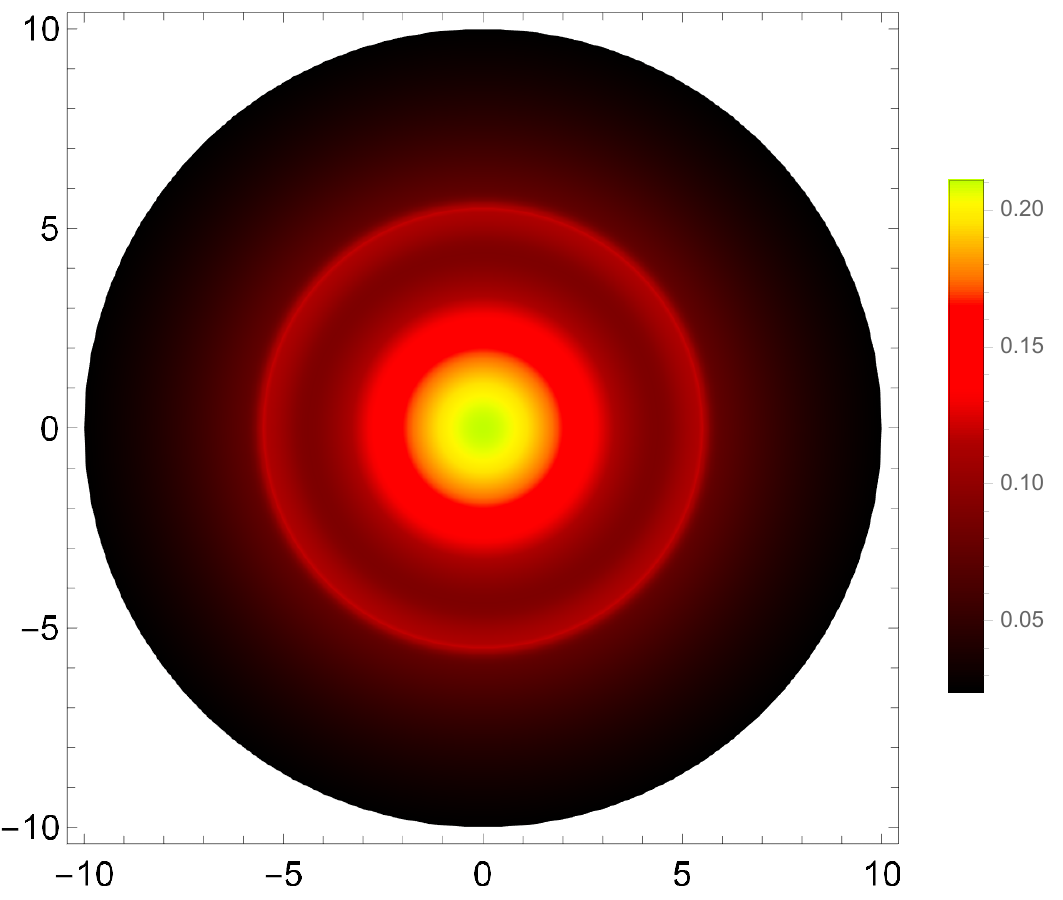}
    \includegraphics[scale=0.5]{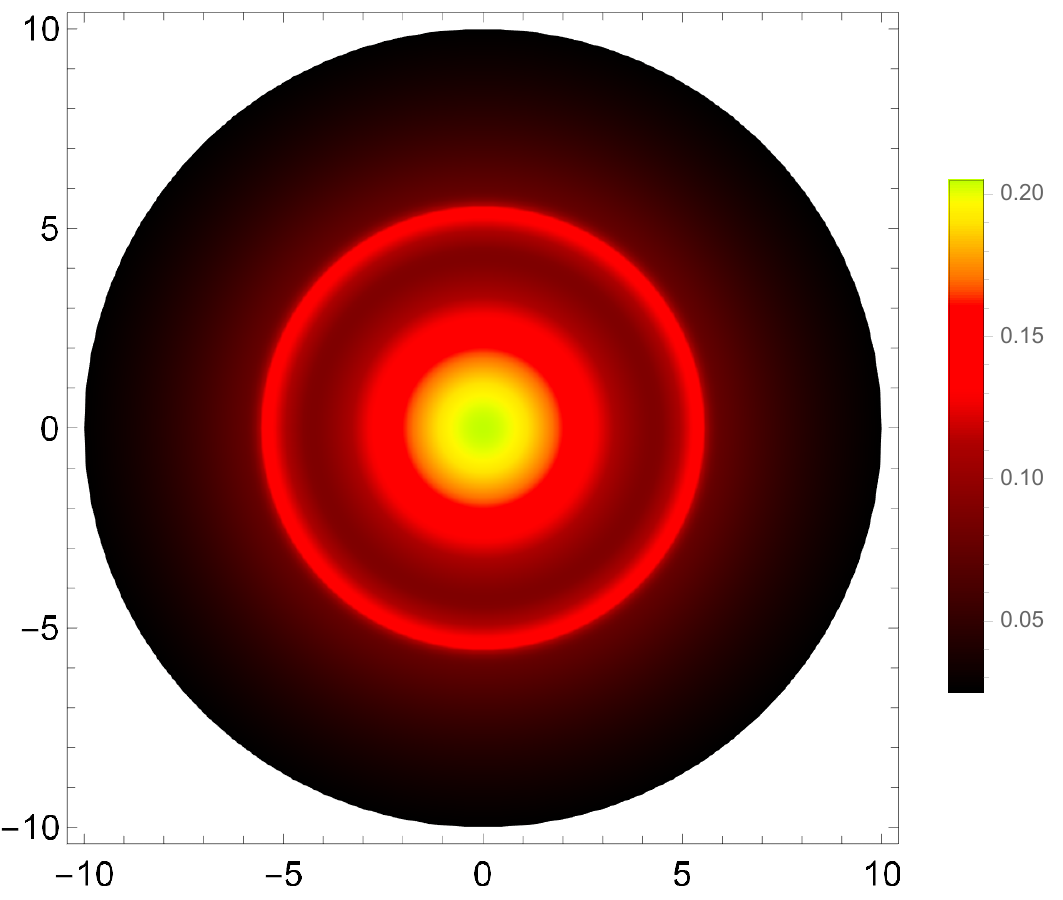}\\
    \includegraphics[scale=0.5]{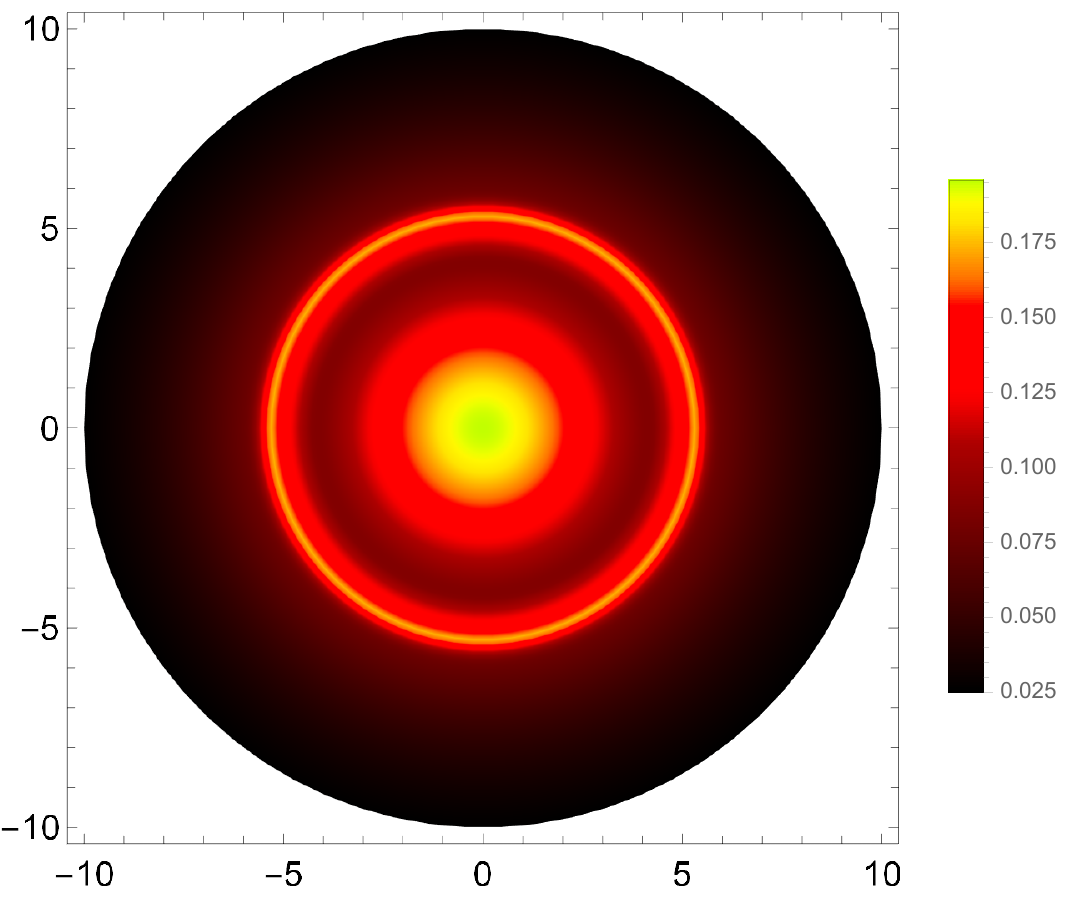}
    \includegraphics[scale=0.5]{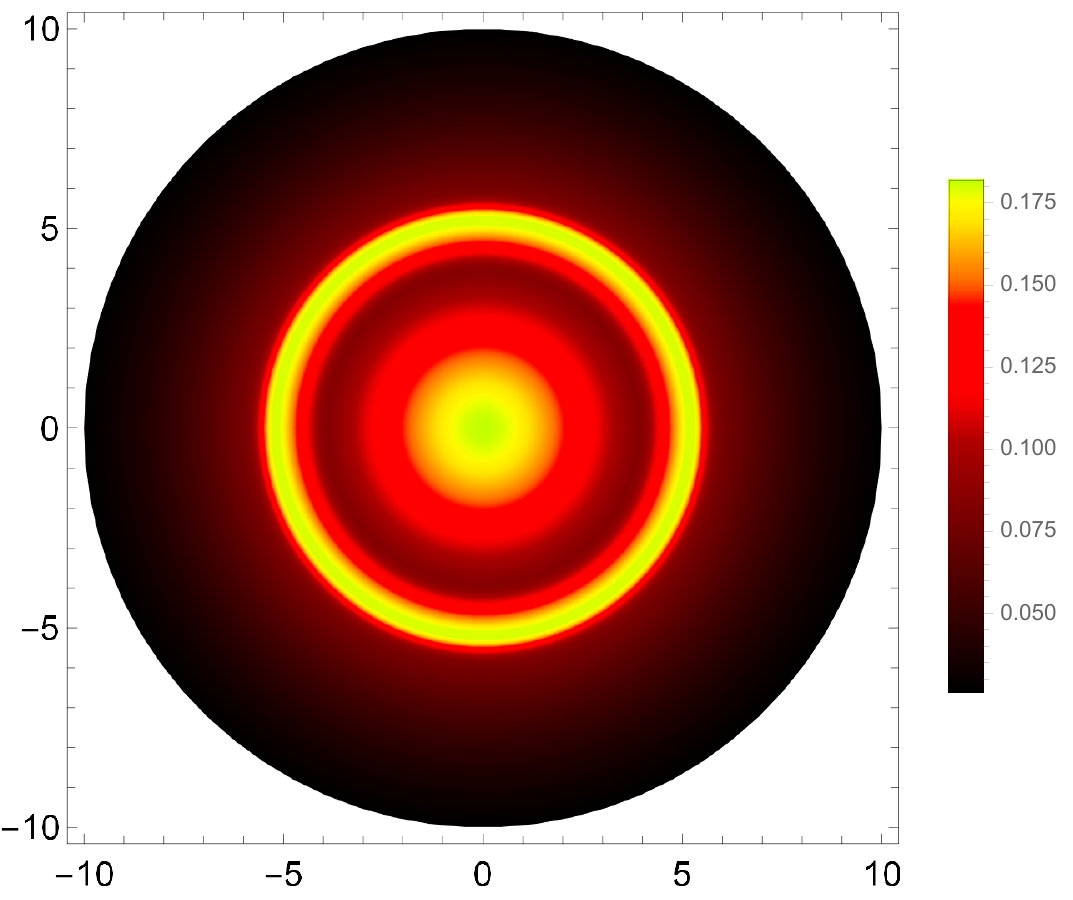}
    \includegraphics[scale=0.5]{plot_r4_rs3_C.pdf}
    \caption{Shadows images with the Centre accretion disk model for configurations with $R=4M$ and $r_\Sigma=4M$ to $r_\Sigma=3M$ in steps of $0.2M$, starting in the top left corner down to the bottom right corner.}
    \label{fig:pt2_shadows_C}
\end{figure*}

\begin{figure*}[h]
    \centering
    \includegraphics[scale=0.63]{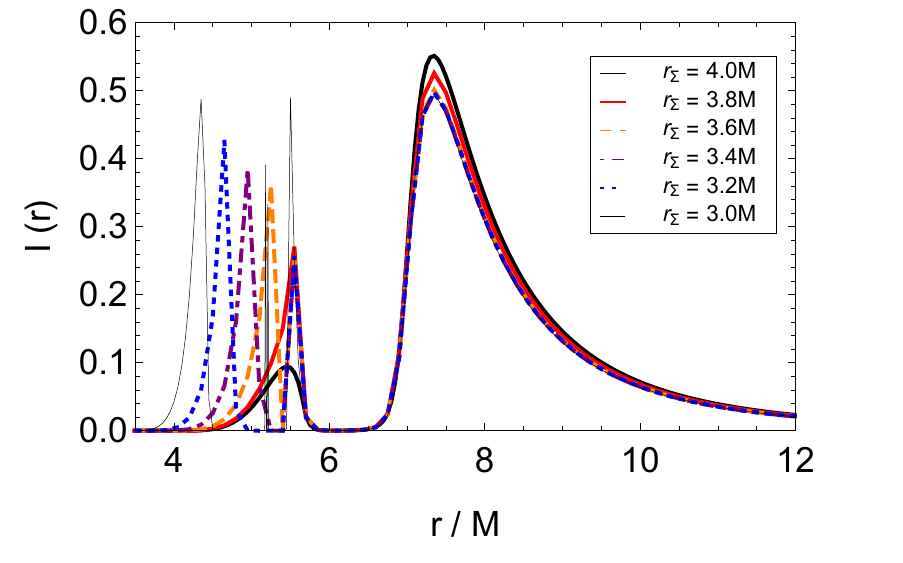}
    \includegraphics[scale=0.63]{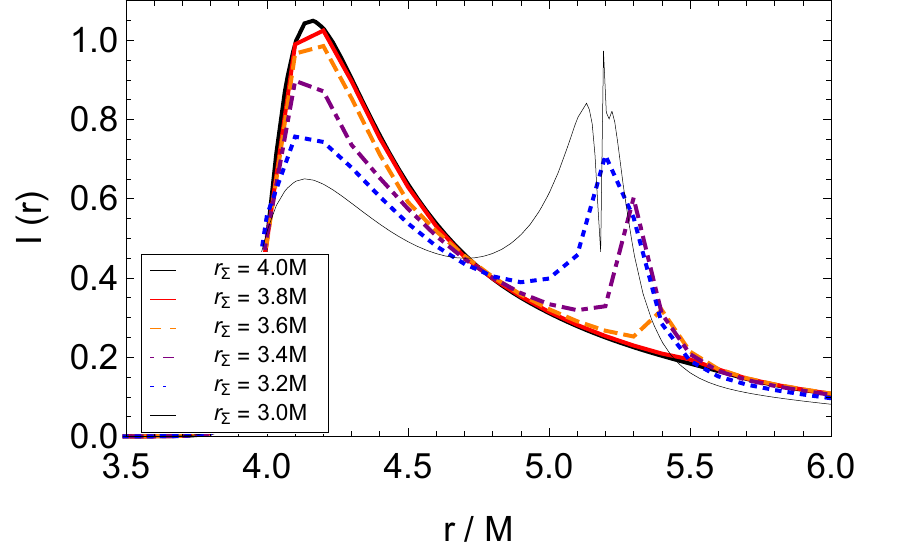}
    \includegraphics[scale=0.63]{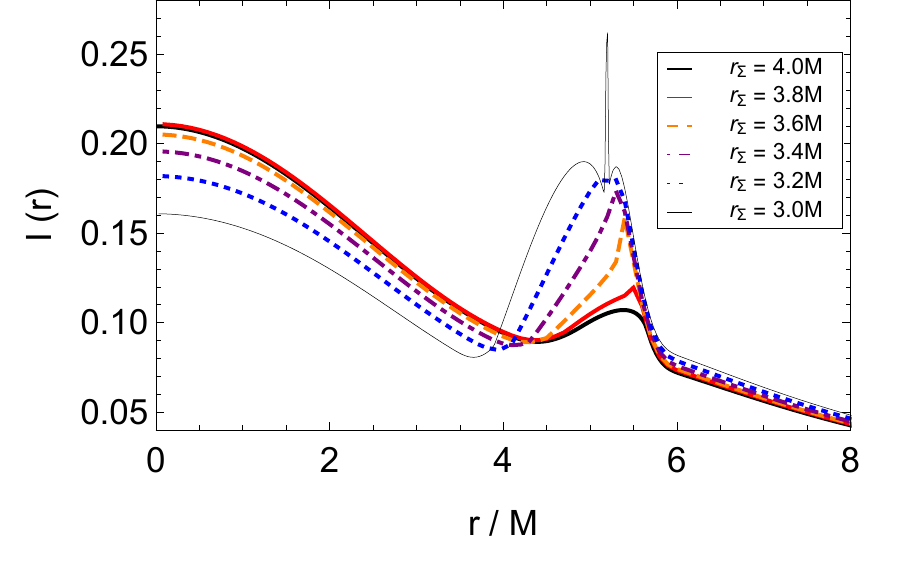}
    \caption{Observed intensity profiles $I_o$ as a function of the normalized radial coordinate $r/M$ with the ISCO (left panel), LR (middle panel), and Centre (right panel) accretion disk models, for configurations with $R=4M$ and $r_\Sigma=4M$ to $r_\Sigma=3M$ in steps of $0.2M$.}
    \label{fig:pt2_intensity}
\end{figure*}

Let us now analyze how the appearance of a LR qualitatively affects the observational properties of these configurations in the presence of a thin-shell, e.g., the transition that occurs between the configurations $S_{44}$ and $S_{43}$ in the previous section. For this purpose, we consider four extra configurations with $R=4M$ and $r_\Sigma=\{3.8M;3.6M;3.4M;3.2M\}$, and repeat the ray-tracing analysis from the previous sections. The analysis in this section can be thought of as a study on how the collapse of the exterior layers of the relativistic fluid of the star into a thin-shell of increasing mass affects their observational properties, while the inner density of the star is maintained constant. The images produced for these configurations are given in Fig. \ref{fig:pt2_shadows_ISCO} for the ISCO disk model, Fig. \ref{fig:pt2_shadows_LR} for the LR disk model, and Fig. \ref{fig:pt2_shadows_C} for the Centre disk model. The corresponding intensity profiles are given in Fig. \ref{fig:pt2_intensity}.

One observes that the effects of increasing the compacticity of these configurations by decreasing $r_\Sigma$ while maintaining $R=4M$ constant are similar to those obtained previously by decreasing $R=r_\Sigma$ simultaneously, but the magnitude of the effects, in particular the gravitational redshift, is visually smaller. Indeed, the broadening and the first splitting of the secondary image is visible for the ISCO and LR disk models, but the second split which previously occurred when the LR formed is absent. Furthermore, although there is a dimming in the primary image in the Centre disk model, the gravitational redshift is not strong enough to produce a shadow-like feature. This is caused by the uneven distribution of mass in these configurations. Indeed, the internal density $\rho$ of the configurations throughout the analysis in this section is maintained constant while it is the mass of the thin-shell that increases, as opposed to the previous analysis with $R=r_\Sigma$ for which the density of the relativistic fluid increases throughout the whole interior solution. The transition analyzed in this section can thus be thought of as an alternative intermediate step in the transformation between $S_{44}$ and $S_{33}$, which is completed by the transition analyzed in the following section.

\subsection{Development of a thin-shell in the presence of a LR}

\begin{figure*}[h]
    \centering
    \includegraphics[scale=0.5]{plot_r4_rs3_ISCO.pdf}
    \includegraphics[scale=0.5]{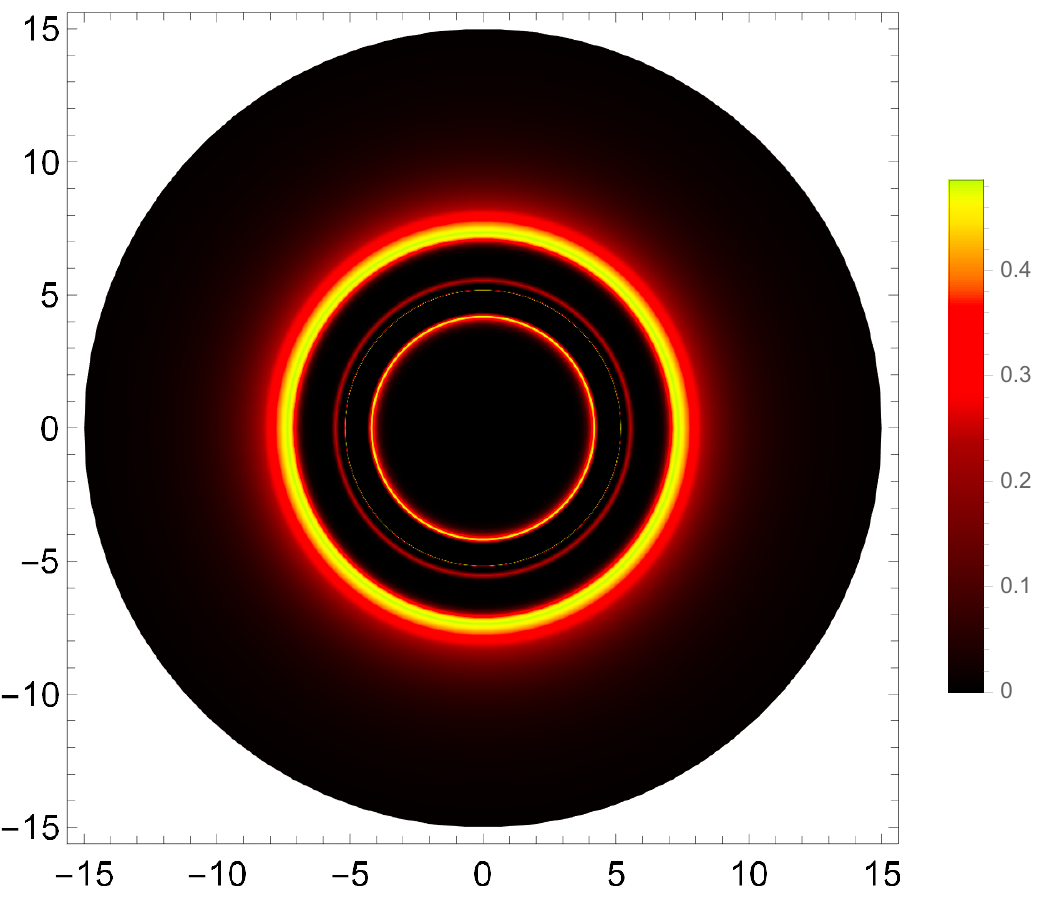}
    \includegraphics[scale=0.5]{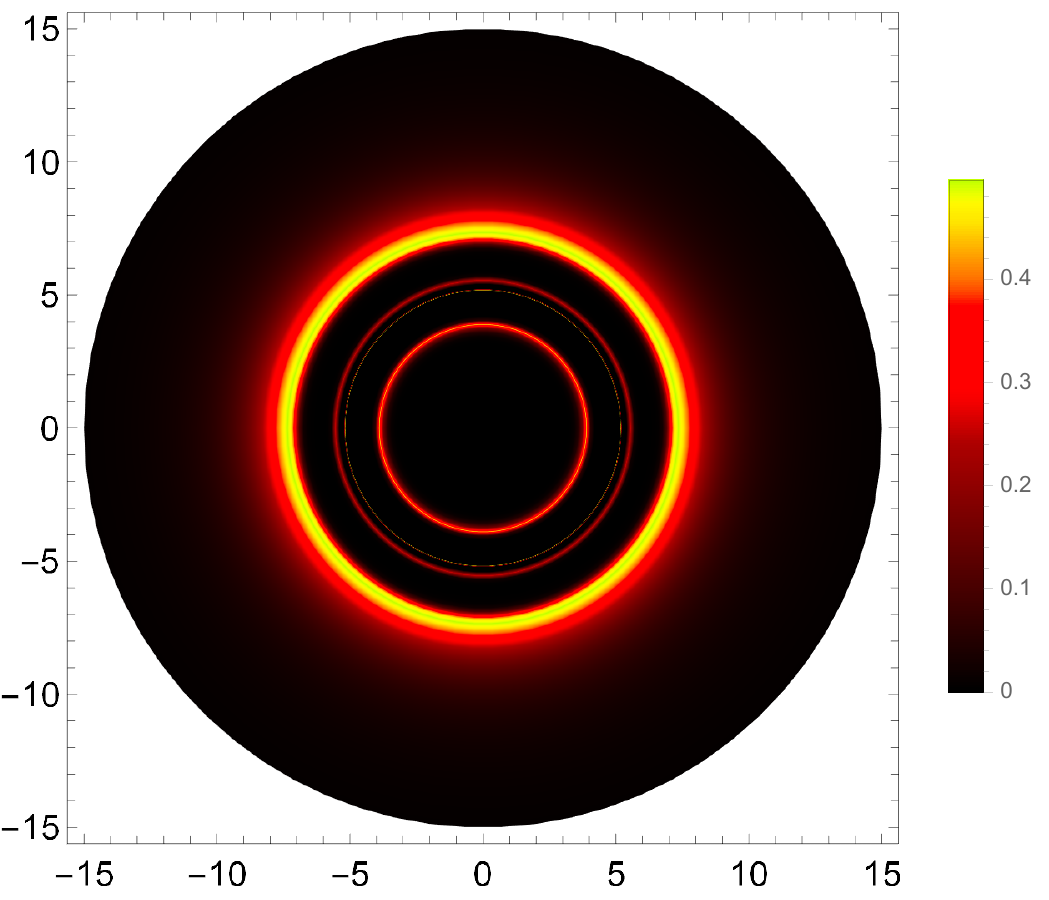}\\
    \includegraphics[scale=0.5]{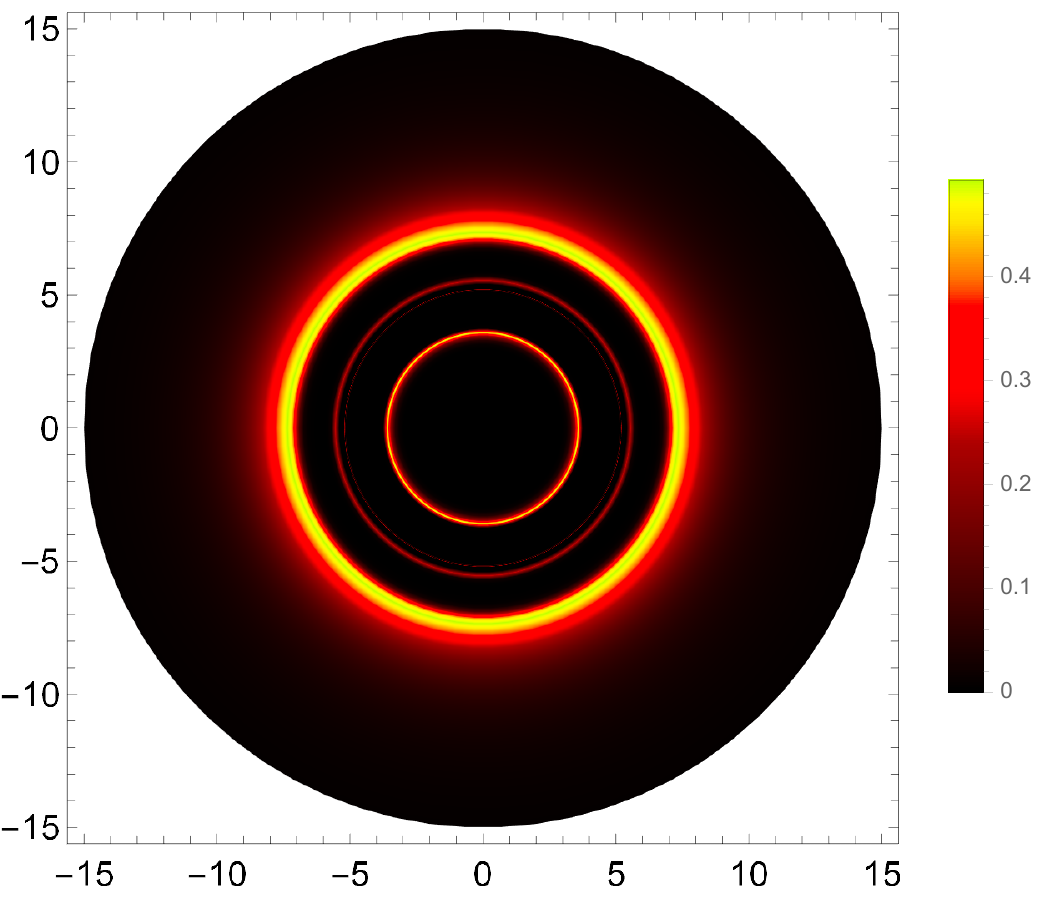}
    \includegraphics[scale=0.5]{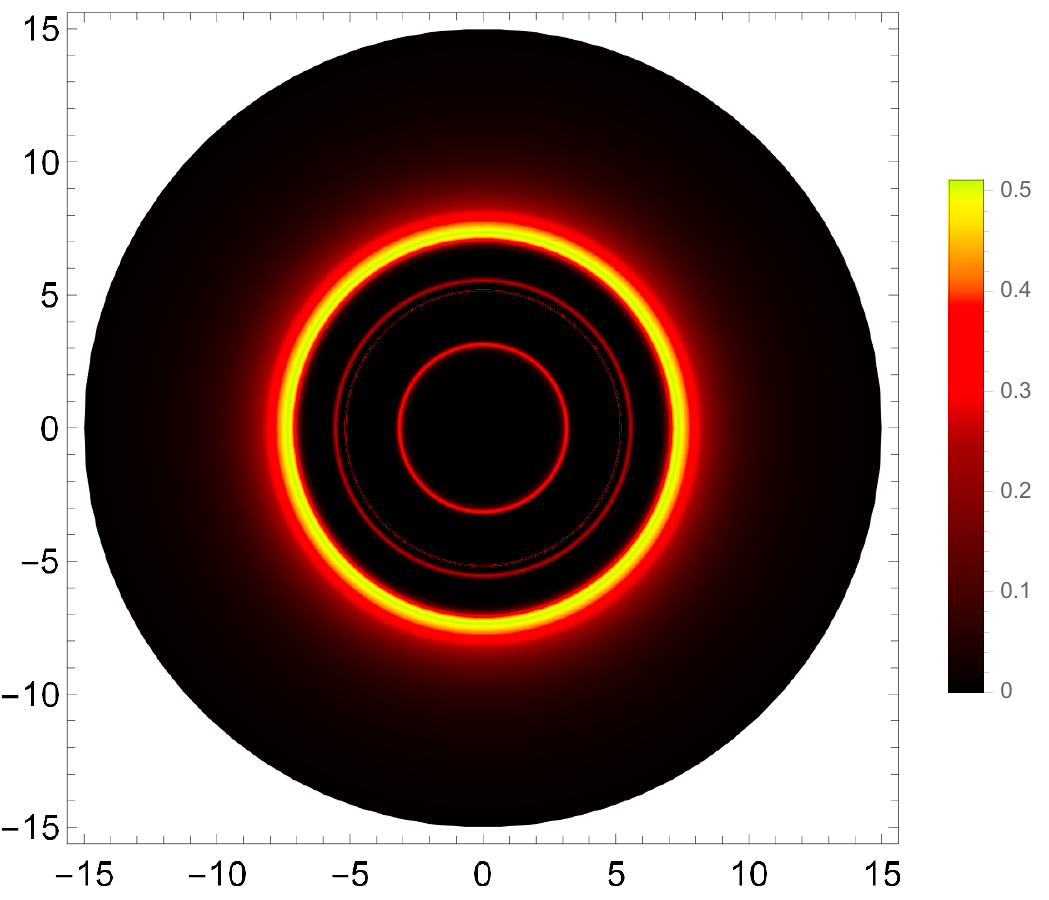}
    \includegraphics[scale=0.5]{plot_r3_rs3_ISCO.pdf}
    \caption{Shadows images with the Centre accretion disk model for configurations with $r_\Sigma=4M$ and $R=4M$ to $R=3M$ in steps of $0.2M$, starting in the top left corner down to the bottom right corner.}
    \label{fig:pt3_shadows_ISCO}
\end{figure*}

\begin{figure*}[h]
    \centering
    \includegraphics[scale=0.5]{plot_r4_rs3_LR.pdf}
    \includegraphics[scale=0.5]{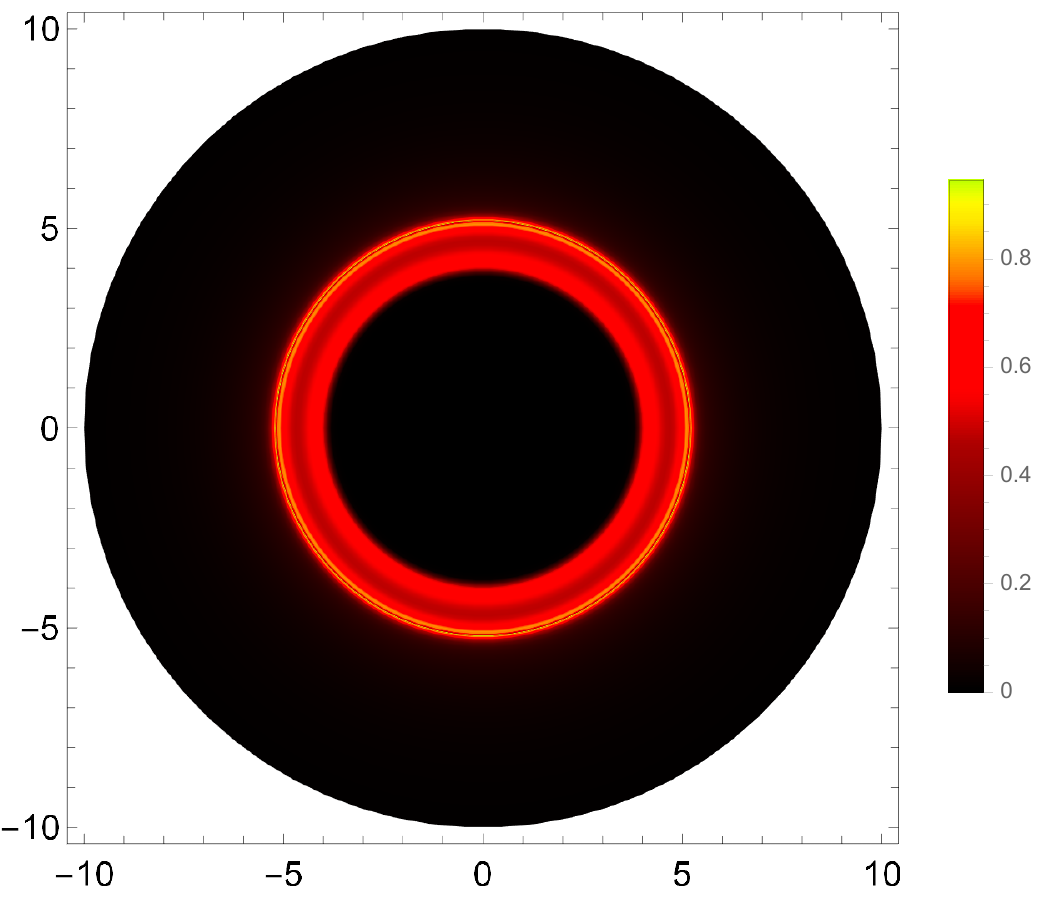}
    \includegraphics[scale=0.5]{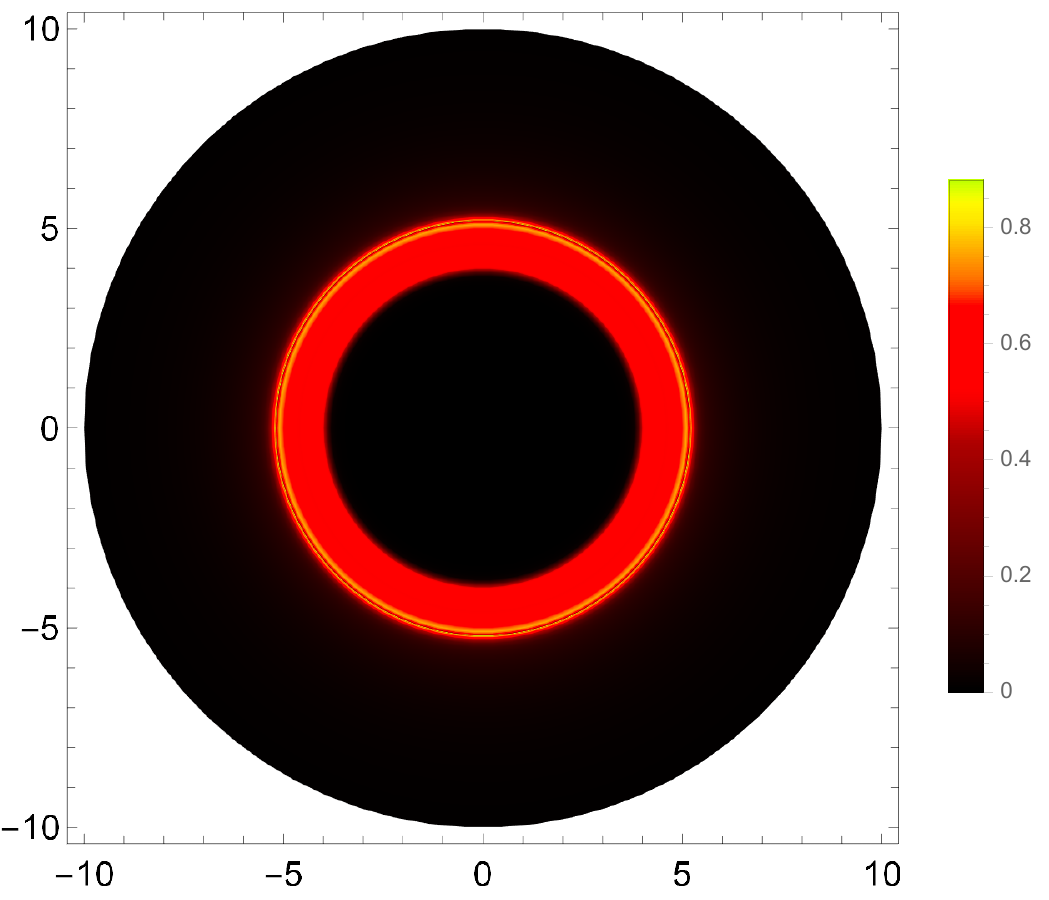}\\
    \includegraphics[scale=0.5]{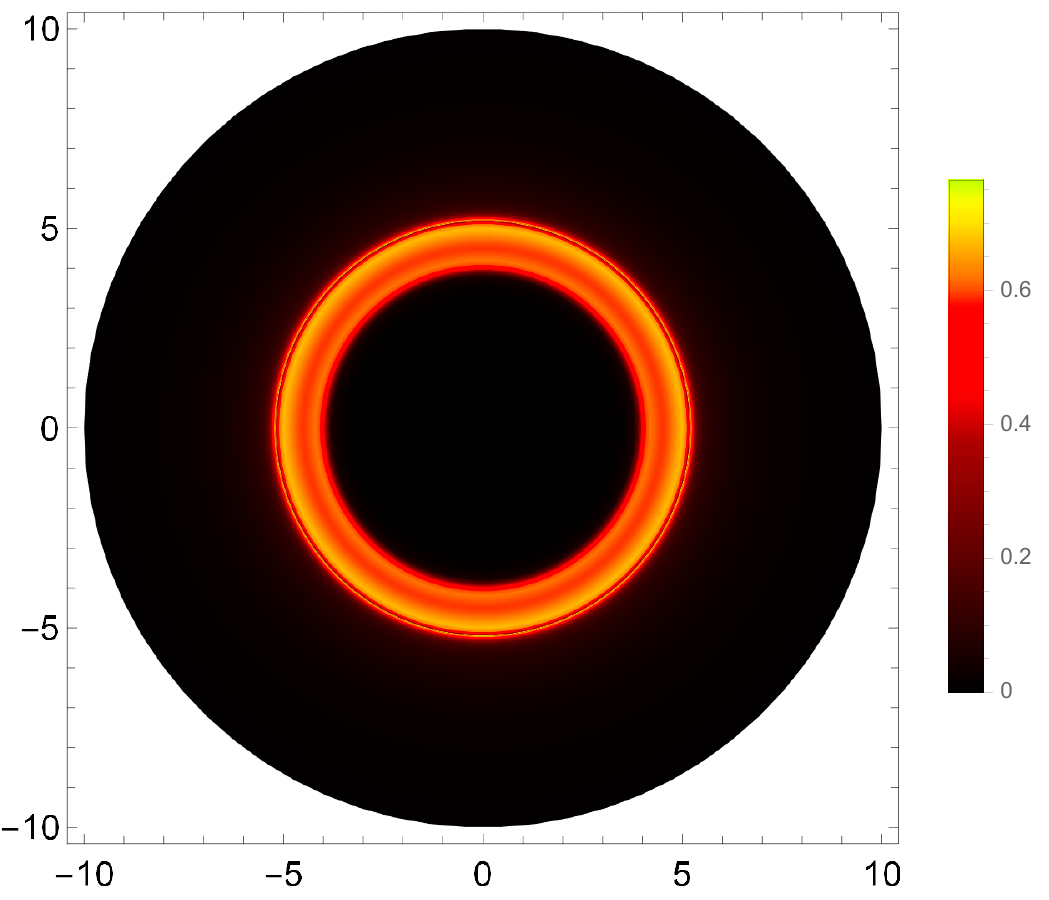}
    \includegraphics[scale=0.5]{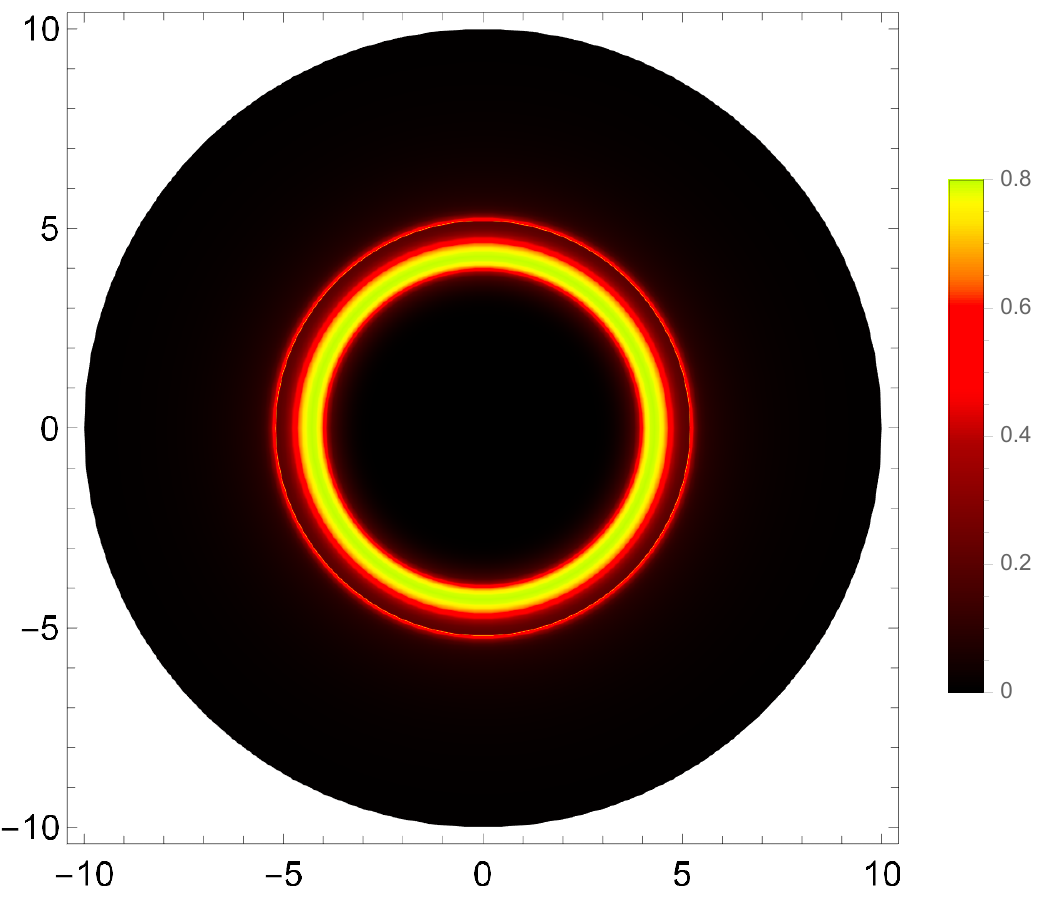}
    \includegraphics[scale=0.5]{plot_r3_rs3_LR.pdf}
    \caption{Shadows images with the Centre accretion disk model for configurations with $r_\Sigma=4M$ and $R=4M$ to $R=3M$ in steps of $0.2M$, starting in the top left corner down to the bottom right corner.}
    \label{fig:pt3_shadows_LR}
\end{figure*}

\begin{figure*}[h]
    \centering
    \includegraphics[scale=0.5]{plot_r4_rs3_C.pdf}
    \includegraphics[scale=0.5]{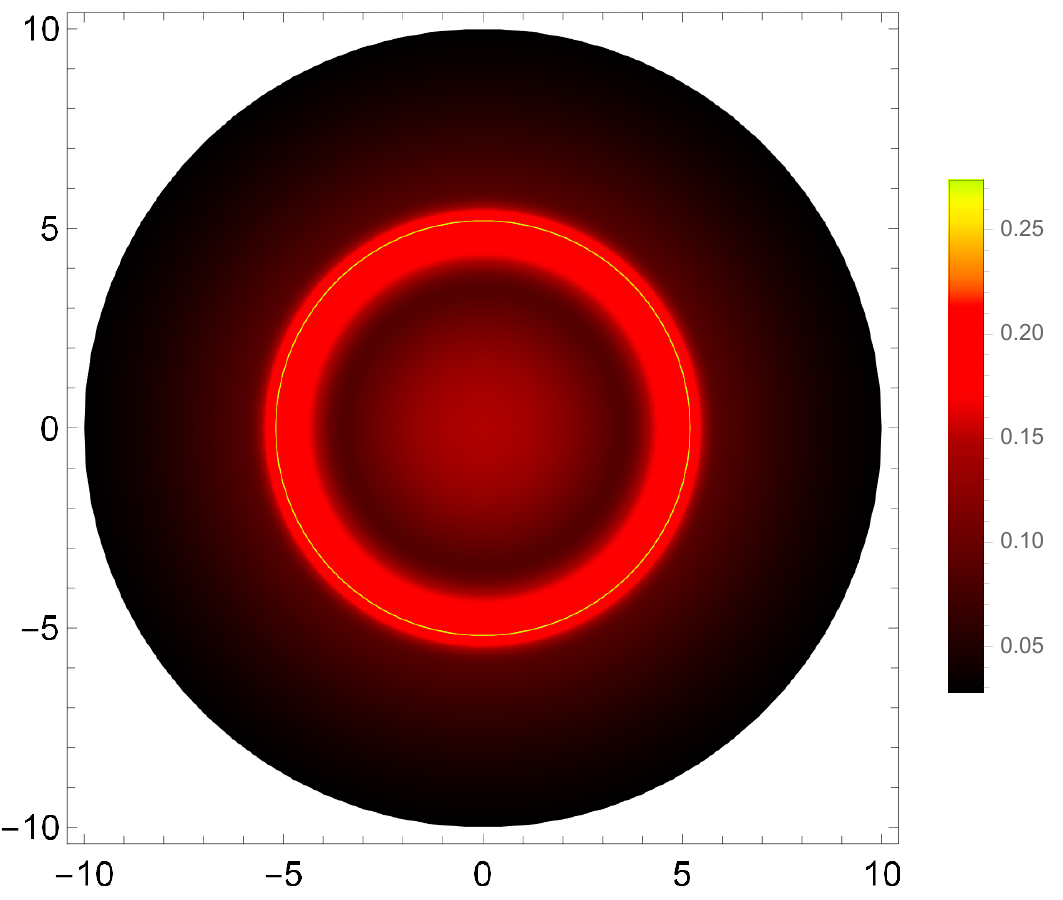}
    \includegraphics[scale=0.5]{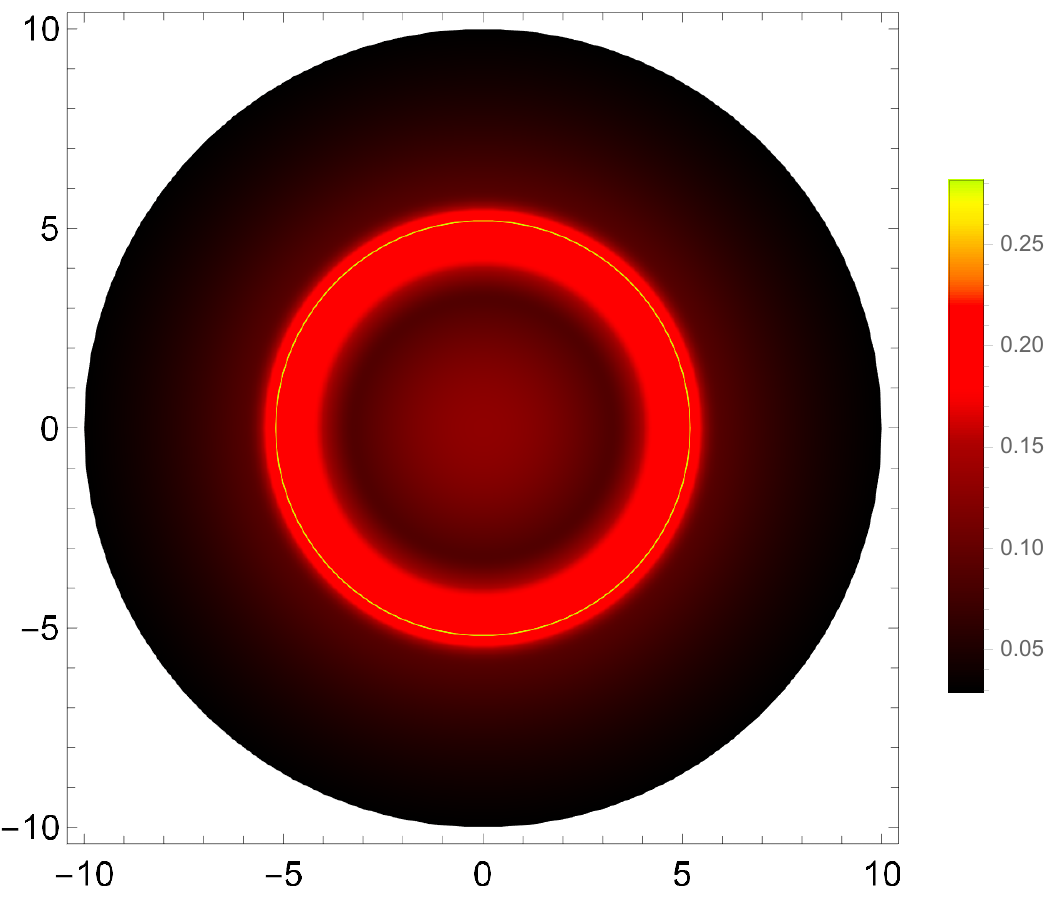}\\
    \includegraphics[scale=0.5]{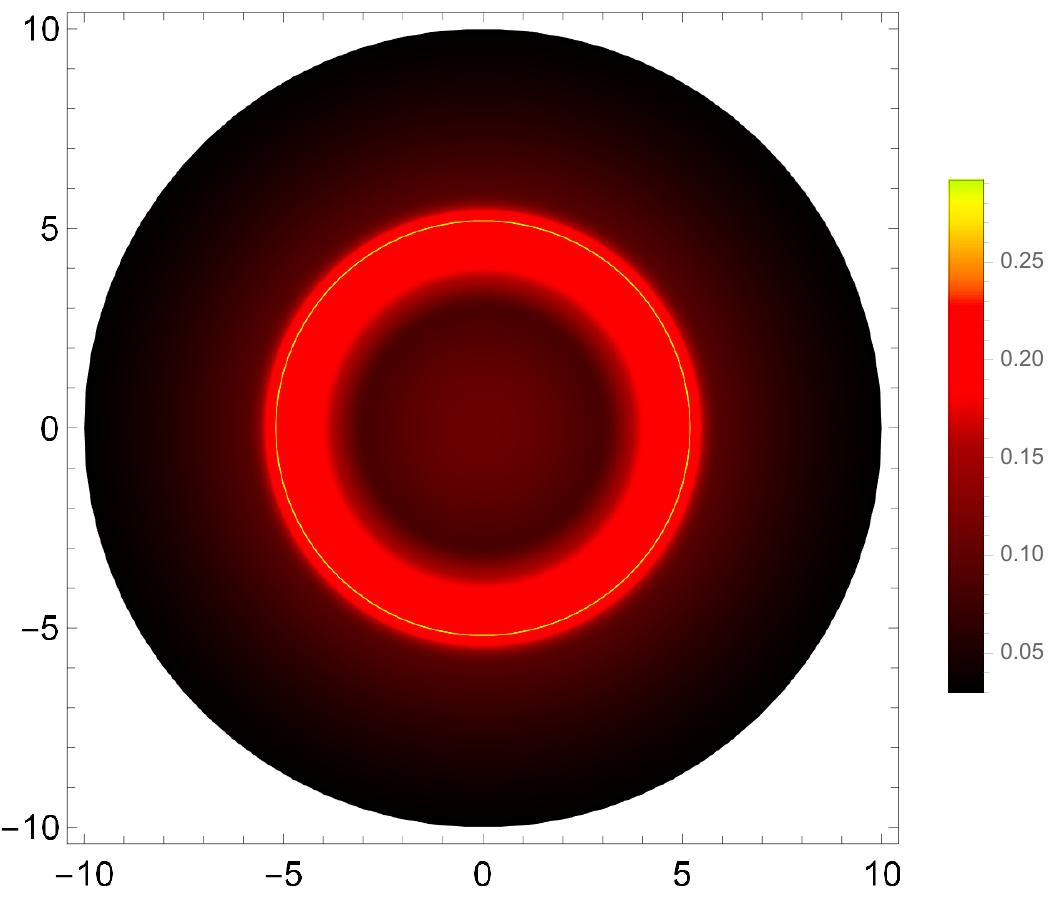}
    \includegraphics[scale=0.5]{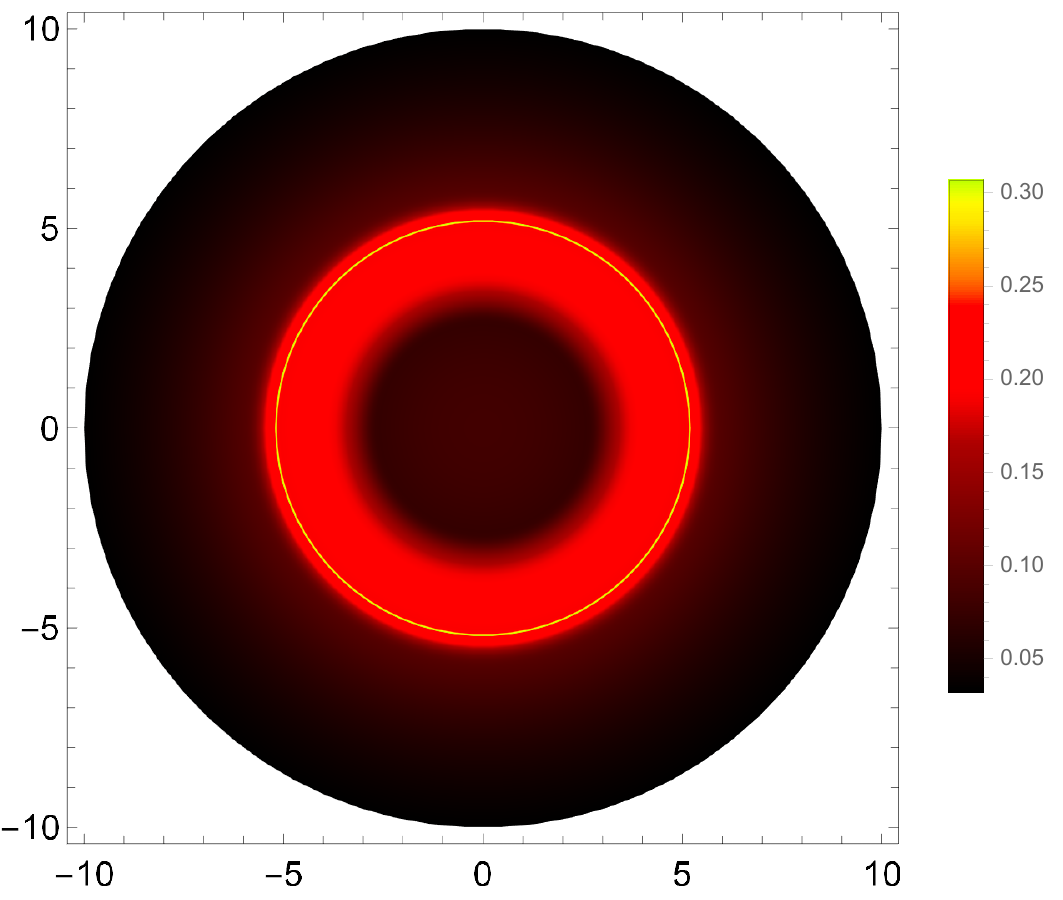}
    \includegraphics[scale=0.5]{plot_r3_rs3_C.pdf}
    \caption{Shadows images with the Centre accretion disk model for configurations with $r_\Sigma=4M$ and $R=4M$ to $R=3M$ in steps of $0.2M$, starting in the top left corner down to the bottom right corner.}
    \label{fig:pt3_shadows_C}
\end{figure*}

\begin{figure*}[h]
    \centering
    \includegraphics[scale=0.63]{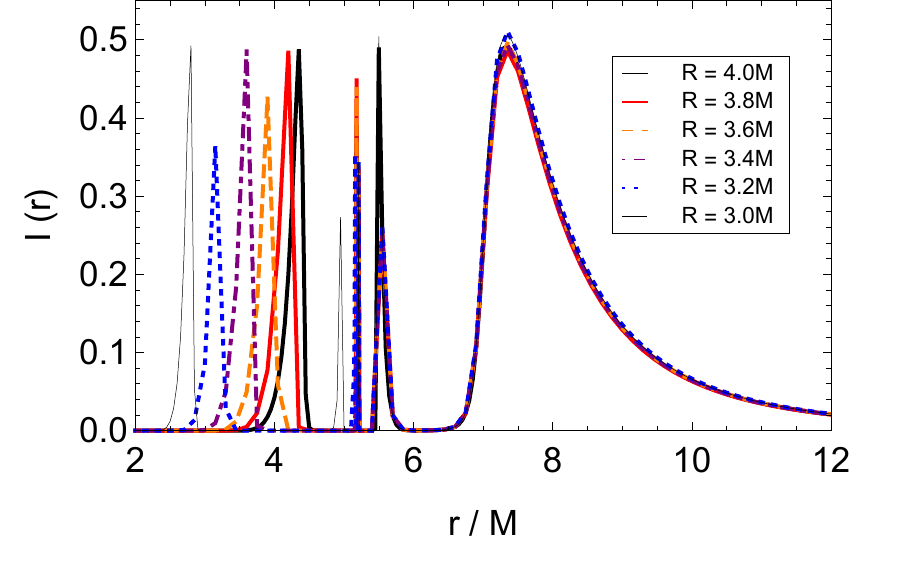}
    \includegraphics[scale=0.63]{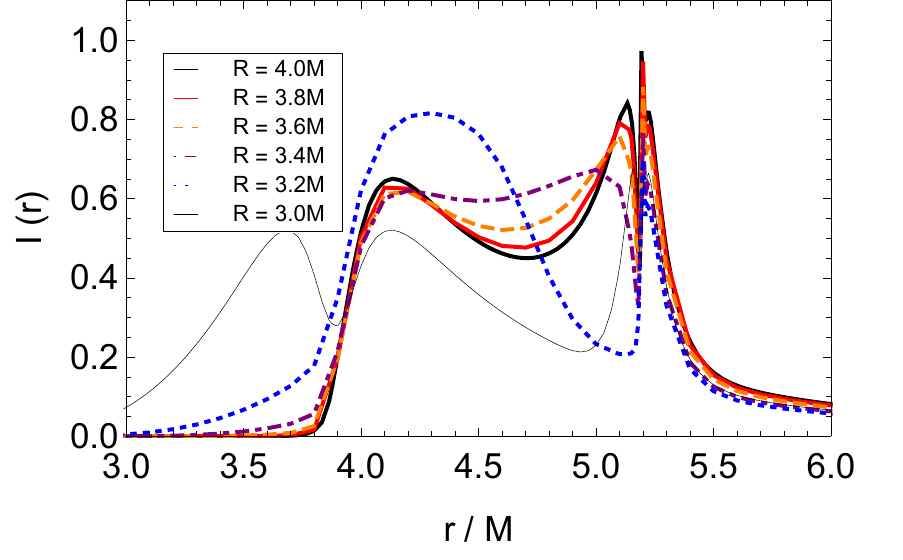}
    \includegraphics[scale=0.63]{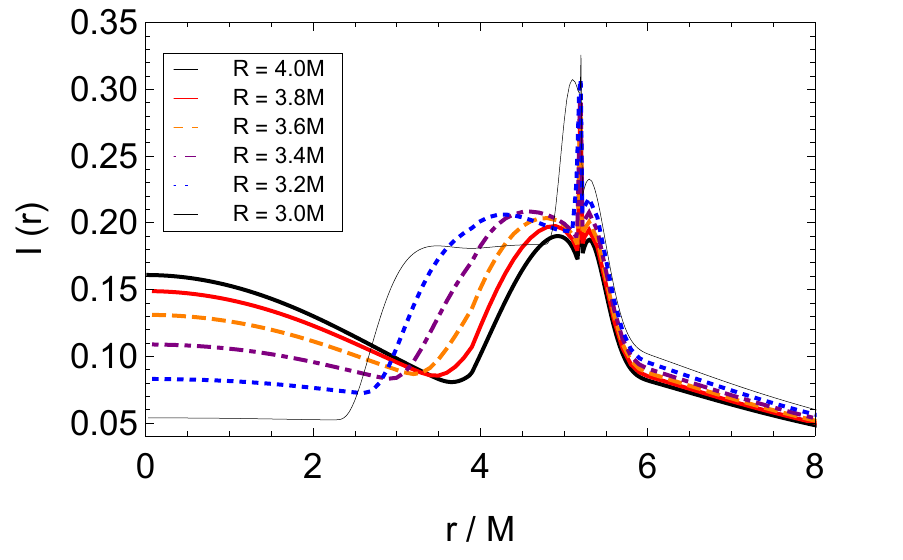}
    \caption{Observed intensity profiles $I_o$ as a function of the normalized radial coordinate $r/M$ with the ISCO (left panel), LR (middle panel), and Centre (right panel) accretion disk models, for configurations with $r_\Sigma=4M$ and $R=4M$ to $R=3M$ in steps of $0.2M$.}
    \label{fig:pt3_intensity}
\end{figure*}

Finally, we now analyze how the appearence of a thin-shell affects the observational properties of our configurations in the presence of a LR, i.e., the transition that occurs between the configurations $S_{43}$ and $S_{33}$. To do it, we consider four new configurations with $r_\Sigma=3M$ and $R=\{3.8M;3.6M;3.4M;3.2M\}$, and again repeat the ray-tracing analysis of the previous sections. The analysis in this sections corresponds to a study on how the transfer of mass from the thin-shell to the interior fluid affects the observational properties of the star, while the mass and compacticity of the solutions are maintained constant. The images produced for these configurations are given in Fig. \ref{fig:pt3_shadows_ISCO} for the ISCO disk model, Fig. \ref{fig:pt3_shadows_LR} for the LR disk model, and Fig. \ref{fig:pt3_shadows_C} for the Centre disk model. The corresponding intensity profiles are given in Fig. \ref{fig:pt3_intensity}.

In the previous section, we stated that the analyzed transition between $S_{44}$ and $S_{43}$ could be taken as an alternative intermediate route for the transition between $S_{44}$ and $S_{33}$. The results of this section confirm this statement in the sense that as the radius $R$ decreases, the remaining qualitative changes observed in the transition between $S_{44}$ and $S_{33}$ but not observed in the transition between $S_{44}$ and $S_{43}$ finally occur: a third secondary image appears at $R=3.2M$ for the ISCO and LR disk models, and the effects of gravitational redshift are strong enough to significantly dim the primary image in the Centre disk model. The results of this section clarify the statement made previously regarding how the distribution of mass in the star affects its observational properties. Indeed, as one decreases the mass of the star and consequently increases the density of the inner fluid, the effects of the gravitational redshift increase until the intensity of the primary image is negligible in comparison to the other components, producing a shadow-like feature.


\end{document}